%% file: Revised_Manuscript__Clean_Version_.tex
\documentclass[3p,times]{elsarticle}
\usepackage{amsthm}
\usepackage{amsmath}
\usepackage{cases}
\usepackage{amscd}
\usepackage{amsfonts}

\usepackage{amssymb}
\usepackage{graphicx}
\usepackage{float}
\usepackage{booktabs}
\usepackage{threeparttable}
\makeatletter
\long\def\papertable@makecaption#1#2{%
  \vskip\abovecaptionskip
  {\small\raggedright #1: #2\par}%
  \vskip\belowcaptionskip}
\newcommand{\papertablestyle}{%
  \small\setlength{\tabcolsep}{4pt}\renewcommand{\arraystretch}{1.10}%
  \let\@makecaption\papertable@makecaption}
\makeatother

\usepackage[colorlinks=true, linkcolor=blue, citecolor=blue, urlcolor=blue]{hyperref}
\usepackage{cleveref}
\crefname{figure}{Fig.}{Figs.}
\Crefname{figure}{Fig.}{Figs.}
\usepackage{color}
\allowdisplaybreaks
\makeatletter
\newcommand\figcaption{\def\@captype{figure}\caption}
\newcommand\tabcaption{\def\@captype{table}\caption}
\makeatother
\usepackage{multirow}
\usepackage{mathrsfs}
\biboptions{sort}

\newcommand{\beq}{
\begin{equation}}
  \newcommand{\eeq}{
\end{equation}}
\newcommand{\bea}{
\begin{eqnarray}}
  \newcommand{\eea}{
\end{eqnarray}}
\newcommand{\beas}{
\begin{eqnarray*}}
  \newcommand{\eeas}{
\end{eqnarray*}}

\newcommand{\uvec}{\boldsymbol{u}}
\newcommand{\xvec}{\boldsymbol{x}}
\newcommand{\fvec}{\boldsymbol{f}}

\newcommand{\Dt}{\Delta T}

\newcommand{\Retau}{Re_{\tau}}
\newcommand{\Rey}{Re}
\providecommand{\sep}{\unskip\hspace{0.5em};\hspace{0.5em}}

\RequirePackage{xcolor}
\AtBeginDocument{\setlength{\emergencystretch}{4em}}

\usepackage{tikz}
\usetikzlibrary{arrows.meta,positioning,calc}
\usepackage{algorithm,algpseudocode}
\usepackage{placeins,needspace}

\usepackage{array,tabularx}

\AtBeginDocument{\hypersetup{citecolor=.,linkcolor=.,urlcolor=.}}

\begin{document}

\begin{frontmatter}

\title{Long-time turbulence prediction at fine time steps using multi-stride neural operators}

\author[aff1,aff2]{Guanyu Pan}
\ead{202421511213@smail.xtu.edu.cn}

\author[aff1]{Huiyu Yang}
\ead{12432341@mail.sustech.edu.cn}

\author[aff1]{Yunpeng Wang}
\ead{wangyp@sustech.edu.cn}

\author[aff3]{Zikun Xu}
\ead{25087988g@connect.polyu.hk}

\author[aff2]{Nianyu Yi\corref{cor1}}
\ead{yinianyu@xtu.edu.cn}

\author[aff1]{Jianchun Wang\corref{cor1}}
\ead{wangjc@sustech.edu.cn}

\cortext[cor1]{Corresponding authors.}

\affiliation[aff1]{organization={Department of Mechanics and Aerospace Engineering, Southern University of Science and Technology},
            city={Shenzhen},
            postcode={518055},
            country={China}}

\affiliation[aff2]{organization={School of Mathematics and Computational Science, Xiangtan University},
            city={Xiangtan},
            postcode={411105},
            country={China}}

\affiliation[aff3]{organization={Department of Applied Mathematics, The Hong Kong Polytechnic University},
            city={Hong Kong},
            postcode={999077},
            country={China}}

\begin{abstract}
Long-time turbulence prediction at fine time steps aims to capture the temporal evolution of the flow while reproducing long-time flow statistics. However, fine time steps require a large number of autoregressive steps, increasing computational cost and allowing prediction errors to accumulate. Moreover, accurate short-term field predictions do not guarantee accurate long-time statistics. To address these difficulties, we propose a multi-stride neural operator for three-dimensional turbulence prediction, which uses implicit factorized Transformer (IFF) predictors as its backbone and adds a factorized-attention cross-stride fusion (FA-CSF) module. All predictors take the same current velocity field as input, and the fusion module reuses their velocity outputs and latent features to generate velocity fields at intermediate time steps through interpolation and learned correction. The corrected velocity field at the largest stride is fed back as the input for the next autoregressive step, decoupling the autoregressive step from the output time step. In the first training stage, stride-conditioned IFF (SC-IFF) predictors are trained at power-of-two strides using three stride-conditioning strategies. In the second stage, the predictors are frozen and the fusion module is trained to refine long autoregressive rollouts using both field-data and statistical losses. Numerical experiments on turbulent channel flow and forced homogeneous isotropic turbulence (HIT) assess long-time field predictions, flow statistics, and computational cost. The results show improved long-time prediction performance in both channel flow and HIT with multi-stride prediction and further improvements in several statistics after refinement. At a sampling interval of 10 direct numerical simulation (DNS) time steps, the relative $L^2$ errors of fixed-stride IFF exceed 100\% for the root-mean-square (RMS) velocity fluctuation profiles and Reynolds shear stress profile in channel flow, and for the temporal evolution of the velocity and vorticity RMS values and the time-averaged energy spectrum in HIT. In contrast, SC-IFF with independent stride-specific predictors and FA-CSF refinement keeps the long-time statistics close to the filtered DNS data over the prediction times considered. Relative to fixed-stride IFF, it reduces the errors in the RMS velocity fluctuation profiles and Reynolds shear stress profile by more than 98\% in channel flow and the error in the time-averaged energy spectrum by more than 99\% in HIT. Multi-stride prediction with FA-CSF also accelerates trajectory generation at fine time steps, running 3.69 times faster than autoregressive prediction with fixed-stride IFF and 2.41 times faster than hierarchical multi-stride prediction. These findings demonstrate the potential of multi-stride prediction and trajectory refinement for efficient long-time turbulence prediction at fine time steps.
\end{abstract}


\end{frontmatter}

\renewcommand{\theHequation}{\thesection.\arabic{equation}}
\section{Introduction}\label{sec:introduction}

Turbulent flows exhibit strong multi-scale interactions in both space and time, and high-fidelity simulations, including direct numerical simulation (DNS) and large-eddy simulation (LES), remain computationally expensive for long-time integrations and large parameter sweeps \citep{Pope2000TurbulentFlows}.
These costs motivate data-driven surrogates that seek to accelerate spatiotemporal prediction while retaining coherent structures and flow statistics \citep{BruntonNoackKoumoutsakos2020MLFluid,DuraisamyIaccarinoXiao2019AgeData,BeckKurz2021Perspective}.
Surrogates that predict at fine time steps are particularly important for tasks requiring frequent updates of the flow state, such as Lagrangian particle tracking \citep{YeungPope1988ParticleTracking}, data assimilation \citep{FossellaBiferale2025MSDA}, and model predictive flow control \citep{Bieker2020DeepMPC}.

In recent years, neural operators have become a common choice for such surrogates. They learn mappings between function spaces and provide a framework for partial differential equation (PDE) surrogate modeling, including time advancement of complex flows \citep{KovachkiLiLiuAzizzadenesheliBhattacharyaStuartAnandkumar2023,LuJinPangZhangKarniadakis2021}.
Fourier neural operators (FNO) \citep{LiKovachkiAzizzadenesheliLiuBhattacharyaStuartAnandkumar2021} have been applied to LES of three-dimensional turbulence \citep{Li2022Fourier,Luo2024CompressibleRT}, complementing other neural-network approaches to LES \citep{ParkChoi2021Toward,GuanChattopadhyaySubelHassanzadeh2022Stable,Zhao2025LESnets}.
Attention mechanisms have also been incorporated into neural operators for turbulence simulation \citep{Peng2022AttentionEnhanced,Peng2023LinearAttentionFNO}, and Transformer-based surrogates have been developed for high-dimensional PDE solutions, including turbulent flows \citep{HaoWangSuYingDongLiuChengSongZhu2023GNOT,Li2023Scalable,WuLuoWangWangLong2024Transolver,Li2024TransformerLES,DuKrishnapriyan2025EddyFormer,LaiChenYangWangWangXu2026DynFormer}.
Building on the factorized Transformer (FactFormer) \citep{Li2023Scalable}, implicit factorized Transformer (IFactFormer, abbreviated as IFF in this paper) architectures share parameters across latent iterations to improve the stability of long-time prediction of isotropic turbulence on coarse LES grids \citep{Yang2024An}, and have extended stable long-time prediction to turbulent channel flows \citep{Yang2026Implicit}.

However, these IFF models achieved stable long-time prediction with relatively large time steps; when the time step becomes very fine, the long-time predictions of neural operators have been found to degrade \citep{Zou2025Uncertainty}.
Such degradation arises because neural-operator surrogates of three-dimensional turbulence are usually applied autoregressively: a model learns the map over a fixed time interval and is applied repeatedly, with each prediction serving as the next input \citep{Li2022Fourier,Wang2024Prediction,Yang2026Implicit}.
Because the model is trained with reference fields as inputs but applied to its own predictions, prediction errors are fed back and can grow over many steps in chaotic turbulent flows \citep{McCabe2023Towards}.
Recent studies have therefore moved beyond single-step accuracy, developing neural-operator architectures and training strategies for long-time turbulence prediction \citep{GonzalezDemoulinBernard2023Towards,Li2023LongTermIUFNO,Wang2024Prediction}, analyzing the stability of autoregressive neural operators \citep{McCabe2023Towards}, and modeling space-time turbulence statistics \citep{WuZhangHe2025NOForcing}.
In sequence prediction of one-dimensional text or two-dimensional images, each element of a sequence is low-dimensional, so recurrent neural networks can be trained on long sequences in which their own predictions are fed back as inputs, allowing them to learn to recover from errors that develop during long autoregressive generation \citep{Bengio2015ScheduledSampling,Lamb2016ProfessorForcing}.
In contrast, neural operators for three-dimensional turbulence are commonly trained to predict one step ahead \citep{Wang2024Prediction,Yang2024An,Yang2026Implicit,Zou2025Uncertainty}, and backpropagation through long predicted sequences of such high-dimensional fields is memory-intensive \citep{Chen2025Neural}.
Errors that develop over long autoregressive rollouts are therefore difficult to correct during training.

Beyond fine time steps, any long-time turbulence prediction with a fixed time step faces a trade-off in choosing that time step.
A smaller time step requires more autoregressive steps for a given prediction time, increasing both the computational cost and the accumulated error \citep{McCabe2023Towards}.
A larger time step requires fewer autoregressive steps but weakens the correlation between successive snapshots, making the single-step mapping harder to learn \citep{Zou2025Uncertainty}.
In forced isotropic turbulence, the long-time predictions of neural operators have been found to degrade at both excessively small and large time intervals \citep{Zou2025Uncertainty}.
A fixed time step is therefore necessarily a compromise between the number of autoregressive steps and the difficulty of the single-step mapping.

Learning at multiple strides offers a way to relax this trade-off, where the stride denotes the number of time steps advanced by a single prediction.
Large strides advance the flow with few autoregressive steps and capture its slow, large-scale temporal evolution, whereas fine strides resolve fast, small-scale temporal dynamics and provide outputs at fine time steps \citep{Liu2022Hierarchical}.
Hierarchical time-steppers exploit this idea by combining predictors trained at different step sizes \citep{Liu2022Hierarchical}; continuous-time formulations \citep{Linot2023Stabilized,Chen2025Neural,Abueidda2026Time} and time-conditioned networks such as Stormer and Poseidon \citep{Nguyen2024Stormer,Herde2024Poseidon} can likewise predict with different time step sizes.
In these approaches, however, fields at intermediate fine time steps still require additional network evaluations: hierarchical prediction repeatedly applies finer-stride predictors to predicted states \citep{Liu2022Hierarchical}, and ModAFNO reconstructs intermediate fields from the states at both ends of an interval with a separate interpolation model \citep{Leinonen2024ModAFNO}.
These intermediate fields are thus obtained at extra cost, without reusing the information already computed at the selected strides.

Besides computational efficiency, long-time prediction requires the flow statistics to be reproduced, which short-term training alone does not guarantee \citep{Chen2025Neural}.
Architectural modifications \citep{McCabe2023Towards}, iterative refinement \citep{LippeEtAl2023}, and physics-based correction \citep{HuangPerdikaris2025PhysicsCorrect} have been proposed to reduce the errors that accumulate during long rollouts, mainly by improving the predicted fields.
The neural dynamical operator instead combines short-term trajectories with long-time statistics during training, because short-term data alone could not ensure that the long-time statistics were matched \citep{Chen2025Neural}.
Constraining long-time statistics, however, requires predicted trajectories that are long enough for the statistics to be evaluated, and backpropagating through such trajectories of three-dimensional fields is memory-intensive \citep{Chen2025Neural}.
Spectral-Refiner reduces the cost of improving a trained operator by fine-tuning only a spectral convolution layer with the remaining parameters fixed \citep{Cao2025SpectralRefiner}, which suggests that a pretrained predictor can be frozen while a lightweight module is trained.
Inspired by these studies, we seek to efficiently generate stable turbulence predictions at fine time steps while closely reproducing long-time flow statistics.

We develop a two-stage framework that combines multi-stride prediction with trajectory refinement at fine time steps.
In the first stage, we train stride-conditioned IFF (SC-IFF) predictors (built on IFactFormer-m \citep{Yang2026Implicit}) at a set of power-of-two training strides, either sharing parameters across strides or using separate parameters for each stride.
In the second stage, we freeze these predictors and train a factorized-attention cross-stride fusion (FA-CSF) module, which reuses their predictions at adjacent strides and their latent features to generate velocity fields at intermediate time steps.
The corrected state at the largest stride initializes the next autoregressive step, so that the output time step is decoupled from the autoregressive step.
Only the FA-CSF parameters are optimized, using field errors and statistical discrepancies evaluated over long autoregressive rollouts, so parameter gradients and optimizer states are required only for this small module.

We evaluate the proposed approach on two benchmarks commonly used in neural-operator studies of three-dimensional turbulence prediction: turbulent channel flow \citep{Wang2024Prediction,Yang2026Implicit} and forced homogeneous isotropic turbulence (HIT) \citep{Li2022Fourier,Yang2024An}.
The snapshots are separated by 50 and 10 DNS time steps, finer than the 200 DNS time steps used in previous studies of the same flows \citep{Yang2024An,Yang2026Implicit}.
We compare multi-stride prediction with fixed-stride IFF, hierarchical prediction \citep{Liu2022Hierarchical}, and conventional LES models \citep{MoinSquiresCabotLee1991Dynamic,NicoudDucros1999WALE}, focusing on field evolution, agreement of long-time statistics with filtered DNS (fDNS) data, and computational cost.

\Needspace{7\baselineskip}
\paragraph{Contributions}
\begin{itemize}
  \item We extend IFF to multi-stride prediction by training it at power-of-two strides, and compare three stride-conditioning strategies: input concatenation, independent predictors for each stride, and feature-wise linear modulation (FiLM).
  \item We propose FA-CSF, a lightweight module that fuses the outputs and latent features of the predictors for adjacent strides to generate the velocity fields at all intermediate time steps. With a single evaluation of the predictor at each training stride, outputs are obtained at every fine time step while the flow is advanced autoregressively only with the largest stride.
  \item We introduce a two-stage training scheme that freezes the trained SC-IFF predictors and optimizes only FA-CSF with field-data and statistical losses on long autoregressive rollouts, so that the corrections are optimized for their effect on subsequent predictions and long-time statistics.
\end{itemize}

The remainder of the paper is organized as follows.
Section \ref{sec:method} introduces the learning problem and the two-stage stride-conditioned prediction framework.
Section \ref{sec:experiments} presents numerical experiments for long-time prediction at fine time steps.
Section \ref{sec:conclusion} summarizes the findings and their scope.

\section{Methodology}\label{sec:method}
\setcounter{equation}{0}
\subsection{Governing equations and prediction problem}\label{sec:prediction_problem}

We consider incompressible turbulent flows governed by the nondimensional Navier--Stokes equations \citep{Pope2000TurbulentFlows}
\begin{equation}\label{eq:ns}
  \begin{aligned}
    \nabla\cdot\uvec &= 0,\\
    \partial_t \uvec + \uvec\cdot\nabla \uvec &= -\nabla p + \nu \Delta \uvec + \fvec,
  \end{aligned}
\end{equation}
where \(\uvec(\xvec,t)=(u,v,w)^T\) is the velocity field, \(p(\xvec,t)\) is the pressure, \(\nu\) is the nondimensional viscosity coefficient, and \(\fvec\) denotes a (possibly problem-dependent) forcing.
The nonlinear advection term \(\uvec\cdot\nabla\uvec\) transfers kinetic energy between spatial scales, producing the energy cascade, while viscosity dissipates energy at small scales.
Pressure enforces the divergence-free constraint \(\nabla\cdot\uvec=0\). Its Poisson equation follows from the divergence of the momentum equation and is supplemented by boundary conditions \citep{Pope2000TurbulentFlows}.

We express the governing equations in nondimensional form using characteristic velocity and length scales \(U_{\rm ref}\) and \(L_{\rm ref}\), with \(L_{\rm ref}/U_{\rm ref}\) as the time scale. The nondimensional viscosity coefficient is then \(\nu=\nu_{\rm phys}/(U_{\rm ref}L_{\rm ref})=1/\Rey\), where \(\nu_{\rm phys}\) is the dimensional kinematic viscosity.
For wall-bounded turbulence, we also use the friction Reynolds number \(\Retau:=u_\tau\delta/\nu\), where \(u_\tau\) and \(\delta\) denote the friction velocity and channel half-height nondimensionalized by \(U_{\rm ref}\) and \(L_{\rm ref}\), respectively.

Our numerical experiments focus on (i) turbulent channel flow in a planar channel of half-height \(\delta\), driven by a constant mean pressure gradient (or an equivalent uniform body force), with periodic boundary conditions in the streamwise/spanwise directions and no-slip wall boundary conditions at \(y=\pm\delta\), and (ii) forced homogeneous isotropic turbulence (HIT) in a triply periodic box \citep{Pope2000TurbulentFlows}.
In channel flow, the walls introduce wall-normal inhomogeneity and near-wall anisotropy. This spatial dependence is retained in the wall-normal statistical profiles used below.
For HIT, large-scale forcing \(\fvec\) sustains a statistically stationary, homogeneous, and isotropic state.
Boundary conditions and driving are fixed within each benchmark and reflected in the filtered DNS (fDNS) data used for training and evaluation; they are not supplied as separate model inputs \citep{Li2022Fourier,Wang2024Prediction,Yang2026Implicit}.

From an LES viewpoint, applying a spatial filter to \eqref{eq:ns} yields
\begin{equation}\label{eq:les}
  \begin{aligned}
    \nabla\cdot\bar{\uvec} &= 0,\\
    \partial_t \bar{\uvec} + \bar{\uvec}\cdot\nabla \bar{\uvec} &= -\nabla \bar{p} + \nu \Delta \bar{\uvec} - \nabla\cdot\boldsymbol{\tau}_{\mathrm{sgs}} + \bar{\fvec},
  \end{aligned}
\end{equation}
where \(\bar{\uvec}\), \(\bar{p}\), and \(\bar{\fvec}\) denote the filtered velocity, pressure, and forcing, respectively, and \(\boldsymbol{\tau}_{\mathrm{sgs}} := \overline{\uvec\otimes\uvec}-\bar{\uvec}\otimes\bar{\uvec}\) is the subgrid-scale (SGS) stress tensor with components \(\tau^{\mathrm{sgs}}_{ij}\).
The SGS stress accounts for the part of the filtered nonlinear momentum flux that cannot be expressed by the product of the filtered velocities alone.
A conventional eddy-viscosity closure approximates the deviatoric SGS stress as
\[
\tau^{\rm sgs}_{ij}-\frac13\tau^{\rm sgs}_{kk}\delta_{ij}
\approx-2\nu_t\bar S_{ij},\qquad
\bar S_{ij}=\frac12(\partial_i\bar u_j+\partial_j\bar u_i).
\]
Here \(\bar S_{ij}\) is the resolved strain-rate tensor, \(\nu_t\) is the modeled eddy viscosity, and \(\delta_{ij}\) is the Kronecker delta; repeated Cartesian indices are summed. In the Smagorinsky model, \(\nu_t=(C_s\Delta_{\rm f})^2|\bar S|\), with \(|\bar S|=(2\bar S_{ij}\bar S_{ij})^{1/2}\), filter width \(\Delta_{\rm f}\), and dimensionless coefficient \(C_s\) \citep{Pope2000TurbulentFlows,Smagorinsky1963}.
In contrast, we do not prescribe an explicit SGS model and instead learn the coarse-grained discrete-time flow map directly from fDNS snapshots.
We omit the overbar in the discrete-time notation below; all learning targets remain filtered fields.

Let \(\mathcal{H}\) denote the space of fDNS velocity fields on the coarse computational grid.
The dataset contains snapshots sampled at a uniform sampling interval \(\Dt\), written as \(\uvec^n=\uvec(t_0+n\Dt)\in\mathcal{H}\), where \(n\) is the snapshot index and \(t_0\) the initial time.
A trajectory contains \(N_{\rm snap}\) stored fields with \(n=0,\ldots,N_{\rm snap}-1\); its \(N_x\times N_y\times N_z\) grid has \(N_{\rm grid}=N_xN_yN_z\) spatial points.
Here \(N_d\) denotes the number of grid points in direction \(d\in\{x,y,z\}\).

\Needspace{4\baselineskip}
For a fixed relative stride \(s\in\mathbb N^+\), the learning task is to predict the velocity field \(\uvec^{n+s}\) from \(\uvec^n\), corresponding to a time interval \(s\Dt\). We denote the learned prediction operator by \(\mathcal G_s:\mathcal H\to\mathcal H\), with
\begin{equation}\label{eq:fixed_stride_prediction}
\widehat{\uvec}^{\,n+s}=\mathcal G_s(\uvec^n).
\end{equation}
Fixed-stride autoregressive prediction is a common setting in learned turbulence forecasting \citep{Li2022Fourier,Yang2026Implicit}. For long-time prediction at fine time steps, we extend this fixed-stride task to a multi-stride formulation in Section~\ref{sec:stride_conditioning}.

\subsection{Implicit factorized Transformer backbone}\label{sec:backbone}

Our prediction operators use the IFF architecture. Building on FactFormer \citep{Li2023Scalable}, Yang et al.\ introduced parameter sharing across latent iterations for long-time turbulence prediction \citep{Yang2024An} and later replaced chained axial transformations with parallel ones \citep{Yang2026Implicit}. We adopt this latter variant (IFactFormer-m) as our backbone; hereafter, IFF refers to this variant.

The factorized attention makes this architecture computationally attractive for three-dimensional velocity fields. On an \(N_x\times N_y\times N_z\) grid, explicitly evaluating full pairwise attention costs \(\mathcal O((N_xN_yN_z)^2)\) at fixed feature width. FactFormer replaces the full attention kernel with three axial kernels: their construction costs \(\mathcal O(N_x^2+N_y^2+N_z^2)\), while applying them to the three-dimensional feature field costs \(\mathcal O(N_xN_yN_z(N_x+N_y+N_z))\) \citep{Li2023Scalable}. This avoids forming a matrix of interactions between every pair of grid points while retaining spatial coupling along all three axes.

With fine time steps, generating a long trajectory with a fixed-stride predictor requires a large number of autoregressive steps, increasing computational cost and allowing prediction errors to accumulate. Long-time statistics can therefore deteriorate even when single-step prediction errors are small \citep{Zou2025Uncertainty}. Section~\ref{sec:stride_conditioning} introduces stride conditioning so that IFF can be trained at multiple strides, learning flow evolution over both short and longer time spans. Section~\ref{sec:fusion} adds a cross-stride fusion module that reuses these predictions and their latent features to obtain velocity fields at intermediate time steps with fewer backbone evaluations. With the backbone frozen, this module is trained on autoregressive rollouts to correct field errors and discrepancies in long-time statistics.

\begin{figure}[!t]
\centering
\input{figure/method/iff_backbone.tex}
\caption{Architecture of the IFF backbone and parallel axial integration (PAI). (a) Velocity prediction through encoding, implicit iterations with shared parameters, and decoding. (b) PAI module and its expanded \(x\)-axis branch. Blue denotes feature transformations, violet positional encoding, amber attention and aggregation, and gray normalization, averaging and concatenation.}
\label{fig:native_backbone}
\end{figure}

\Needspace{8\baselineskip}
\subsubsection{Overall architecture and implicit iteration}\label{sec:backbone-iteration}

The backbone uses a parallel factorized attention and shared latent iteration structure consistent with that of Yang et al.~\citep{Yang2026Implicit}. We present the backbone here as the basis for the stride conditioning and cross-stride fusion extensions developed below.

Figure~\ref{fig:native_backbone}(a) summarizes the IFF prediction at stride \(s\). The three velocity components are mapped to a higher-dimensional feature space, refined by implicit iteration, and projected back to velocity space, where the predicted increment updates the three-dimensional velocity field. For the current field \(\uvec=\uvec^n\), the input is first normalized:
\begin{equation}\label{eq:input_standardization}
\widetilde{\uvec}=\frac{\uvec-\boldsymbol\mu^{u}}{\boldsymbol\sigma^{u}}.
\end{equation}
Here \(\boldsymbol\mu^{u}\) and \(\boldsymbol\sigma^{u}\) are the componentwise velocity means and standard deviations over the training trajectories, time, and spatial grid. The division is componentwise, and the same input statistics are used at every stride.

The lifting module maps \(\widetilde{\uvec}\) to \(U_{\rm lift}\) through learned pointwise channel maps with a GELU nonlinearity, increasing the channel dimension from three to \(C\) while preserving the spatial grid. A coordinate embedding \(U_{\rm pos}\), obtained from Gaussian Fourier features of the spatial coordinates \citep{Tancik2020FourierFeatures}, supplies positional information on the same grid. Their sum forms the initial latent field:
\begin{equation}\label{eq:input_lift}
U^{(0)}=U_{\rm lift}+U_{\rm pos}
\in\mathbb R^{N_x\times N_y\times N_z\times C}.
\end{equation}

The latent field is refined through implicit iteration with shared parameters. Each update uses a parallel axial integration (PAI) module, which combines factorized attention with feature mixing, as detailed in Section~\ref{sec:backbone-kernels}:
\begin{equation}\label{eq:shared_iteration}
  U^{(\ell+1)} = U^{(\ell)} + \frac{1}{L}\,\operatorname{PAI}\!\left(U^{(\ell)}\right),
  \qquad \ell=0,\dots,L-1.
\end{equation}
The same PAI parameters are used in all \(L\) iterations, and the factor \(1/L\) scales each update before its addition to the latent field.

The projection module first applies group normalization (GN) to the final latent field \(U^{(L)}\). The normalized features \(\widetilde U_{\rm out}=\operatorname{GN}(U^{(L)})\) then pass through pointwise channel maps to yield a normalized vector field \(\widetilde{\boldsymbol r}_s\) on the three-dimensional grid. This field estimates the normalized finite-time average velocity-change rate \(\boldsymbol r_s=(\uvec^{n+s}-\uvec^n)/(s\Dt)\). Denormalization gives
\begin{equation}\label{eq:rate_denormalization}
\widehat{\boldsymbol r}_s
=\boldsymbol\mu_s^{r}+\boldsymbol\sigma_s^{r}\odot\widetilde{\boldsymbol r}_s.
\end{equation}
Here \(\boldsymbol\mu_s^{r}\) and \(\boldsymbol\sigma_s^{r}\) are the training-set mean and standard deviation of \(\boldsymbol r_s\) for stride \(s\), and \(\odot\) denotes elementwise multiplication. For channel flow, these statistics are averaged over trajectories, time, and the homogeneous \(x,z\) directions, retaining the wall-normal dependence of each component; they are broadcast over the homogeneous directions when applied to \(\widetilde{\boldsymbol r}_s\). The predicted velocity field is then
\begin{equation}\label{eq:native_prediction}
\widehat{\uvec}^{\,n+s}=\uvec+s\Dt\,\widehat{\boldsymbol r}_s.
\end{equation}
The term \(s\Dt\,\widehat{\boldsymbol r}_s\) is the predicted velocity increment over the interval \(s\Dt\), which IFF adds to the current field to obtain the future velocity field.

\subsubsection{Parallel axial integration}\label{sec:backbone-kernels}

Figure~\ref{fig:native_backbone}(b) expands the PAI module used in each implicit iteration. At iteration \(\ell\), the latent field \(U^{(\ell)}\) undergoes layer normalization (LN) over channels, giving \(\widetilde U=\operatorname{LN}(U^{(\ell)})\). Two learned linear maps give the kernel features \(Z\) and value features~\(V\):
\[
Z=\widetilde U W_{\rm in}\in\mathbb R^{N_x\times N_y\times N_z\times C},
\qquad
V=\widetilde U W_V\in\mathbb R^{N_x\times N_y\times N_z\times C_v}.
\]
Here \(Z\) retains the latent channel width \(C\), while \(V\) has \(C_v\) value channels per attention head. With channel vectors written as rows, the learned matrices \(W_{\rm in}\in\mathbb R^{C\times C}\) and \(W_V\in\mathbb R^{C\times C_v}\) act by right multiplication at every grid point.

To construct attention kernels along \(x,y,z\), the three branches apply separate learned linear maps \(W_x,W_y,W_z\in\mathbb R^{C\times C}\) to \(Z\). Each branch then averages its transformed features over the other two spatial directions. With \(i,j,k\) indexing grid positions along \(x,y,z\), respectively, \(Z_{ijk}\in\mathbb R^{1\times C}\) is the feature row at one grid point. The resulting axial features are
\begin{subequations}\label{eq:axis_features}
\begin{align}
U_x(i)&=\operatorname{MLP}_x\,\Biggl(\operatorname{LN}\,\biggl(
\frac{1}{N_yN_z}\sum_{j=1}^{N_y}\sum_{k=1}^{N_z} Z_{ijk}W_x\biggr)\Biggr),\label{eq:axis_features_x}\\*[3pt]
U_y(j)&=\operatorname{MLP}_y\,\Biggl(\operatorname{LN}\,\biggl(
\frac{1}{N_xN_z}\sum_{i=1}^{N_x}\sum_{k=1}^{N_z} Z_{ijk}W_y\biggr)\Biggr),\label{eq:axis_features_y}\\*[3pt]
U_z(k)&=\operatorname{MLP}_z\,\Biggl(\operatorname{LN}\,\biggl(
\frac{1}{N_xN_y}\sum_{i=1}^{N_x}\sum_{j=1}^{N_y} Z_{ijk}W_z\biggr)\Biggr).\label{eq:axis_features_z}
\end{align}
\end{subequations}
Here \(\operatorname{MLP}_x\), \(\operatorname{MLP}_y\), and \(\operatorname{MLP}_z\) are feed-forward networks (multilayer perceptrons, MLPs) with GELU activation, applied to the channel vector at each axial position. The resulting axial features have shapes \(U_x\in\mathbb R^{N_x\times C}\), \(U_y\in\mathbb R^{N_y\times C}\), and \(U_z\in\mathbb R^{N_z\times C}\).

\Needspace{8\baselineskip}
Each branch maps its axial features to query and key matrices with \(C_q\) channels per head. The learned channel maps \(W_{Q,x},W_{K,x},W_{Q,y},W_{K,y},W_{Q,z},W_{K,z}\) all have shape \(C\times C_q\). Rotary positional encoding (RoPE) rotates these vectors according to the corresponding axial coordinates, allowing their inner products to represent relative positional dependence \citep{Su2024RoFormer,Li2023Scalable}. The three kernels are formed as
\begin{subequations}\label{eq:axial_kernel}
\begin{alignat}{3}
Q_x&=\operatorname{RoPE}_x(U_xW_{Q,x}),\qquad &
K_x&=\operatorname{RoPE}_x(U_xW_{K,x}),\qquad &
\kappa_x&=Q_xK_x^T,\label{eq:axial_kernel_x}\\*[3pt]
Q_y&=\operatorname{RoPE}_y(U_yW_{Q,y}),\qquad &
K_y&=\operatorname{RoPE}_y(U_yW_{K,y}),\qquad &
\kappa_y&=Q_yK_y^T,\label{eq:axial_kernel_y}\\*[3pt]
Q_z&=\operatorname{RoPE}_z(U_zW_{Q,z}),\qquad &
K_z&=\operatorname{RoPE}_z(U_zW_{K,z}),\qquad &
\kappa_z&=Q_zK_z^T.\label{eq:axial_kernel_z}
\end{alignat}
\end{subequations}
The query and key matrices have shapes \(Q_x,K_x\in\mathbb R^{N_x\times C_q}\), \(Q_y,K_y\in\mathbb R^{N_y\times C_q}\), and \(Q_z,K_z\in\mathbb R^{N_z\times C_q}\). Their products give the kernel matrices \(\kappa_x\in\mathbb R^{N_x\times N_x}\), \(\kappa_y\in\mathbb R^{N_y\times N_y}\), and \(\kappa_z\in\mathbb R^{N_z\times N_z}\). In particular, \(\kappa_x(i,i')=[Q_xK_x^T]_{ii'}\) is the inner product between the query at target position \(i\) and the key at source position \(i'\); it weights the contribution from \(i'\) to \(i\). The entries \(\kappa_y(j,j')\) and \(\kappa_z(k,k')\) provide the corresponding aggregation weights along \(y\) and \(z\), respectively.

The value features are then aggregated along each axis using the corresponding kernel matrix:
\begin{subequations}\label{eq:parallel_aggregation}
\begin{align}
V^{(x)}_{ijk}&=\sum_{i'=1}^{N_x}\kappa_x(i,i')V_{i'jk},\label{eq:parallel_aggregation_x}\\*
V^{(y)}_{ijk}&=\sum_{j'=1}^{N_y}\kappa_y(j,j')V_{ij'k},\label{eq:parallel_aggregation_y}\\*
V^{(z)}_{ijk}&=\sum_{k'=1}^{N_z}\kappa_z(k,k')V_{ijk'}.\label{eq:parallel_aggregation_z}
\end{align}
\end{subequations}
For example, Eq.~\eqref{eq:parallel_aggregation_x} collects the feature vectors \(V_{i'jk}\) along \(x\) at fixed \(j,k\), yielding \(V^{(x)}_{ijk}\) at the target grid point. Each aggregated field has shape \(N_x\times N_y\times N_z\times C_v\). The three branches independently apply their respective kernels to \(V\), allowing the directional aggregations to be evaluated in parallel.

The three directional outputs from all attention heads are concatenated along the channel dimension. Before channel mixing, instance normalization centers and rescales each channel using its mean and variance over the spatial grid. A pointwise feed-forward network mixes these features and maps them back to \(C\) channels, yielding the latent update \(\operatorname{PAI}(U^{(\ell)})\) in Eq.~\eqref{eq:shared_iteration}.

\subsection{Three IFF variants for multi-stride prediction}\label{sec:stride_conditioning}

For a fixed total prediction time, a smaller stride requires more autoregressive steps and can amplify error accumulation, whereas a larger stride reduces the number of autoregressive steps but may increase the error of each prediction. We therefore extend the fixed-stride task in Section~\ref{sec:prediction_problem} to multiple time intervals, allowing training to constrain flow evolution over both short and longer intervals.

Rather than training at every stride, we select power-of-two strides,
\begin{equation}\label{eq:native_strides}
\mathcal S=\{1,2,4,\ldots,s_{\max}\}=\{2^0,2^1,\ldots,2^{\log_2 s_{\max}}\},
\end{equation}
where \(s_{\max}\ge2\) is a power of two. These training strides cover successive temporal scales with a factor-of-two spacing. Each \(\mathcal G_s\) predicts a future field on the same spatial grid; the corresponding time interval is \(s\Dt\). Writing the stride explicitly as an input, we represent this family by \(\mathcal G:\mathcal H\times\mathcal S\to\mathcal H\), with \(\mathcal G(\uvec^n,s)=\mathcal G_s(\uvec^n)\).

Figure~\ref{fig:stride_parameterizations} compares three stride-conditioned IFF (SC-IFF) variants for multi-stride prediction. SC-IFF-Concat appends a stride channel to the input, SC-IFF-Indep selects a separate predictor for each training stride, and SC-IFF-FiLM modulates intermediate features using a stride embedding. Each retains the IFF backbone and predicts a rate that determines the velocity increment added to the current field, as in Eq.~\eqref{eq:native_prediction}.

\begin{figure}[!t]
\centering
\input{figure/method/stride_parameterizations.tex}
\caption{Stride conditioning in the three SC-IFF variants. (a) Concat appends the normalized stride as an input channel. (b) Indep uses a separate IFF predictor for each training stride. (c) FiLM applies stride-dependent scales and shifts to the normalized PAI input and the group-normalized final latent field.}
\label{fig:stride_parameterizations}
\end{figure}

\Needspace{10\baselineskip}
\subsubsection{SC-IFF-Concat: Append a stride channel to the input}\label{sec:concat}

Appending time information to the input provides a simple way to condition a predictor. In LESnets, Zhao et al.~\citep{Zhao2025LESnets} concatenate spatial and temporal coordinates with the velocity field. SC-IFF-Concat provides a simple baseline by encoding the prediction stride in an additional input channel. The normalized stride \(\rho(s)=s/s_{\max}\) is broadcast over the spatial grid and concatenated with the normalized velocity along the channel dimension:
\begin{equation}\label{eq:concat}
\widetilde{\uvec}_{\rm Concat}(\xvec)
=\operatorname{Concat}\bigl(\widetilde{\uvec}(\xvec),\rho(s)\bigr).
\end{equation}
The lifting module therefore receives four channels---the three velocity components and the normalized stride---and maps them to the latent width \(C\). The subsequent latent iterations and output projection are unchanged from the IFF backbone in Section~\ref{sec:backbone-iteration}.

Training stride-conditioned predictors requires examples at multiple strides, which we construct by reindexing the original stored trajectories. For a trajectory sampled at \(\Dt\), stride \(s\) separates the snapshots into \(s\) sequences,
\begin{equation}\label{eq:stride_phase_sequences}
\bigl(\uvec^{\tau_0},\uvec^{\tau_0+s},\uvec^{\tau_0+2s},\ldots\bigr),
\qquad \tau_0=0,\ldots,s-1.
\end{equation}
Here \(\tau_0\) is the starting snapshot index, and successive fields are separated by \(s\Dt\). In the example shown in Figure~\ref{fig:stride_phase_sequences}, \(s=1\) retains the original trajectory, whereas \(s=2\) separates its even- and odd-indexed fields into two sequences sampled every \(2\Dt\). Repeating the separation at \(s=4\) gives four sequences sampled every \(4\Dt\).

\begin{figure}[!t]
\centering
\input{figure/method/stride_phase_sequences.tex}
\caption{Construction of stride-specific sequences for \(s=1,2,4\). Cubes denote stored three-dimensional velocity fields, aligned by snapshot index \(n\); rows at each stride have starting indices \(\tau_0=0,\ldots,s-1\).}
\label{fig:stride_phase_sequences}
\end{figure}

For each stride \(s\), we form input--output pairs from one velocity field and a sequence of \(M\) reference outputs separated by \(s\Dt\), where \(M\ge1\) is the number of autoregressive prediction steps used during training:
\begin{equation}\label{eq:stride_training_examples}
\underbrace{(\uvec^n,s)}_{\text{input}}
\quad\longmapsto\quad
\underbrace{(\uvec^{n+s},\uvec^{n+2s},\ldots,\uvec^{n+Ms})}_{\text{reference outputs}},
\qquad 0\le n<N_{\rm snap}-Ms.
\end{equation}
Let \(\mathcal D_s\) collect these input--output pairs from all training trajectories. We set \(\widehat{\uvec}^{\,n}=\uvec^n\) and generate the predicted output sequence autoregressively:
\begin{equation}\label{eq:sc_training_predictions}
\widehat{\uvec}^{\,n+ms}
=\mathcal G(\widehat{\uvec}^{\,n+(m-1)s},s;\Theta),
\qquad m=1,\ldots,M.
\end{equation}
At each subsequent step, the predicted field serves as the next input, and gradients are propagated through all \(M\) steps. For three-dimensional flow prediction, unrolled training requires substantial GPU memory, so we typically set \(M=1\) or \(2\).
Here \(\Theta\) denotes all trainable parameters of the single SC-IFF-Concat network, shared across the training strides. For a minibatch \(\mathcal B_s\subset\mathcal D_s\) at a fixed stride \(s\), the stride-specific data loss averages the relative field error over all \(M\) outputs and all samples:
\begin{equation}\label{eq:stride_data_loss}
\mathcal L_{\mathrm{data},s}(\Theta;\mathcal B_s)
=\frac{1}{M|\mathcal B_s|}
\sum_{\mathcal B_s}\sum_{m=1}^{M}
\frac{\|\widehat{\uvec}^{\,n+ms}-\uvec^{n+ms}\|_2}
{\|\uvec^{n+ms}\|_2}.
\end{equation}
Here \(|\mathcal B_s|\) is the batch size, and the norm includes all grid points and velocity components.

Since SC-IFF-Concat accepts different strides within one shared predictor, we pool the stride-specific datasets into \(\mathcal D=\bigcup_{s\in\mathcal S}\mathcal D_s\) and shuffle the samples for joint training. In a mixed minibatch \(\mathcal B\subset\mathcal D\), let \(\mathcal B_s\subset\mathcal B\) collect the samples with stride \(s\), so that \(|\mathcal B|=\sum_{s\in\mathcal S}|\mathcal B_s|\). The training loss averages these sequence losses over all samples in the batch:
\begin{equation}\label{eq:mixed_stride_data_loss}
\mathcal L_{\mathrm{data}}(\Theta;\mathcal B)
=\sum_{\substack{s\in\mathcal S\\|\mathcal B_s|>0}}
\frac{|\mathcal B_s|}{|\mathcal B|}
\mathcal L_{\mathrm{data},s}(\Theta;\mathcal B_s).
\end{equation}

\Needspace{6\baselineskip}
\subsubsection{SC-IFF-Indep: Select a separate predictor for each training stride}\label{sec:independent}

Another natural approach is to train a separate predictor at each power-of-two stride. Liu et al.~\citep{Liu2022Hierarchical} train neural time-steppers at different time step sizes and combine them hierarchically for long-time prediction. We adopt this idea for turbulence prediction at fine time steps: SC-IFF-Indep assigns each training stride \(s\in\mathcal S\) an independent IFF predictor with trainable parameters \(\theta_s\). The requested stride selects its predictor, giving \(\mathcal G(\uvec,s;\Theta)=\mathcal G_s(\uvec;\theta_s)\), where \(\Theta=\{\theta_s:s\in\mathcal S\}\) collects all predictor parameters.

Because \(\mathcal S\) contains only power-of-two strides, this formulation requires only \(\lvert\mathcal S\rvert\) predictors rather than one for every stride. To improve training efficiency, we use transfer learning across these scales. We first train the predictor at the largest stride \(s_{\max}\), then use its learned weights to initialize the predictor at the next smaller stride, \(s_{\max}/2\). We repeat this process toward smaller strides. For example, \(s_{\max}=4\) gives the transfer-learning sequence \(\mathcal G_4\to\mathcal G_2\to\mathcal G_1\). This initialization allows a training budget to be concentrated on \(\mathcal G_{s_{\max}}\), with fewer optimization updates used to adapt the predictors at smaller strides.

Each predictor is trained on minibatches \(\mathcal B_s\subset\mathcal D_s\) using the stride-specific data loss in Eq.~\eqref{eq:stride_data_loss}:
\begin{equation}\label{eq:independent_data_objective}
\min_{\theta_s}\;
\mathcal L_{\mathrm{data},s}(\theta_s;\mathcal B_s),
\qquad s=s_{\max},s_{\max}/2,s_{\max}/4,\ldots,1.
\end{equation}

If each stride-specific IFF predictor has the same width and depth as the backbone in SC-IFF-Concat, SC-IFF-Indep stores approximately \(\lvert\mathcal S\rvert\) times as many parameters. These predictors are trained one at a time, so peak training memory remains comparable to that of SC-IFF-Concat. The extra checkpoint storage is small relative to the memory occupied by the three-dimensional feature tensors during training. We use the same batch size and comparable overall training budgets for these schemes.

\Needspace{10\baselineskip}
\subsubsection{SC-IFF-FiLM: Modulate features with the stride}\label{sec:film}

Feature-wise linear modulation (FiLM) provides a general way to incorporate conditioning information into a neural network \citep{Perez2018FiLM}. It generates channelwise scales and shifts from that information and applies them to intermediate features. In weather forecasting, Stormer uses a vision Transformer backbone with adaptive layer normalization in each block to condition on the forecast interval \citep{Nguyen2024Stormer}. Poseidon introduces lead-time-conditioned layer normalization into its scalable operator Transformer (scOT), a multiscale architecture built from SwinV2 blocks \citep{Herde2024Poseidon}.

We apply this conditioning idea to the IFF architecture to obtain SC-IFF-FiLM, allowing one shared predictor to adapt its features to the requested stride. Specifically, stride-dependent modulation is applied after layer normalization at the PAI input and after group normalization in the output projection. The stride thus influences both axial feature aggregation and the conversion of the final latent field into the predicted rate.

We represent the requested stride by five scalar features:
\begin{equation}\label{eq:stride_code}
\boldsymbol c(s)=\left[\rho(s),\frac{\log_2s}{\log_2s_{\max}},
\sin(\pi\rho(s)),\cos(\pi\rho(s)),1\right]^T.
\end{equation}
The first component, \(\rho(s)=s/s_{\max}\), expresses the stride relative to the largest training stride. The logarithmic component increases by the same amount whenever the stride doubles. The sine--cosine pair provides a smooth nonlinear representation of the normalized stride, while the final component is a stride-independent constant. An MLP-based conditioning network with trainable parameters \(\Theta_{\mathrm{cond}}\) maps these features to two pairs of modulation coefficients:
\begin{equation}\label{eq:stride_coefficients}
\bigl(\boldsymbol\gamma_s^{\mathrm{FA}},\boldsymbol\beta_s^{\mathrm{FA}},
\boldsymbol\gamma_s^{\mathrm{out}},\boldsymbol\beta_s^{\mathrm{out}}\bigr)
=\operatorname{Cond}\bigl(\boldsymbol c(s);\Theta_{\mathrm{cond}}\bigr).
\end{equation}
The coefficient pairs \((\boldsymbol\gamma_s^{\mathrm{FA}},\boldsymbol\beta_s^{\mathrm{FA}})\) and \((\boldsymbol\gamma_s^{\mathrm{out}},\boldsymbol\beta_s^{\mathrm{out}})\) provide channelwise modulation of the normalized PAI input \(\widetilde U\) and the normalized output-projection features \(\widetilde U_{\rm out}\), respectively.

At the PAI input, modulation replaces \(\widetilde U\) before the kernel features \(Z\) and value features \(V\) are formed. In the output projection, modulation follows group normalization and precedes the channel maps that produce \(\widetilde{\boldsymbol r}_s\). These transformations are
\begin{subequations}\label{eq:film}
\begin{align}
\operatorname{FiLM}_s^{\mathrm{FA}}\bigl(\widetilde U\bigr)
&=[1+\boldsymbol\gamma_s^{\mathrm{FA}}]\odot\widetilde U
+\boldsymbol\beta_s^{\mathrm{FA}},\label{eq:film_attention}\\*
\operatorname{FiLM}_s^{\mathrm{out}}\bigl(\widetilde U_{\rm out}\bigr)
&=[1+\boldsymbol\gamma_s^{\mathrm{out}}]\odot\widetilde U_{\rm out}
+\boldsymbol\beta_s^{\mathrm{out}}.\label{eq:film_output}
\end{align}
\end{subequations}
The same PAI coefficient pair is reused in all \(L\) implicit iterations for a given stride. The conditioning network is initialized to output zero coefficients, so modulation starts with unit scales and zero shifts.

We train the IFF backbone and conditioning network jointly, collecting their parameters in \(\Theta\). As in SC-IFF-Concat, training uses mixed-stride minibatches \(\mathcal B\subset\mathcal D\) and the data loss \(\mathcal L_{\mathrm{data}}(\Theta;\mathcal B)\) in Eq.~\eqref{eq:mixed_stride_data_loss}, with the same number of optimization updates and the same batch size per update.

\Needspace{8\baselineskip}
\subsection{Cross-stride fusion for outputs at intermediate time steps and trajectory refinement}\label{sec:fusion}

Training at power-of-two strides reduces the number of strides used for supervision, but does not directly provide predictions at the intermediate time steps. Generating velocity fields at all intermediate time steps requires predictions at all strides in \(\mathcal T=\{1,2,3,\ldots,s_{\max}-1,s_{\max}\}\), while the predictors are trained only at the power-of-two strides in \(\mathcal S\). Hierarchical prediction \citep{Liu2022Hierarchical} fills the intervening times by applying successively finer-stride predictors to previously predicted states. This approach still requires repeated predictor evaluations across multiple levels of the hierarchy.

Beyond computational efficiency, long-time prediction also requires accurate reproduction of the flow statistics. Fitting the predictors to short windows of a finite dataset constrains their field predictions, but does not directly constrain the long-time statistics of the autoregressive rollouts they generate. For chaotic flows, accurately predicting future velocity fields and reproducing long-time statistics are distinct learning objectives \citep{Jiang2023Invariant,Chen2025Neural}. Extending training to long sequences can expose statistical discrepancies, but backpropagating through many three-dimensional fields has a substantial memory cost \citep{Chen2025Neural}.

We introduce factorized-attention cross-stride fusion (FA-CSF) to reconstruct velocity fields at intermediate time steps from the reusable predictions at the training strides and refine the statistics of the predicted flow. Starting from the same current field, the SC-IFF predictors are evaluated once at their respective strides; a smaller trainable fusion module uses their outputs and latent features to reconstruct and correct the fields at all requested strides. The corrected velocity field at \(s_{\max}\) becomes the input for the next autoregressive step. This feedback generates an autoregressive rollout, using model predictions after the initial field.

The SC-IFF predictor training described in Section~\ref{sec:stride_conditioning} forms the first stage of the training procedure (Stage~1). We then freeze the trained SC-IFF predictors and optimize only the fusion module (Stage~2), using field errors and statistical discrepancies evaluated along these autoregressive rollouts.

\begin{figure}[!t]
\centering
\input{figure/method/cross_stride_fusion.tex}
\caption{Cross-stride reconstruction with FA-CSF. (a) Schematic of linear interpolation and learned correction for \(s_{\max}=8\) and requested stride \(s=6\). (b) Computation using the selected predictors, a parameter-free interpolation path, and spatial and temporal correction branches. Blue and green frames denote IFF predictors and FA-CSF; snowflakes mark frozen parameters.}
\label{fig:cross_stride_fusion}
\end{figure}

\subsubsection{Interpolation and learned correction}\label{sec:adjacent_selection}\label{sec:fa_reconstruction}

For a current field \(\uvec^n\), we seek predictions at every requested stride \(s\in\mathcal T=\{1,2,\ldots,s_{\max}\}\) using the predictors trained at strides in \(\mathcal S\). Inspired by mixture-of-experts models \citep{Jacobs1991Adaptive,Shazeer2017Outrageously,Wang2025MixtureOfExperts}, we use a fixed stride-based router to select up to two predictors according to the requested stride. The selected predictors use the current field to generate velocity predictions and latent features at their training strides for FA-CSF correction. As shown in Figure~\ref{fig:cross_stride_fusion}, FA-CSF combines their velocity predictions and latent features through factorized self-attention to learn a correction to a linearly interpolated base field. We denote the trainable correction parameters by \(\phi\), while \(\Theta\) denotes the predictor parameters.

We first construct this base field by linear interpolation between adjacent training strides:
\begin{equation}\label{eq:adjacent_strides}
 s_-\le s\le s_+,\qquad s_+=2s_-,\qquad s_-,s_+\in\mathcal S.
\end{equation}
Here \(s_-\) and \(s_+\) denote the lower and upper strides of the selected adjacent pair. We apply \(\mathcal G_{s_-}\) and \(\mathcal G_{s_+}\) to the current field \(\uvec^n\) and linearly interpolate their predictions:
\begin{equation}\label{eq:linear_reference}
\widehat{\uvec}_{\rm lin}^{\,n+s}
=\frac{s_+-s}{s_+-s_-}\mathcal G_{s_-}(\uvec^n)
+\frac{s-s_-}{s_+-s_-}\mathcal G_{s_+}(\uvec^n),
\qquad s\in\mathcal T.
\end{equation}

We learn an additive velocity correction to the interpolated field, using autoregressive rollouts to train it to better reproduce the reference flow statistics.

To construct the correction, we use the normalized rates \(\widetilde{\boldsymbol r}_{s_-}\) and \(\widetilde{\boldsymbol r}_{s_+}\), together with the final latent fields \(U_{s_-}^{(L)}\) and \(U_{s_+}^{(L)}\). After normalizing the latent fields over channels and projecting them pointwise, we concatenate the two projected latent fields, the two normalized rates, and the normalized current velocity \(\widetilde{\uvec}\) along the channel dimension to form \(U_{\rm route}\). These fields share the same \(N_x\times N_y\times N_z\) spatial grid.

To capture spatial interactions in \(U_{\rm route}\), FA-CSF uses a lightweight PAI module with the factorized attention described in Section~\ref{sec:backbone-kernels}, forming the spatial branch. Alongside it, a temporal branch introduces the requested stride so that the correction can adapt to the time interval \(s\Dt\). The spatial branch first applies a pointwise linear map with GELU activation, then aggregates the resulting features along the three spatial axes through PAI to obtain \(U_{\rm sp}\). The temporal branch uses an MLP to embed the stride features \(\boldsymbol c(s)\) from Eq.~\eqref{eq:stride_code}. This embedding is broadcast over the spatial grid and concatenated with \(U_{\rm route}\) along the channel dimension; a pointwise linear map with GELU activation then gives \(U_{\rm tm}\). Both branches produce \(C_{\rm f}\)-channel feature fields, with \(U_{\rm sp},U_{\rm tm}\in\mathbb R^{N_x\times N_y\times N_z\times C_{\rm f}}\).

\Needspace{6\baselineskip}

We add the two feature fields, apply layer normalization over channels, and use a pointwise linear map from \(C_{\rm f}\) to three channels to obtain the normalized rate correction \(\widetilde{\boldsymbol r}_s^{\mathrm c}\), where the superscript \(\mathrm c\) denotes correction:
\begin{equation}\label{eq:fusion_branch_combination}
\widetilde{\boldsymbol r}_s^{\mathrm c}
=\operatorname{Linear}\Bigl(\operatorname{LN}\bigl(U_{\rm sp}+U_{\rm tm}\bigr)\Bigr).
\end{equation}

We multiply \(\widetilde{\boldsymbol r}_s^{\mathrm c}\) by the componentwise rate standard deviation \(\boldsymbol\sigma_s^r\) and the time interval \(s\Dt\). Adding the resulting velocity increment to the interpolated field gives the final velocity prediction \(\widehat{\uvec}^{\,n+s}\):

\begin{equation}\label{eq:reconstruction}
\widehat{\uvec}^{\,n+s}
=\widehat{\uvec}_{\rm lin}^{\,n+s}
+s\Dt\,\boldsymbol\sigma_s^{r}\odot
\widetilde{\boldsymbol r}_s^{\mathrm c}.
\end{equation}

For each \(s\in\mathcal T\), the rate standard deviation \(\boldsymbol\sigma_s^r\) is computed from the training trajectories. The final three-channel map is initialized to zero, so the model initially reproduces the linearly interpolated field.

\Needspace{8\baselineskip}
\subsubsection{Autoregressive trajectory generation}\label{sec:trajectory_prediction}\label{sec:workflow-stage2}

The construction above gives the velocity at any sampling time within the next \(s_{\max}\Dt\). We denote the complete model, from the SC-IFF predictions through interpolation and learned correction, by \(\mathcal F\):
\begin{equation}\label{eq:full_operator}
\widehat{\uvec}^{\,n+s}=\mathcal F(\uvec^n,s;\Theta,\phi),
\qquad s\in\mathcal T.
\end{equation}
With the predictor parameters \(\Theta^*\) fixed after Stage~1, we initialize \(\widehat{\uvec}^{\,0}=\uvec^0\). For each autoregressive step starting at \(n\), the model uses the current field \(\widehat{\uvec}^{\,n}\) as a common input and evaluates all requested strides \(s\in\mathcal T\) together. One forward call returns the next \(s_{\max}\) velocity fields:
\begin{equation}\label{eq:fused_rollout}
\bigl(\widehat{\uvec}^{\,n+1},\ldots,\widehat{\uvec}^{\,n+s_{\max}}\bigr)
=\bigl(\mathcal F(\widehat{\uvec}^{\,n},s;\Theta^*,\phi)\bigr)_{s=1}^{s_{\max}}.
\end{equation}
The final field \(\widehat{\uvec}^{\,n+s_{\max}}\) becomes the input for the next autoregressive step. Figure~\ref{fig:trajectory_prediction} illustrates the prediction and feedback process. We repeat this procedure until the requested time window contains \(N_{\rm win}\) predicted fields, giving an output every \(\Dt\) and feeding back a new current field every \(s_{\max}\Dt\). We refer to each such advance by \(s_{\max}\Dt\) as an autoregressive step: all outputs within it are generated from the same current field without intermediate feedback.

\begin{figure}[!htbp]
\centering
\input{figure/method/autoregressive_trajectory.tex}
\caption{Generation of velocity fields at every time step over two autoregressive steps of length \(s_{\max}\Delta T\). (a) Fixed-stride IFF. (b) Hierarchical SC-IFF. (c) SC-IFF + FA-CSF. Arcs denote IFF evaluations; blue and green filled circles denote predictor and corrected outputs, respectively. Ringed circles are fed back, and dashed lines separate successive autoregressive steps of length \(s_{\max}\Delta T\) in (b,c).}
\label{fig:trajectory_prediction}
\end{figure}
\FloatBarrier

\Needspace{8\baselineskip}
\subsubsection{Training with field-data and statistical losses}\label{sec:trajectory_objective}

FA-CSF is trained on autoregressive rollouts generated by Eq.~\eqref{eq:fused_rollout}. We extract windows \((\uvec^0,\uvec^1,\ldots,\uvec^{N_{\rm win}})\) from the training trajectories and collect them in \(\mathcal D_{\rm traj}\). Each window forms an input--output pair: \(\uvec^0\) initializes the model, and the following \(N_{\rm win}\) fields provide the reference output trajectory.

Each training window spans several autoregressive steps, allowing the statistical losses to constrain deviations that develop during autoregressive prediction. These losses train FA-CSF to better reproduce the mean flow, fluctuation intensity, and energy distribution over the resulting trajectory.

\paragraph{Trajectory data loss}
The trajectory loss averages the relative field error in Eq.~\eqref{eq:stride_data_loss} over all predicted times, including the corrected states at the end of each autoregressive step:
\begin{equation}\label{eq:stage2_data}
\mathcal L_{\rm traj}=\frac{1}{N_{\rm win}}\sum_{n=1}^{N_{\rm win}}
\frac{\|\widehat{\uvec}^{\,n}-\uvec^n\|_2}{\|\uvec^n\|_2}.
\end{equation}

\paragraph{Mean and second-order velocity statistics}
Let \(\mathcal W=\{1,\ldots,N_{\rm win}\}\) denote the predicted time indices. Brackets denote arithmetic averages over the listed spatial directions and then the time window. With fluctuations \(u_i^{\prime n}=u_i^n-\mu_i\), the mean velocity and covariance are
\begin{equation}\label{eq:window_statistics}
\begin{aligned}
\mu_i(y)&=\langle u_i^n\rangle_{x,z,\mathcal W},&
R_{ij}(y)&=\langle u_i^{\prime n}u_j^{\prime n}\rangle_{x,z,\mathcal W},
&&\text{channel flow},\\
\mu_i&=\langle u_i^n\rangle_{x,y,z,\mathcal W},&
R_{ij}&=\langle u_i^{\prime n}u_j^{\prime n}\rangle_{x,y,z,\mathcal W},
&&\text{HIT}.
\end{aligned}
\end{equation}
Here \(i,j\in\{1,2,3\}\) index the velocity components \((u,v,w)\). Channel statistics retain their dependence on \(y\), whereas HIT statistics are global. The component variance is \(R_{ii}\), without summation over \(i\), and its square root is the root-mean-square (RMS) fluctuation.

Predicted statistics, marked with hats, use the same definitions applied to \(\widehat{\uvec}^{\,n}\); predicted fluctuations are centered by their own mean \(\widehat\mu_i\). We normalize the statistical errors using fixed velocity scales \(\sigma_i^{\rm stat}>0\) computed from the training data. For channel flow, \(\sigma_i^{\rm stat}\) is the RMS velocity fluctuation of component \(i\) computed over the whole channel; for HIT, they are the velocity standard deviations $\sigma_i^u$.

For channel flow, the losses compare the mean profile, component variances, and signed Reynolds shear stress \(R_{12}=\langle u'v'\rangle_{x,z,\mathcal W}\)~\cite{Wang2024Prediction,Yang2026Implicit}:
\begin{subequations}\label{eq:profile_losses}
\begin{align}
\mathcal L_{\rm mean}
&=\left\langle\left(\frac{\widehat\mu_i-\mu_i}{\sigma_i^{\rm stat}}\right)^2\right\rangle_{y,i},\label{eq:profile_mean}\\*
\mathcal L_{\rm var}
&=\left\langle\left(\frac{\widehat R_{ii}-R_{ii}}{(\sigma_i^{\rm stat})^2}\right)^2\right\rangle_{y,i},\label{eq:profile_variance}\\*
\mathcal L_{\rm stress}
&=\left\langle\left(\frac{\widehat R_{12}-R_{12}}{\sigma_1^{\rm stat}\sigma_2^{\rm stat}}\right)^2\right\rangle_y.\label{eq:profile_stress}
\end{align}
\end{subequations}
For HIT, the mean and variance losses use the same expressions with averaging only over components. The stress loss includes all three off-diagonal covariances:
\begin{equation}\label{eq:hit_stress_loss}
\mathcal L_{\rm stress}=\frac13\sum_{1\le i<j\le3}
\left(\frac{\widehat R_{ij}-R_{ij}}{\sigma_i^{\rm stat}\sigma_j^{\rm stat}}\right)^2.
\end{equation}

\Needspace{12\baselineskip}
\paragraph{Spectral shape}
The variance losses constrain fluctuation intensity. To also constrain its distribution over spatial scales, we compare normalized velocity spectra. In the definitions below, each discrete Fourier transform (DFT) is normalized by the number of transformed grid points.

For channel flow, the one-sided streamwise and spanwise spectra give the variance carried by each discrete mode:
\begin{subequations}\label{eq:channel_spectra}
\begin{align}
P_i^{(x)}(k,y)&=a_k^{(x)}\left\langle
\left|\operatorname{DFT}_x[u_i^{\prime n}](k,y,z)\right|^2
\right\rangle_{z,\mathcal W},\label{eq:channel_spectrum_x}\\
P_i^{(z)}(k,y)&=a_k^{(z)}\left\langle
\left|\operatorname{DFT}_z[u_i^{\prime n}](x,y,k)\right|^2
\right\rangle_{x,\mathcal W}.\label{eq:channel_spectrum_z}
\end{align}
\end{subequations}
For direction \(d\in\{x,z\}\), \(k=0,\ldots,\lfloor N_d/2\rfloor\) is the mode index. The factor \(a_k^{(d)}\) accounts for the omitted negative modes: it equals one at zero and the even-grid Nyquist mode, and two elsewhere. The spectrum therefore satisfies \(\sum_k P_i^{(d)}(k,y)=R_{ii}(y)\), and its normalized shape at the interior planes, where \(R_{ii}(y)>0\), is
\begin{equation}\label{eq:spectral_shape}
\pi_i^{(d)}(k,y)=\frac{P_i^{(d)}(k,y)}{R_{ii}(y)}.
\end{equation}
Applying the same normalization to the predicted spectrum and variance gives \(\widehat\pi_i^{(d)}\); both shapes sum to one over modes.

To reduce the contribution of planes with negligible reference fluctuations, we use \(\lambda_i^{\rm ref}(y)=R_{ii}(y)/(R_{ii}(y)+\alpha_i)\), where \(\alpha_i>0\) is a fixed variance scale from the training data. The unit-mean weights \(\lambda_{\rm band}^{(d)}(k)\) emphasize higher spatial modes. We then penalize logarithmic ratios of the predicted and reference spectral shapes:
\begin{equation}\label{eq:spectral_loss}
\mathcal L_{\rm shape}
=\frac12\sum_{d\in\{x,z\}}\left\langle
\lambda_i^{\rm ref}(y)\lambda_{\rm band}^{(d)}(k)
\left[\log\frac{\widehat\pi_i^{(d)}(k,y)}{\pi_i^{(d)}(k,y)}\right]^2
\right\rangle_{k,y,i}.
\end{equation}

For HIT, we sum the three-dimensional Fourier energy over spherical shells in the periodic cubic domain. Let \(\mathcal K_k\) contain the resolved integer wavevectors \(\boldsymbol k'\) satisfying \(k-\tfrac12\le\|\boldsymbol k'\|_2<k+\tfrac12\). The energy per unit mass in shell \(k\), averaged over the time window, is
\begin{equation}\label{eq:hit_spectrum}
E(k)=\frac12\sum_{i=1}^{3}
\left\langle\sum_{\boldsymbol k'\in\mathcal K_k}
\left|\operatorname{DFT}_{xyz}[u_i^n](\boldsymbol k')\right|^2
\right\rangle_{\mathcal W}.
\end{equation}
Summing all shells, including \(k=0\), gives \(\tfrac12\langle|\uvec^n|^2\rangle_{x,y,z,\mathcal W}\). The spectral-shape loss uses the \(N_{\rm shell}\) nonzero resolved shells, normalized by their total energy:
\begin{equation}\label{eq:hit_spectral_shape}
\pi(k)=\frac{E(k)}{\sum_{k'=1}^{N_{\rm shell}}E(k')},
\qquad k=1,\ldots,N_{\rm shell}.
\end{equation}
The predicted shape \(\widehat\pi(k)\) is defined from \(\widehat E(k)\) using its own total nonzero-mode energy. With weights \(\lambda_{\rm band}(k)\) for the different wavenumber ranges, the loss is
\begin{equation}\label{eq:hit_spectral_loss}
\mathcal L_{\rm shape}
=\left\langle \lambda_{\rm band}(k)
\left[\log\frac{\widehat\pi(k)}{\pi(k)}\right]^2\right\rangle_k.
\end{equation}

\paragraph{Energy levels within the trajectory}
Whole-window statistics do not resolve the variation of energy within a trajectory. We divide the trajectory window \(\mathcal W\) into consecutive segments \(\mathcal W_b\), each containing \(N_{\rm seg}\) snapshots, with \(b=1,\ldots,N_{\rm win}/N_{\rm seg}\). We then compare the energy levels of the predicted and reference trajectories within each segment. For channel flow, the mean profile in segment \(b\) is \(\mu_{i,b}(y)=\langle u_i^n\rangle_{x,z,\mathcal W_b}\). We compute the component second moment from fluctuations about this profile for channel flow, and from the velocity itself for HIT:
\begin{equation}\label{eq:segment_energy}
e_{i,b}=\begin{cases}
\left\langle\left(u_i^n-\mu_{i,b}(y)\right)^2
\right\rangle_{x,y,z,\mathcal W_b},&\text{channel flow},\\[3pt]
\left\langle (u_i^n)^2\right\rangle_{x,y,z,\mathcal W_b},&\text{HIT}.
\end{cases}
\end{equation}
For channel flow, \(\tfrac12\sum_i e_{i,b}\) is a nodal average of fluctuation energy per unit mass, with equal weight at each point of the nonuniform \(y\) grid. For HIT, the uniform grid makes it the volume-averaged total kinetic energy per unit mass.

We compute \(\widehat e_{i,b}\) on the same time segments of the predicted trajectory, using its own segment means for channel flow. The energy loss compares these second moments through a squared logarithmic ratio:
\begin{equation}\label{eq:segment_energy_loss}
\mathcal L_{\rm energy}
=\left\langle\left[\log\frac{\widehat e_{i,b}+\alpha_i}
{e_{i,b}+\alpha_i}\right]^2\right\rangle_{i,b}.
\end{equation}
The fixed variance scale \(\alpha_i\) limits the sensitivity of this loss to discrepancies in low-energy segments.

\paragraph{Stage 2 objective}
We combine the statistical losses with nonnegative coefficients \(\lambda\):
\begin{equation}\label{eq:statistical_loss}
\mathcal L_{\rm stat}
=\lambda_{\rm mean}\mathcal L_{\rm mean}
+\lambda_{\rm var}\mathcal L_{\rm var}
+\lambda_{\rm stress}\mathcal L_{\rm stress}
+\lambda_{\rm shape}\mathcal L_{\rm shape}
+\lambda_{\rm energy}\mathcal L_{\rm energy}.
\end{equation}
For a minibatch \(\mathcal B_{\rm traj}\subset\mathcal D_{\rm traj}\), the Stage~2 objective is
\begin{equation}\label{eq:stage2_loss}
\mathcal L_{\mathrm{stage\,2}}(\phi;\mathcal B_{\rm traj})
=\frac{1}{|\mathcal B_{\rm traj}|}
\sum_{(\uvec^0,\uvec^1,\ldots,\uvec^{N_{\rm win}})\in\mathcal B_{\rm traj}}
\bigl(\lambda_{\rm traj}\mathcal L_{\rm traj}+\mathcal L_{\rm stat}\bigr).
\end{equation}
Each trajectory and its reference are evaluated separately before averaging over the minibatch. The loss coefficients are specified in Section~\ref{sec:experimental-protocol}.

Only the FA-CSF parameters \(\phi\) are updated; the predictor parameters \(\Theta^*\) remain fixed. Gradients are retained through the frozen predictors and successive autoregressive steps, so FA-CSF is optimized for the effect of its corrections on subsequent fields and on the statistics of the predicted flow.

\FloatBarrier
\Needspace{8\baselineskip}
\subsection{Two-stage training of SC-IFF and FA-CSF}\label{sec:workflow}\label{sec:native_training}\label{sec:workflow-stage1}

This section summarizes the two-stage training procedure for the SC-IFF predictors and FA-CSF. Stage~1 trains the SC-IFF variants at a few discrete time scales, specified by the power-of-two training strides \(\mathcal S\). Stage~2 freezes the trained SC-IFF predictors and uses their outputs and latent features to learn stride-dependent corrections through FA-CSF. Training on the resulting trajectories at fine time steps constrains their long-time flow statistics. Figure~\ref{fig:two_stage_framework} shows the two-stage workflow, and Algorithms~\ref{alg:native_training} and~\ref{alg:fusion_training} detail the training steps.

\paragraph{Stage 1: SC-IFF predictor training}
We reindex the training trajectories to construct the datasets \(\mathcal D_s\) for the training strides \(s\in\mathcal S\), following Eq.~\eqref{eq:stride_training_examples}. SC-IFF-Concat and SC-IFF-FiLM each train a single predictor on mixed-stride minibatches drawn from the pooled datasets. SC-IFF-Indep trains a separate predictor on each stride-specific dataset.

The data loss \(\mathcal L_{\mathrm{stage\,1}}\) is \(\mathcal L_{\rm data}\) for Concat and FiLM, and \(\mathcal L_{{\rm data},s}\) for each Indep predictor, as defined in Eqs.~\eqref{eq:mixed_stride_data_loss} and~\eqref{eq:stride_data_loss}. Indep training proceeds from the largest stride to the smallest, using the parameters selected on validation trajectories at one stride to initialize the next smaller-stride predictor. Concat and FiLM are validated through hierarchical prediction. For Indep, validation covers all reindexed sequences at the largest stride; at each smaller stride, the candidate is evaluated together with the predictors already selected at larger strides. The validation-selected set \(\Theta^*\) contains the parameters of one predictor for Concat or FiLM, or of all stride-specific predictors for Indep. These parameters are passed to Stage~2.

\paragraph{Stage 2: FA-CSF training}
With \(\Theta^*\) fixed, we train FA-CSF on training windows from \(\mathcal D_{\rm traj}\). Each window provides one initial velocity field and a reference trajectory. The frozen predictors supply the outputs and latent features used to interpolate and correct the fields at the requested strides. The corrected state at the end of each autoregressive step then initializes the next one, generating a longer trajectory through Eq.~\eqref{eq:fused_rollout}. Reference fields are used to evaluate the losses along this trajectory.

The objective \(\mathcal L_{\mathrm{stage\,2}}\) in Eq.~\eqref{eq:stage2_loss} combines trajectory field errors with the statistical losses defined in Section~\ref{sec:trajectory_objective}. Because each window spans several autoregressive steps, these losses constrain statistical deviations that develop as the predictions evolve. Only the FA-CSF parameters \(\phi\) are updated, with gradients retained through the frozen predictors and successive autoregressive steps. Validation trajectories are used to select the FA-CSF parameters.

\begin{figure}[!htbp]
\centering
\input{figure/method/training_prediction_workflow.tex}
\caption{Two-stage training of SC-IFF and FA-CSF. Stage~1 trains predictors on stride-specific datasets. Stage~2 freezes their selected parameters and trains FA-CSF on autoregressive rollouts. Blue and green frames denote the predictors and FA-CSF; snowflakes mark frozen parameters. Red dashed arrows denote gradients, retained through the frozen predictors and successive autoregressive steps.}
\label{fig:two_stage_framework}
\end{figure}

\FloatBarrier
\Needspace{6\baselineskip}
\paragraph{Validation and model selection}\label{sec:model_selection}
Both stages select parameters from autoregressive validation trajectories, assessing their spatial statistics and temporal evolution through the energy-spectrum discrepancy \(\mathcal E_{\rm spec}\) and the temporal-autocorrelation error \(\mathcal E_{\rm ac}\). Each validation sequence starts from its reference initial field; all subsequent fields are predicted, and the initial field is excluded from the scores.

Following the use of spectral covariance by Novati et al.~\citep{Novati2021Automating}, we compare time-mean energy spectra while accounting for correlations between wavenumbers. For channel flow, the absolute directional spectra are \(E^{(d)}(k)=\tfrac12\sum_i\langle P_i^{(d)}(k,y)\rangle_y\), \(d\in\{x,z\}\), using Eq.~\eqref{eq:channel_spectra}; HIT uses the shell spectrum in Eq.~\eqref{eq:hit_spectrum}. Let \(\boldsymbol E\) and \(\widehat{\boldsymbol E}\) collect the reference and predicted time-mean spectra over the retained modes. We define
\begin{equation}\label{eq:selection_spectrum}
\boldsymbol\delta_E=\log\widehat{\boldsymbol E}-\log\boldsymbol E,
\qquad
\mathcal E_{\rm spec}
=\left(\frac{\boldsymbol\delta_E^T\boldsymbol\Sigma_{\rm spec}^{-1}\boldsymbol\delta_E}
{N_{\rm spec}}\right)^{1/2},
\end{equation}
where \(N_{\rm spec}\) is the number of retained modes. The regularized covariance \(\boldsymbol\Sigma_{\rm spec}\) is estimated from instantaneous log spectra of the validation reference fields and held fixed across candidates. Spectra are averaged in time before taking logarithms. In channel flow, applying Eq.~\eqref{eq:selection_spectrum} to the \(x\)- and \(z\)-direction spectra gives \(\mathcal E_{\rm spec}^{(x)}\) and \(\mathcal E_{\rm spec}^{(z)}\), respectively; we combine them as \(\mathcal E_{\rm spec}=[((\mathcal E_{\rm spec}^{(x)})^2+(\mathcal E_{\rm spec}^{(z)})^2)/2]^{1/2}\).

Temporal autocorrelation measures how the velocity field remains correlated as the time lag increases, providing a complementary diagnostic of turbulent evolution~\citep{Zou2025Uncertainty}. For an evaluation window containing \(N_{\rm win}\) fields, we average over all available pairs at each lag:
\begin{equation}\label{eq:selection_acf}
f_{\rm ac}(h\Dt)=\frac{N_{\rm win}}{N_{\rm win}-h}
\begin{cases}
\displaystyle\frac{\sum_{n=1}^{N_{\rm win}-h}
\left\langle\uvec^{\prime n}\!\cdot\uvec^{\prime(n+h)}\right\rangle_{x,y,z}}
{\sum_{n=1}^{N_{\rm win}}
\left\langle|\uvec^{\prime n}|^2\right\rangle_{x,y,z}},&\text{channel flow},\\[2ex]
\displaystyle\frac{\sum_{n=1}^{N_{\rm win}-h}
\left\langle\uvec^{n}\!\cdot\uvec^{n+h}\right\rangle_{x,y,z}}
{\sum_{n=1}^{N_{\rm win}}
\left\langle|\uvec^{n}|^2\right\rangle_{x,y,z}},&\text{HIT},
\end{cases}
\end{equation}
where \(0\le h<N_{\rm win}\). Channel flow uses fluctuations about the sequence's own time-mean profile, defined in Eq.~\eqref{eq:window_statistics}, whereas HIT uses the velocity field without mean subtraction. A stride-\(s\) sequence has lag \(hs\Dt\). The predicted autocorrelation \(\widehat f_{\rm ac}\) is computed from the predicted sequence using its own normalization and, for channel flow, its own mean profile. We measure its discrepancy from the reference by
\begin{equation}\label{eq:selection_ac_error}
\mathcal E_{\rm ac}
=\frac{\|\widehat f_{\rm ac}-f_{\rm ac}\|_2}
{\|f_{\rm ac}\|_2},
\end{equation}
where \(\|\cdot\|_2\) denotes the \(L^2\) norm over the common physical time-lag interval used for each comparison, evaluated by trapezoidal integration.

Both scores are computed separately for each validation trajectory and then averaged with equal weight. For reindexed sequences, scores are first averaged within their original trajectory. Lower values indicate closer agreement with the reference statistics. Among candidates with complete, finite validation results, we retain the Pareto candidates: no other candidate is at least as good in both scores and strictly better in one. Among these candidates, we minimize
\begin{equation}\label{eq:selection_rule}
\max\left\{
\frac{\mathcal E_{\rm spec}}{\mathcal E_{\rm spec,min}},\;
\frac{\mathcal E_{\rm ac}}{\mathcal E_{\rm ac,min}}
\right\},
\end{equation}
where the denominators are the respective minimum scores across the eligible candidates. This balances the two measures by limiting the worse normalized discrepancy. Exact ties are resolved by the short-window validation error and then the earlier checkpoint. The two scores are used for selection, not as additional training losses; test trajectories are evaluated only after the parameters have been selected.

\begingroup
\makeatletter
\setlength{\@fptop}{0pt}
\setlength{\@fpsep}{16pt}
\setlength{\@fpbot}{0pt plus 1fil}
\makeatother
\begin{algorithm}[!htbp]
\caption{Stage 1: SC-IFF predictor training}\label{alg:native_training}
\small
\begin{algorithmic}[1]
\Require Training and validation trajectories; training strides \(\mathcal S\); chosen SC-IFF variant
\State Fit normalization statistics on the training trajectories and construct \(\mathcal D_s\) for each \(s\in\mathcal S\).
\If{the variant is SC-IFF-Concat or SC-IFF-FiLM}
  \State Initialize one predictor with parameters \(\Theta\); pool \(\mathcal D=\bigcup_{s\in\mathcal S}\mathcal D_s\).
  \State Optimize \(\Theta\) on mixed-stride minibatches \(\mathcal B\subset\mathcal D\) using Eq.~\eqref{eq:mixed_stride_data_loss}; save candidate checkpoints.
  \State Evaluate each checkpoint by hierarchical prediction on validation trajectories using \(\mathcal E_{\rm spec}\) and \(\mathcal E_{\rm ac}\).
  \State Select \(\Theta^*\) from these checkpoints using Eq.~\eqref{eq:selection_rule}.
\Else\Comment{SC-IFF-Indep}
  \For{\(s=s_{\max},s_{\max}/2,\ldots,1\)}
    \If{\(s=s_{\max}\)}
      \State Initialize the largest-stride predictor with parameters \(\theta_s\).
    \Else
      \State Initialize \(\theta_s\gets\theta_{2s}^*\).
    \EndIf
    \State Use the normalization statistics for stride \(s\).
    \State Optimize \(\theta_s\) on \(\mathcal B_s\subset\mathcal D_s\) using Eq.~\eqref{eq:stride_data_loss}; save candidate checkpoints.
    \If{\(s=s_{\max}\)}
      \State Evaluate each checkpoint by autoregressive prediction on all reindexed validation sequences.
    \Else
      \State Evaluate each checkpoint by hierarchical prediction with the predictors already selected at larger strides.
    \EndIf
    \State Compute \(\mathcal E_{\rm spec}\) and \(\mathcal E_{\rm ac}\); select \(\theta_s^*\) using Eq.~\eqref{eq:selection_rule}.
  \EndFor
  \State Collect \(\Theta^*=\{\theta_s^*:s\in\mathcal S\}\).
\EndIf
\State Retain \(\Theta^*\) and the normalization statistics for FA-CSF training.
\end{algorithmic}
\end{algorithm}

\begin{algorithm}[!htbp]
\caption{Stage 2: FA-CSF training on autoregressive rollouts}\label{alg:fusion_training}
\small
\begin{algorithmic}[1]
\Require Frozen \(\Theta^*\); initialized \(\phi\); training windows \(\mathcal D_{\rm traj}\); validation trajectories
\State Fit the rate and loss-normalization scales on the training trajectories.
\For{each optimization update}
  \State Draw a minibatch \(\mathcal B_{\rm traj}\subset\mathcal D_{\rm traj}\).
  \For{each window \((\uvec^0,\uvec^1,\ldots,\uvec^{N_{\rm win}})\in\mathcal B_{\rm traj}\)}
    \State Initialize \(\widehat{\uvec}^{\,0}=\uvec^0\) and \(n=0\).
    \While{\(n<N_{\rm win}\)}
      \State Evaluate \(\mathcal G(\widehat{\uvec}^{\,n},s;\Theta^*)\) once for each \(s\in\mathcal S\), retaining the fields, rates, and latent features.
      \State Use the current field and these outputs in FA-CSF to generate \(\widehat{\uvec}^{\,n+1},\ldots,\widehat{\uvec}^{\,n+s_{\max}}\).
      \State Retain the fields within the window; use the corrected state at the end of the autoregressive step as the next input and advance \(n\gets n+s_{\max}\).
    \EndWhile
    \State Compute \(\mathcal L_{\rm traj}\) and \(\mathcal L_{\rm stat}\) against this window's reference trajectory.
  \EndFor
  \State Average the window losses using Eq.~\eqref{eq:stage2_loss} and update only \(\phi\).
  \Statex \hspace{\algorithmicindent}Retain gradients through the frozen predictors and all autoregressive steps.
  \State Save \(\phi\) at the prescribed checkpoint updates.
\EndFor
\State Evaluate the saved checkpoints on validation trajectories using \(\mathcal E_{\rm spec}\) and \(\mathcal E_{\rm ac}\).
\State Select \(\phi^*\) using Eq.~\eqref{eq:selection_rule}, with \(\Theta^*\) unchanged.
\end{algorithmic}
\end{algorithm}
\FloatBarrier
\endgroup

\FloatBarrier\Needspace{9\baselineskip}

\section{Numerical experiments}\label{sec:experiments}\label{sec:experimental-protocol}
\setcounter{equation}{0}

We evaluate the proposed framework on turbulent channel flow at $\Retau\approx180$ and forced homogeneous isotropic turbulence (HIT) at $\Rey_\lambda\approx100$, with emphasis on long-time prediction at fine time steps. The experiments assess velocity-field accuracy, the reproduction of long-time flow statistics, and the computational cost of generating trajectories at fine time steps. We first consider a sampling interval of 50 DNS time steps (DT50), and then extend the evaluation to the finer sampling interval of 10 DNS time steps (DT10). The resulting datasets have finer time steps than those used in previous studies of the same channel-flow and HIT benchmarks, where snapshots were recorded every 200 DNS time steps \citep{Yang2024An,Yang2026Implicit}. Further comparisons examine the roles of multi-stride prediction and learned correction in the observed performance.

In these experiments, we regard an autoregressive prediction as stable over the prediction time considered when it remains bounded and its long-time statistics stay close to the fDNS reference. This is an empirical criterion for the finite prediction times considered here, not a theoretical stability guarantee.

Table~\ref{tab:temporal_protocol} summarizes the datasets and trajectory splits. Here, $\Delta T$ is the nondimensional sampling interval, $N_{\rm snap}$ is the number of stored fields per trajectory, and split sizes count trajectories. The data are partitioned into training, validation, and test sets before constructing stride-dependent samples or training windows. Normalization and loss scales are fitted using the training data, and model weights are selected using the validation set. Each test trajectory is predicted autoregressively from its first stored field without further reference inputs. The selected predictors and complete models are evaluated on the test set, with the initial field excluded from the reported statistics. The weights of all Stage~2 loss terms in Eqs.~\eqref{eq:statistical_loss} and~\eqref{eq:stage2_loss} are set to one.

\begin{table}[!htbp]
\centering
\papertablestyle
\begin{threeparttable}
\caption{Sampling intervals, trajectory lengths and data splits for the channel-flow and HIT datasets.}
\label{tab:temporal_protocol}
\begin{tabular*}{\linewidth}{@{\extracolsep{\fill}}l r r r r r@{}}\toprule
Dataset & $\Delta T$ & $N_{\rm snap}$ & Training & Validation & Test\\\midrule
Channel-DT50 & 0.25 & 2000 & 12 & 4 & 4\\
Channel-DT10 & 0.05 & 4000 & 12 & 4 & 4\\
HIT-DT50 & 0.05 & 1600 & 12 & 4 & 4\\
HIT-DT10 & 0.01 & 3200 & 12 & 4 & 4\\\bottomrule
\end{tabular*}
\end{threeparttable}
\end{table}

IFF is used as the fixed-stride baseline at $s=1$. IFF and the three SC-IFF variants use latent and per-head value channel dimensions \(C=C_v=96\) and 5 shared implicit iterations. The SC-IFF variants use \(\mathcal S_{\rm DT50}=\{1,2,4\}\) on the DT50 datasets and \(\mathcal S_{\rm DT10}=\{1,2,4,8,16,32\}\) on the DT10 datasets. Concat and FiLM share one predictor across strides, whereas Indep stores a separate predictor for each stride. Each variant is also combined with FA-CSF, which uses \(C_{\rm f}=48\) feature channels. For both IFF and FA-CSF, the query/key dimension per attention head is set to \(C_q=3C_v\). The SC-IFF count includes all stored predictors for Indep, while the FA-CSF count gives the additional correction parameters. Table~\ref{tab:parameter_counts} summarizes the model configurations and parameter counts. Only the SC-IFF-Indep parameter count scales with the number of training strides \(|\mathcal S|\); the parameter counts of IFF, Concat, and FiLM are unchanged between DT50 and DT10.

\begin{table}[!htbp]\centering
\papertablestyle
\begin{threeparttable}
\caption{Configurations and parameter counts of IFF, SC-IFF and FA-CSF.}
\label{tab:parameter_counts}
\begin{tabular*}{\linewidth}{@{\extracolsep{\fill}}l r r r r r@{}}\toprule
Model & Channels & Attention heads & Implicit iterations & \multicolumn{2}{r}{Parameters}\\\midrule
IFF & \(C=C_v=96\) & 5 & 5 & & 1,165,059\\
SC-IFF-Concat & \(C=C_v=96\) & 5 & 5 & & 1,165,203\\\cmidrule(l){5-6}
\multirow[t]{2}{*}{SC-IFF-Indep} & \multirow[t]{2}{*}{\(C=C_v=96\)} & \multirow[t]{2}{*}{5} & \multirow[t]{2}{*}{5} & DT50 (\(|\mathcal S_{\rm DT50}|=3\)) & 3,495,177\\
& & & & DT10 (\(|\mathcal S_{\rm DT10}|=6\)) & 6,990,354\\\cmidrule(l){5-6}
SC-IFF-FiLM & \(C=C_v=96\) & 5 & 5 & & 1,168,747\\\midrule
FA-CSF & \(C_{\rm f}=48,\ C_v=32\) & 3 & 1 & & 141,283\\\bottomrule
\end{tabular*}
\end{threeparttable}
\end{table}
\FloatBarrier

In Stage~1, the predictors are trained with AdamW, using a cosine learning-rate schedule from $5\times10^{-4}$ to $5\times10^{-6}$, weight decay $10^{-2}$, and gradient clipping at unit norm. The batch size is 10 for channel flow and 5 for HIT. For both flows, all Stage~1 models, including fixed-stride IFF, use the data loss with \(M=2\) unrolling steps. Table~\ref{tab:common_training_settings} summarizes the training settings and budgets.

In Stage~2, the selected SC-IFF predictors are frozen and only the FA-CSF parameters are trained with the trajectory and statistical objectives. This stage uses AdamW with a constant learning rate of $10^{-4}$ and weight decay $10^{-4}$. Each minibatch contains 4 input--output pairs randomly sampled from the training trajectories. We perform one parameter update per minibatch, with 360 updates for DT50 and 720 for DT10.

\begin{table}[!htbp]\centering
\papertablestyle
\begin{threeparttable}
\caption{Optimization settings and training budgets for Stage~1 predictors and Stage~2 FA-CSF.}
\label{tab:common_training_settings}
\begin{tabular*}{\linewidth}{@{\extracolsep{\fill}}l r r r@{}}\toprule
\multicolumn{2}{@{}l}{Setting} & Stage 1: IFF / SC-IFF & Stage 2: FA-CSF\\\midrule
\multicolumn{2}{@{}l}{Optimizer} & AdamW & AdamW\\
\multicolumn{2}{@{}l}{Weight decay} & $10^{-2}$ & $10^{-4}$\\
\multicolumn{2}{@{}l}{Learning rate} & Cosine, $5\times10^{-4}$ to $5\times10^{-6}$ & Constant, $10^{-4}$\\\cmidrule(l){2-4}
\multicolumn{2}{@{}l}{\multirow[t]{2}{*}{Batch size}} & 10 (channel flow) & \multirow[t]{2}{*}{4 training windows}\\
\multicolumn{2}{@{}l}{} & 5 (HIT) & \\\cmidrule(l){2-4}
\multirow[t]{6}{*}{Training budget} & \multirow[t]{3}{*}{DT50} & 24 epochs on \(\mathcal D_1\) for the fixed-stride baseline & \multirow[t]{3}{*}{\begin{tabular}[t]{@{}r@{}}360 updates\\\(N_{\rm win}=256\)\\\(N_{\rm seg}=8\)\end{tabular}}\\
& & 8 epochs on \(\mathcal D=\bigcup_{s\in\mathcal S_{\rm DT50}}\mathcal D_s\) for Concat/FiLM & \\
& & 14 epochs on \(\mathcal D_4\); 5 epochs on each \(\mathcal D_s\), \(s\in\mathcal S_{\rm DT50}\setminus\{4\}\), for Indep & \\\cmidrule(l){3-4}
& \multirow[t]{3}{*}{DT10} & 48 epochs on \(\mathcal D_1\) for the fixed-stride baseline & \multirow[t]{3}{*}{\begin{tabular}[t]{@{}r@{}}720 updates\\\(N_{\rm win}=1024\)\\\(N_{\rm seg}=32\)\end{tabular}}\\
& & 8 epochs on \(\mathcal D=\bigcup_{s\in\mathcal S_{\rm DT10}}\mathcal D_s\) for Concat/FiLM & \\
& & 18 epochs on \(\mathcal D_{32}\); 6 epochs on each \(\mathcal D_s\), \(s\in\mathcal S_{\rm DT10}\setminus\{32\}\), for Indep & \\\cmidrule(l){2-4}
\multicolumn{2}{@{}l}{Evaluation} & Every epoch & Every 60 updates\\\bottomrule
\end{tabular*}\end{threeparttable}
\end{table}

Table~\ref{tab:training_cost_all_cases} summarizes the training cost on a single NVIDIA A100 PCIe 40~GB GPU. Although SC-IFF-Indep stores more parameters, its training time and GPU memory usage are comparable to those of the shared variants. Sequential transfer learning trains one predictor at a time in Stage~1, while Stage~2 stores parameter gradients and optimizer states only for FA-CSF. Stage~1 update counts differ owing to stride-dependent dataset construction and discarded incomplete minibatches; models use comparable budgets and are selected throughout training using \(\mathcal E_{\rm spec}\) and \(\mathcal E_{\rm ac}\).

\begin{table}[!htbp]\centering
\papertablestyle
\setlength{\tabcolsep}{3pt}
\renewcommand{\arraystretch}{1.18}
\begin{threeparttable}
\caption{Training updates, training time and GPU memory for channel flow and HIT.}
\label{tab:training_cost_all_cases}
\begin{tabular*}{\linewidth}{@{\extracolsep{\fill}}llrrrrrr@{}}
\toprule
& & \multicolumn{3}{c}{Channel flow} & \multicolumn{3}{c}{HIT}\\
\cmidrule(lr){3-5}\cmidrule(l){6-8}
Stage & Model & Updates & \shortstack{Time\\(h)} & \shortstack{GPU Memory\\(GiB)} & Updates & \shortstack{Time\\(h)} & \shortstack{GPU Memory\\(GiB)}\\
\midrule
\multicolumn{8}{@{}l}{\textbf{DT50}}\\[3pt]
\multirow[t]{4}{*}{Stage 1} & Fixed-stride IFF & $57{,}528$ & 12.38 & 36.07 & $92{,}040$ & 28.16 & 34.86\\
 & SC-IFF-Concat & $57{,}464$ & 12.46 & 36.07 & $91{,}888$ & 28.35 & 34.86\\
 & SC-IFF-Indep & $57{,}420$ & 12.51 & 36.07 & $91{,}805$ & 28.23 & 34.86\\
 & SC-IFF-FiLM & $57{,}464$ & 12.58 & 36.73 & $91{,}888$ & 28.34 & 35.50\\
\cmidrule(l){2-8}
\multirow[t]{3}{*}{Stage 2} & SC-IFF-Concat + FA-CSF & $360$ & 7.16 & 6.41 & $360$ & 14.83 & 12.38\\
 & SC-IFF-Indep + FA-CSF & $360$ & 7.17 & 6.42 & $360$ & 14.74 & 12.39\\
 & SC-IFF-FiLM + FA-CSF & $360$ & 7.30 & 6.41 & $360$ & 14.80 & 12.38\\
\midrule
\multicolumn{8}{@{}l}{\textbf{DT10}}\\[3pt]
\multirow[t]{4}{*}{Stage 1} & Fixed-stride IFF & $230{,}256$ & 49.42 & 36.07 & $368{,}400$ & 112.69 & 34.86\\
 & SC-IFF-Concat & $229{,}184$ & 49.26 & 36.07 & $366{,}216$ & 112.04 & 34.86\\
 & SC-IFF-Indep & $228{,}552$ & 50.11 & 36.07 & $364{,}962$ & 113.80 & 34.86\\
 & SC-IFF-FiLM & $229{,}184$ & 49.83 & 36.73 & $366{,}216$ & 113.10 & 35.50\\
\cmidrule(l){2-8}
\multirow[t]{3}{*}{Stage 2} & SC-IFF-Concat + FA-CSF & $720$ & 14.87 & 13.44 & $720$ & 30.53 & 17.03\\
 & SC-IFF-Indep + FA-CSF & $720$ & 14.82 & 13.46 & $720$ & 30.47 & 17.05\\
 & SC-IFF-FiLM + FA-CSF & $720$ & 14.86 & 13.44 & $720$ & 30.60 & 17.03\\
\bottomrule
\end{tabular*}
\end{threeparttable}
\end{table}

As conventional LES references, we use the dynamic Smagorinsky model (DSM)~\citep{MoinSquiresCabotLee1991Dynamic} for both flows and the wall-adapting local eddy-viscosity (WALE) model~\citep{NicoudDucros1999WALE} for channel flow. Their fields are compared with fDNS on the same coarse grids. The WALE coefficient \(C_w\) is set to the reference value of 0.5 given in the original study.

\begingroup
\renewcommand{\topfraction}{0.9}
\renewcommand{\bottomfraction}{0.85}
\renewcommand{\textfraction}{0.08}
\renewcommand{\floatpagefraction}{0.8}
\Needspace{8\baselineskip}
\subsection{Turbulent channel flow}\label{sec:channel_results}
\clubpenalty=10000
\widowpenalty=10000
\brokenpenalty=10000
The channel has dimensions $(L_x,L_y,L_z)=(4\pi,2,4\pi/3)$ and viscosity $\nu=1/4200$. The DNS grid is $192\times129\times64$, and the fDNS fields are stored on a $32\times33\times16$ grid. With a DNS time step of 0.005, DT50 and DT10 correspond to $\Delta T=0.25$ and 0.05, respectively. Each rollout uses the first snapshot as input and predicts the remaining snapshots of a trajectory containing 2000 (DT50) or 4000 (DT10) snapshots.

\label{sec:channel-settings}

We compare the mean streamwise velocity, the RMS profiles of the three velocity fluctuations, the Reynolds shear stress, and the streamwise and spanwise energy spectra. Mean profiles are averaged over time and the homogeneous $x,z$ directions; velocity fluctuations are defined relative to the corresponding mean profiles.

\Needspace{6\baselineskip}
\subsubsection{\textnormal{Turbulent channel flow sampled every 50 DNS time steps (Channel-DT50)}}\label{sec:channel-dt50}
On Channel-DT50, the SC-IFF variants generally reproduce the fDNS flow statistics more closely than IFF, as seen in the mean-flow and Reynolds-stress profiles. Adding FA-CSF to SC-IFF further improves some of the predicted flow statistics.

Table~\ref{tab:channel_dt50_modes} compares DSM, WALE and IFF with the three SC-IFF variants and their FA-CSF models. We report the errors in the mean streamwise velocity profile $\epsilon_\mu$, the combined RMS velocity fluctuation profiles $\epsilon_{\rm rms}$, the Reynolds shear stress profile $\epsilon_{uv}$, the streamwise energy spectrum $\epsilon_{E_x}$, and the spanwise energy spectrum $\epsilon_{E_z}$. For a statistical quantity $q$, the relative $L^2$ error is $\epsilon_q=\|\widehat q-q_{\rm fDNS}\|_2/\|q_{\rm fDNS}\|_2$, where $\widehat q$ is the predicted statistic and $q_{\rm fDNS}$ is its reference value. Profile norms use wall-normal quadrature weights, the RMS error combines all three velocity components, and spectral norms include all nonzero resolved wavenumbers. Statistics are computed over the entire predicted trajectory, excluding the initial snapshot.

\begin{table}[!htbp]\centering
\papertablestyle
\begin{threeparttable}
\caption{Relative $L^2$ errors in time-averaged flow statistics for Channel-DT50.}
\label{tab:channel_dt50_modes}
{
\begin{tabular*}{\linewidth}{@{\extracolsep{\fill}}l r r r r r@{}}\toprule
Model & $\epsilon_\mu$ & $\epsilon_{\rm rms}$ & $\epsilon_{uv}$ & $\epsilon_{E_x}$ & $\epsilon_{E_z}$\\\midrule
DSM & $3.53\times10^{-2}$ & $4.79\times10^{-1}$ & $8.61\times10^{-1}$ & $1.20\times10^{0}$ & $1.06\times10^{0}$\\
WALE & $3.47\times10^{-2}$ & $2.41\times10^{-1}$ & $\boldsymbol{1.01\times10^{-1}}$ & $4.12\times10^{-1}$ & $5.59\times10^{-1}$\\
Fixed-stride IFF & $1.05\times10^{-1}$ & $2.77\times10^{-1}$ & $6.69\times10^{-1}$ & $\boldsymbol{1.92\times10^{-1}}$ & $3.53\times10^{-1}$\\
\midrule
SC-IFF-Concat & $\boldsymbol{7.23\times10^{-3}}$ & $2.96\times10^{-1}$ & $4.68\times10^{-1}$ & $4.84\times10^{-1}$ & $3.49\times10^{-1}$\\
SC-IFF-Indep & $9.15\times10^{-3}$ & $2.48\times10^{-1}$ & $3.01\times10^{-1}$ & $3.03\times10^{-1}$ & $2.98\times10^{-1}$\\
SC-IFF-FiLM & $7.80\times10^{-3}$ & $1.69\times10^{-1}$ & $2.26\times10^{-1}$ & $5.66\times10^{-1}$ & $2.75\times10^{-1}$\\
\midrule
SC-IFF-Concat + FA-CSF & $2.15\times10^{-2}$ & $3.39\times10^{-1}$ & $5.22\times10^{-1}$ & $5.97\times10^{-1}$ & $3.93\times10^{-1}$\\
SC-IFF-Indep + FA-CSF & $2.03\times10^{-2}$ & $4.18\times10^{-1}$ & $5.62\times10^{-1}$ & $5.27\times10^{-1}$ & $5.97\times10^{-1}$\\
SC-IFF-FiLM + FA-CSF & $1.72\times10^{-2}$ & $\underline{\boldsymbol{4.46\times10^{-2}}}$ & $\underline{1.12\times10^{-1}}$ & $\underline{5.61\times10^{-1}}$ & $\underline{\boldsymbol{1.70\times10^{-1}}}$\\
\bottomrule\end{tabular*}}
\end{threeparttable}
\end{table}

Bold entries indicate the smallest error in each column, while underlined entries indicate a reduction after adding FA-CSF to the corresponding SC-IFF variant, based on unrounded values.

Field accuracy and the different statistical objectives can impose competing requirements in multi-objective training. Reducing the combined training loss therefore does not guarantee that every rollout statistic moves closer to fDNS. In this Channel-DT50 case, FA-CSF improves four of the five statistical diagnostics for SC-IFF-FiLM, although the mean-profile error increases. Among the corrected variants, FiLM has the smallest mean-profile, RMS-profile, shear-stress and spanwise-spectrum errors, while Indep has the smallest streamwise-spectrum error. FiLM's corrected RMS-profile and spanwise-spectrum errors are also the lowest among all compared methods.

Figure~\ref{fig:channel_50_FiLM_profiles} visualizes these statistical differences. In the first column, DSM exhibits substantial discrepancies in the fluctuation intensities and Reynolds shear stress. WALE agrees more closely with fDNS, but underpredicts the near-wall mean velocity and shows deviations in the streamwise and wall-normal RMS velocity fluctuation profiles. Although WALE has the correct near-wall eddy-viscosity scaling, the coarse $32\times33\times16$ grid may not adequately resolve the near-wall velocity gradients needed for accurate stress modeling. Among the machine-learning methods, fixed-stride IFF captures the overall shape of the wall-normal RMS velocity fluctuation profile, but shows deviations in the mean velocity, streamwise and spanwise RMS velocity fluctuation profiles, and Reynolds shear stress. In the middle column, SC-IFF reproduces the mean streamwise velocity $\langle u^+\rangle$, streamwise RMS velocity fluctuation $u_{\rm rms}^{\prime+}$ and Reynolds shear stress $\langle u'v'\rangle$ more closely overall. However, the wall-normal and spanwise RMS velocity fluctuation profiles $v_{\rm rms}^{\prime+}$ and $w_{\rm rms}^{\prime+}$ remain below fDNS.

Although SC-IFF-FiLM still underpredicts the wall-normal and spanwise fluctuations in Stage~1, adding FA-CSF and training it in Stage~2 brings all three RMS velocity fluctuation profiles and the Reynolds shear stress into closer agreement with fDNS. By contrast, the wall-normal and spanwise fluctuation deficits of Concat and Indep persist and can increase after correction, as seen by comparing panels (h) with (i) and (k) with (l).

In Figure~\ref{fig:channel_dt50_spectra}, FA-CSF partially restores the high-wavenumber energy underpredicted by SC-IFF-FiLM in both directions. The spanwise-spectrum error decreases, while the overall streamwise-spectrum error changes only slightly. FiLM's corrected streamwise-spectrum error remains above WALE's.

Stride conditioning improves some statistics over fixed-stride IFF on Channel-DT50, with larger gains at the finer sampling interval of Channel-DT10, discussed in the next subsection. The contrasting responses of Concat, Indep and FiLM suggest that Stage~2 correction depends on the Stage~1 prediction. Large statistical biases may be difficult to remove when field and statistical objectives are optimized together over long trajectories.

\begin{figure}[!htbp]\centering\setlength{\abovecaptionskip}{8pt}
\includegraphics[width=\linewidth]{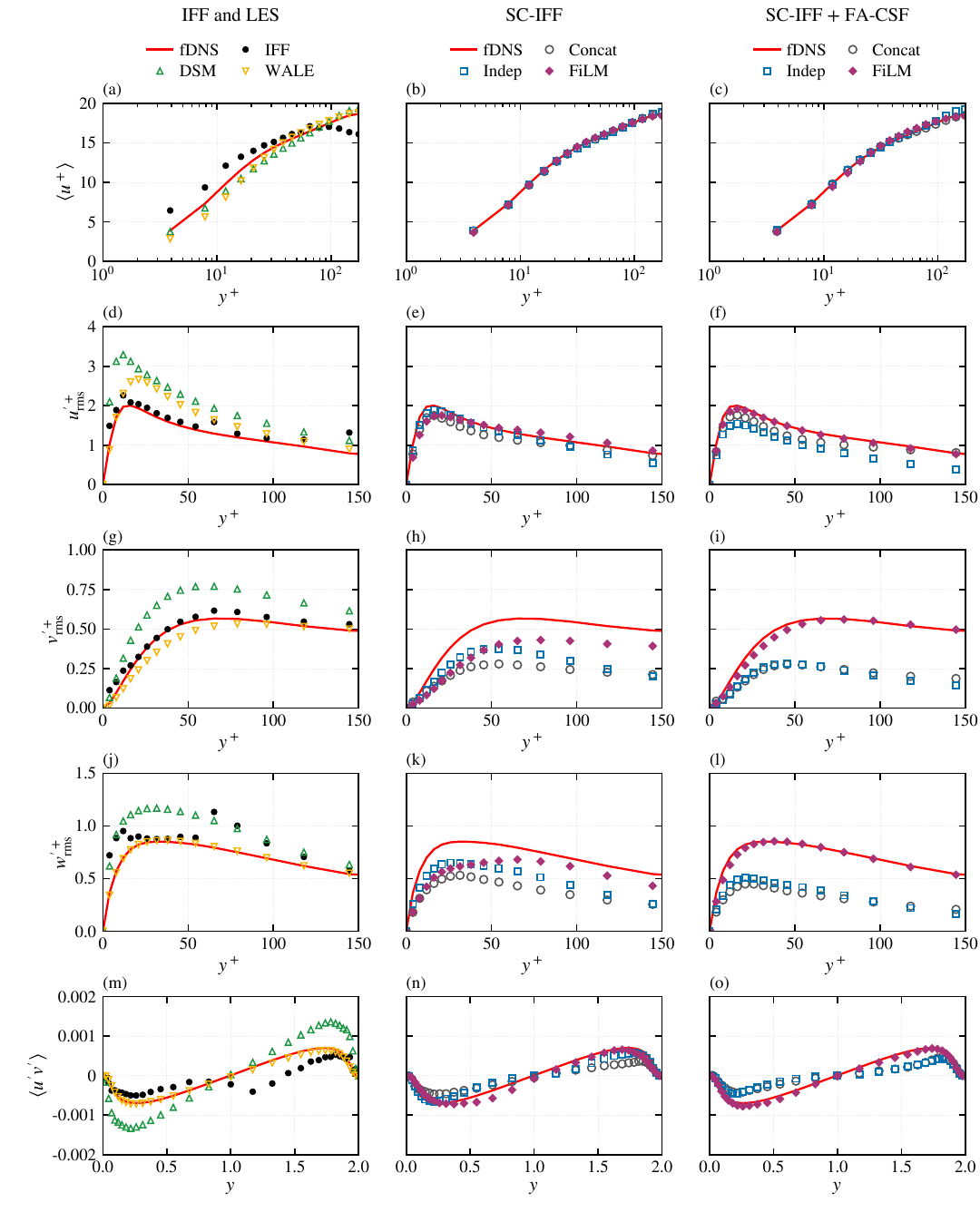}\par
\footnotesize
\caption{Time-averaged profiles for Channel-DT50. (a--c) Mean streamwise velocity; (d--f), (g--i) and (j--l) RMS fluctuations of streamwise, wall-normal and spanwise velocity, respectively; (m--o) Reynolds shear stress.}
\label{fig:channel_50_FiLM_profiles}\end{figure}

\begin{figure}[!htbp]\centering\setlength{\abovecaptionskip}{8pt}
\includegraphics[width=\linewidth]{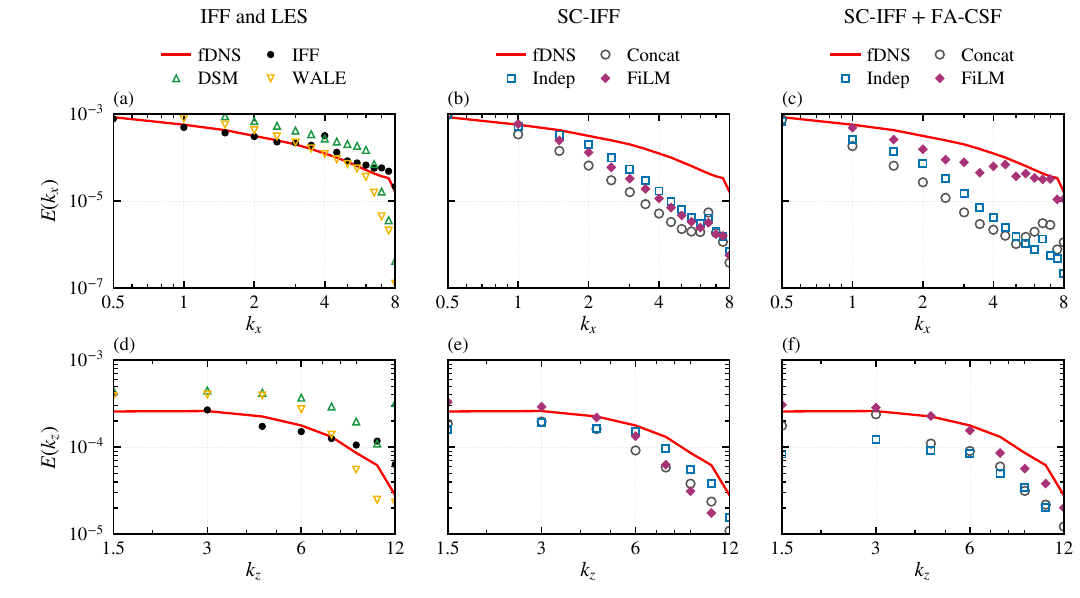}\par
\footnotesize
\caption{Time-averaged energy spectra for Channel-DT50. (a--c) Streamwise; (d--f) spanwise.}
\label{fig:channel_dt50_spectra}\end{figure}

\FloatBarrier
\subsubsection{\textnormal{Turbulent channel flow sampled every 10 DNS time steps (Channel-DT10)}}\label{sec:channel-dt10}
We next examine long-time prediction in Channel-DT10, whose sampling interval is one fifth of that of Channel-DT50. The three SC-IFF variants reproduce the reference flow statistics more closely than IFF. FA-CSF gives more consistent statistical improvements in this case.

In Table~\ref{tab:channel_fusion_summary}, Stage~2 training reduces nearly all statistical errors for Concat, Indep and FiLM. These results show that FA-CSF training effectively improves flow statistics on the Channel-DT10 data with finer time steps.

\begin{table}[!htbp]\centering
\papertablestyle
\begin{threeparttable}
\caption{Relative $L^2$ errors in time-averaged flow statistics for Channel-DT10.}
\label{tab:channel_fusion_summary}
{
\begin{tabular*}{\linewidth}{@{\extracolsep{\fill}}l r r r r r@{}}\toprule
Model & $\epsilon_\mu$ & $\epsilon_{\rm rms}$ & $\epsilon_{uv}$ & $\epsilon_{E_x}$ & $\epsilon_{E_z}$\\\midrule
DSM & $3.05\times10^{-2}$ & $3.97\times10^{-1}$ & $6.24\times10^{-1}$ & $9.87\times10^{-1}$ & $8.38\times10^{-1}$\\
WALE & $3.47\times10^{-2}$ & $2.48\times10^{-1}$ & $7.88\times10^{-2}$ & $3.56\times10^{-1}$ & $6.07\times10^{-1}$\\
Fixed-stride IFF & $1.22\times10^{-1}$ & $1.42\times10^{0}$ & $4.18\times10^{0}$ & $5.67\times10^{-1}$ & $8.93\times10^{-1}$\\
\midrule
SC-IFF-Concat & $1.63\times10^{-2}$ & $1.24\times10^{-1}$ & $1.38\times10^{-1}$ & $2.60\times10^{-1}$ & $2.67\times10^{-1}$\\
SC-IFF-Indep & $7.49\times10^{-3}$ & $4.64\times10^{-2}$ & $9.95\times10^{-2}$ & $8.83\times10^{-2}$ & $1.16\times10^{-1}$\\
SC-IFF-FiLM & $1.55\times10^{-2}$ & $6.72\times10^{-2}$ & $1.38\times10^{-1}$ & $1.52\times10^{-1}$ & $1.07\times10^{-1}$\\
\midrule
SC-IFF-Concat + FA-CSF & \underline{$\boldsymbol{5.15\times10^{-3}}$} & \underline{$2.48\times10^{-2}$} & \underline{$\boldsymbol{3.85\times10^{-2}}$} & \underline{$1.12\times10^{-1}$} & \underline{$7.02\times10^{-2}$}\\
SC-IFF-Indep + FA-CSF & $8.22\times10^{-3}$ & \underline{$2.67\times10^{-2}$} & \underline{$5.77\times10^{-2}$} & \underline{$\boldsymbol{5.25\times10^{-2}}$} & \underline{$9.83\times10^{-2}$}\\
SC-IFF-FiLM + FA-CSF & \underline{$7.82\times10^{-3}$} & \underline{$\boldsymbol{2.44\times10^{-2}}$} & \underline{$3.90\times10^{-2}$} & \underline{$7.73\times10^{-2}$} & \underline{$\boldsymbol{3.56\times10^{-2}}$}\\
\bottomrule\end{tabular*}}
\end{threeparttable}
\end{table}

\begin{figure}[!htbp]\centering\setlength{\abovecaptionskip}{8pt}
\includegraphics[width=\linewidth]{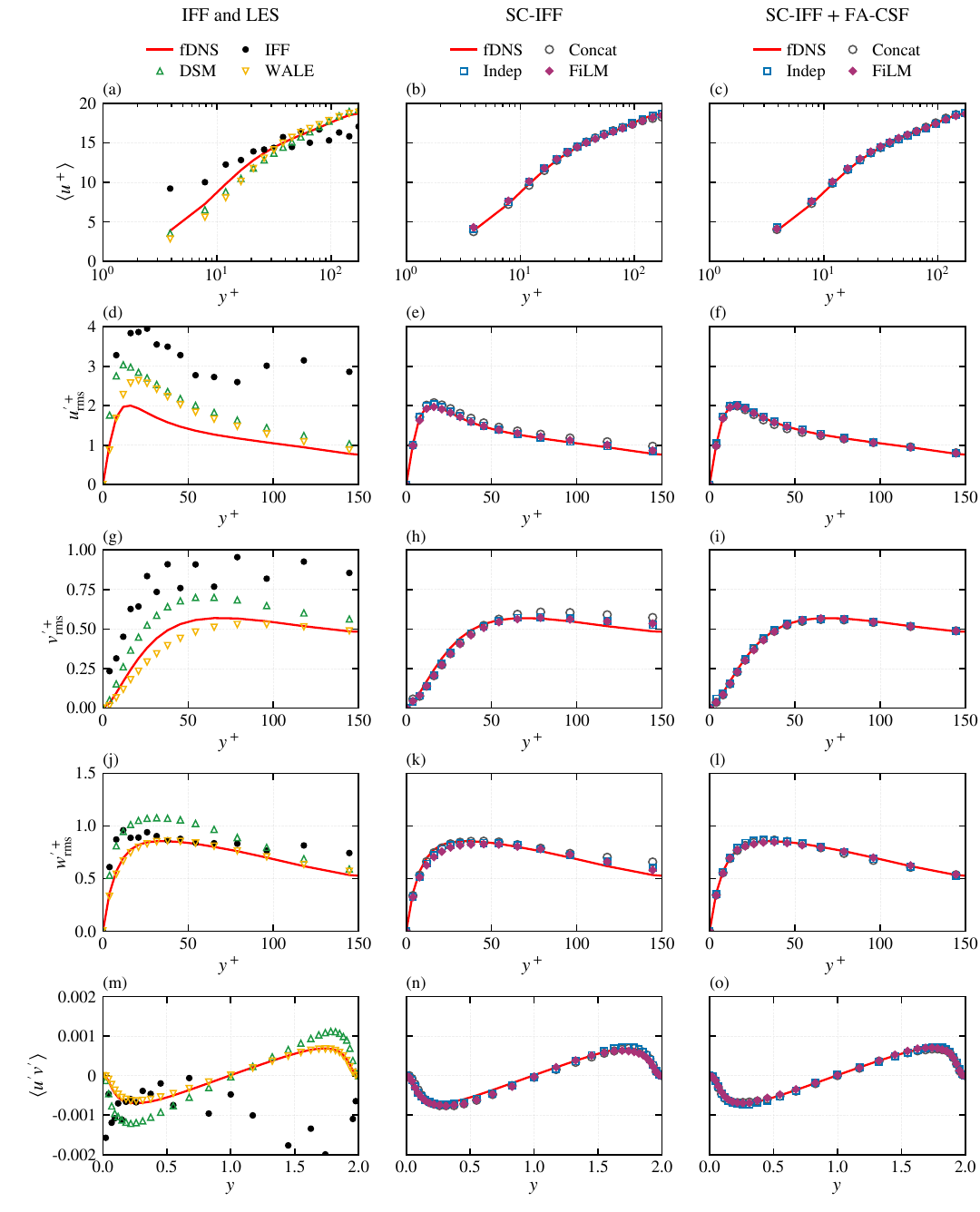}\par
\footnotesize
\caption{Time-averaged profiles for Channel-DT10. (a--c) Mean streamwise velocity; (d--f), (g--i) and (j--l) RMS fluctuations of streamwise, wall-normal and spanwise velocity, respectively; (m--o) Reynolds shear stress.}
\label{fig:channel_10_FiLM_profiles}\end{figure}
\FloatBarrier

In Figure~\ref{fig:channel_10_FiLM_profiles}, DSM and WALE reproduce the mean streamwise velocity more closely than the fluctuation profiles, as in Channel-DT50. DSM shows substantial discrepancies in all three fluctuation components and the Reynolds shear stress. WALE agrees better with the reference spanwise fluctuations and shear stress, but its streamwise and wall-normal RMS velocity fluctuation profiles still deviate noticeably. With repeated fixed-stride prediction, IFF develops much larger profile deviations, with some fluctuation and shear-stress values outside the displayed ranges.

The three SC-IFF variants agree much more closely with fDNS in the middle column. Concat slightly overpredicts the wall-normal fluctuations, particularly away from the wall. After Stage~2 training, this excess is reduced, and both Concat and FiLM reproduce the three fluctuation profiles more closely. The corrected Indep profiles also remain close to the reference.

The spectra in Figure~\ref{fig:channel_dt10_spectra} show substantial deviations for DSM and WALE; IFF also departs from fDNS, particularly at high wavenumbers. SC-IFF follows the reference spectra more closely in both directions. FA-CSF reduces the streamwise- and spanwise-spectrum errors for Concat and FiLM, with corrected FiLM giving the closest spanwise spectrum overall. Some underprediction remains at the highest streamwise wavenumbers.

\begin{figure}[!htbp]\centering\setlength{\abovecaptionskip}{8pt}
\includegraphics[width=\linewidth]{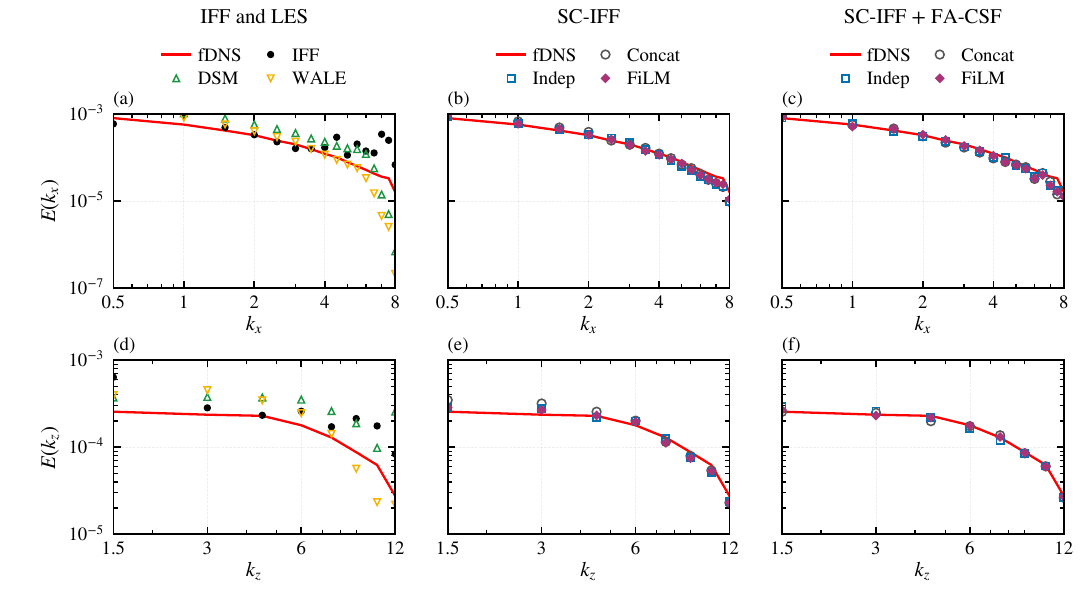}\par
\footnotesize
\caption{Time-averaged energy spectra for Channel-DT10. (a--c) Streamwise; (d--f) spanwise.}
\label{fig:channel_dt10_spectra}\end{figure}

Figure~\ref{fig:channel_dt10_slices} compares streamwise-velocity slices at the 100th, 500th, 2000th and 4000th snapshots. IFF develops nearly streamwise-uniform bands at late times. All three SC-IFF variants and their FA-CSF models retain streamwise variations. The locations and amplitudes of individual structures still differ from fDNS; we therefore focus on reproducing the flow statistics rather than matching every instantaneous feature.

\begin{figure}[!htbp]\centering\setlength{\abovecaptionskip}{8pt}
\includegraphics[width=\linewidth]{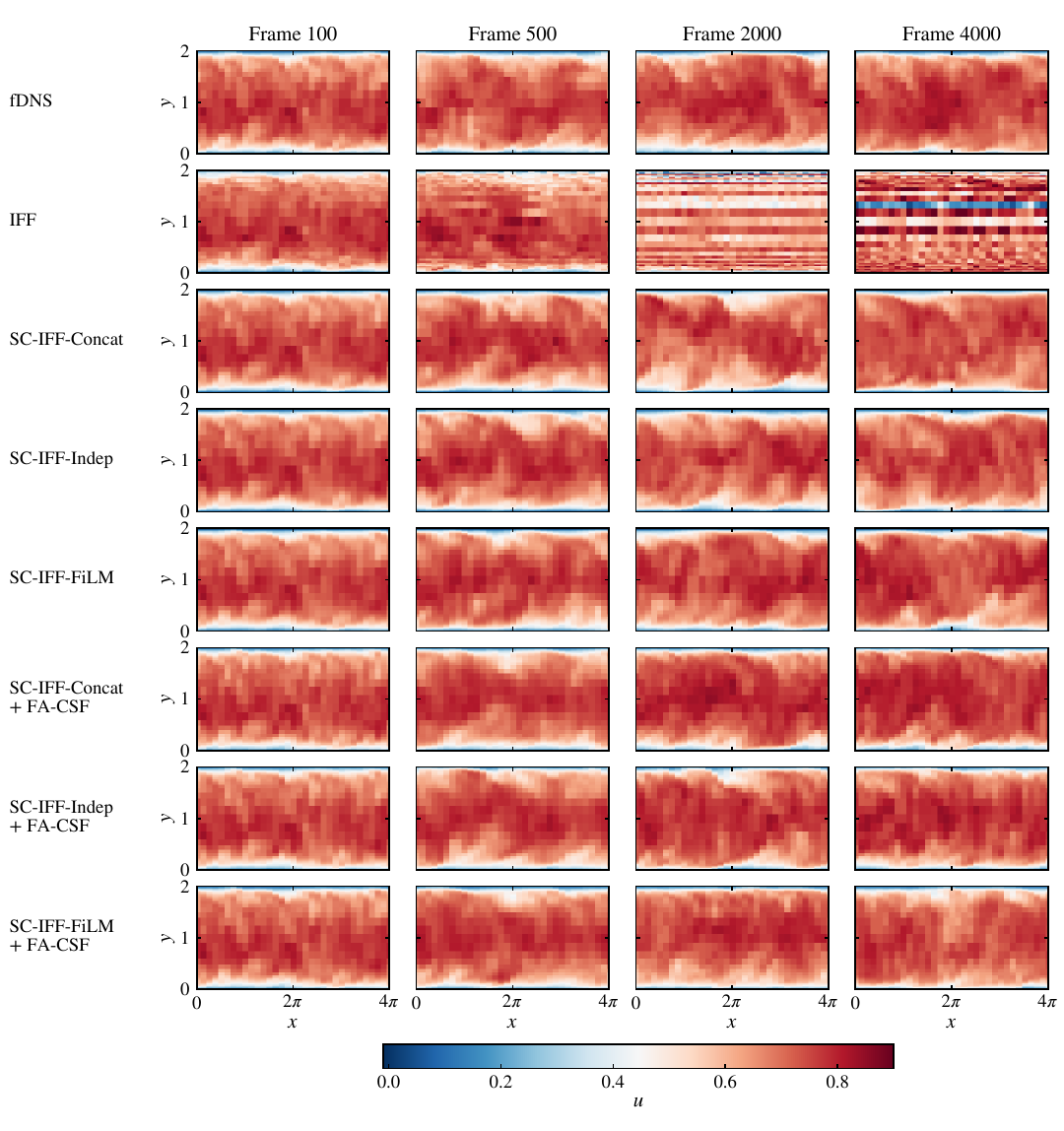}\par
\footnotesize
\caption{Instantaneous streamwise-velocity fields for Channel-DT10 in the $x$--$y$ plane at $z/\delta=2\pi/3$. Rows show fDNS, IFF, the Concat, Indep and FiLM variants of SC-IFF, and their FA-CSF models, in that order. Columns show snapshots 100, 500, 2000 and 4000. All panels use the same color scale.}
\label{fig:channel_dt10_slices}\end{figure}
\FloatBarrier

Figure~\ref{fig:channel_dt10_field_pearson} shows the relative field $L^2$ error and spatial Pearson correlation coefficient over time. The IFF error rises to approximately 0.4, whereas the SC-IFF errors level off at approximately 0.16--0.17, with similarly bounded errors after correction. Although low field error alone does not establish stability, its slow late-time growth for SC-IFF is consistent with the statistical comparisons above. The IFF Pearson correlation falls to approximately 0.51 at the final time, indicating weaker spatial correspondence with the reference streamwise field. For all three SC-IFF variants and their FA-CSF models, the correlation stays between approximately 0.90 and 1.00, indicating substantially stronger spatial similarity.

Small strides increase the number of autoregressive steps and can amplify error accumulation, whereas larger strides may increase per-step prediction error. On Channel-DT10, fixed-stride IFF requires 3999 autoregressive steps for a trajectory of 4000 snapshots, while SC-IFF uses about 125 autoregressive steps of $32\Delta T$ and finer-stride predictions to retain outputs every $\Delta T$. The higher IFF errors on DT10 than DT50 (Tables~\ref{tab:channel_dt50_modes} and~\ref{tab:channel_fusion_summary}), despite the shorter total prediction time, are consistent with this explanation. Multi-stride prediction improves the stability of the tested rollouts and the overall agreement with fDNS statistics, but statistical discrepancies remain. By freezing the pretrained predictors and training only the small FA-CSF module, Stage~2 enables refinement using field-data and statistical losses on longer autoregressive rollouts with repeated feedback.

\begin{figure}[!htbp]\centering\setlength{\abovecaptionskip}{8pt}
\includegraphics[width=\linewidth]{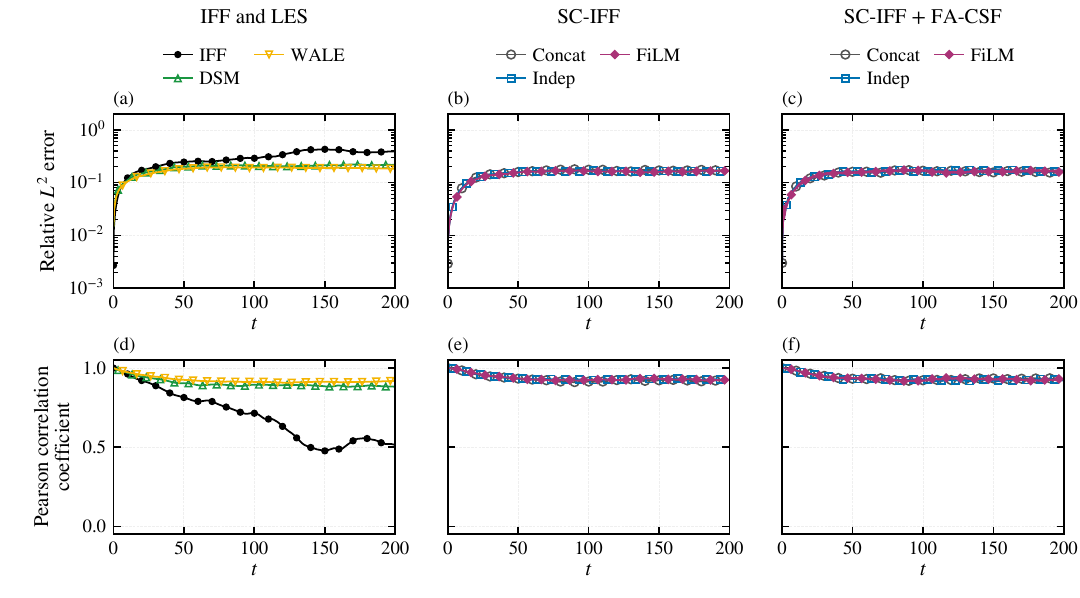}\par
\footnotesize
\caption{Time histories of prediction errors for Channel-DT10. (a--c) Relative velocity-field $L^2$ error; (d--f) spatial Pearson correlation of streamwise velocity with fDNS.}
\label{fig:channel_dt10_field_pearson}\end{figure}

\FloatBarrier
\subsubsection{\textnormal{Statistical losses, correction architecture and computational cost}}\label{sec:channel-controls}
Table~\ref{tab:channel_correction_controls} compares Stage~2 configurations for the same frozen SC-IFF-Indep predictors; ``$-$'' denotes hierarchical prediction without interpolation or correction, and $\epsilon_{L^2}$ denotes the relative $L^2$ error of the predicted velocity field. Without learned correction, hierarchical prediction and linear interpolation between adjacent training strides yield comparable statistical errors. Compared with linear interpolation, FA-CSF trained with field-data and statistical losses lowers the RMS-profile, shear-stress and streamwise-spectrum errors, but increases the mean-profile and spanwise-spectrum errors.

Under otherwise identical training settings, adding statistical losses to FA-CSF reduces all five statistical errors relative to training with the field-data loss alone: those in the mean profile, RMS velocity fluctuation profiles, Reynolds shear stress, and streamwise and spanwise energy spectra. The model trained with the field-data loss alone nevertheless has a lower relative field $L^2$ error, showing that lower field error does not necessarily imply more accurate long-time statistics.

\begin{table}[!htbp]\centering
\papertablestyle
\begin{threeparttable}
\caption{Relative $L^2$ errors in velocity fields and flow statistics for Stage~2 ablations of the correction module and the loss function with frozen SC-IFF-Indep predictors on Channel-DT10.}
\label{tab:channel_correction_controls}
{
\begin{tabular*}{\linewidth}{@{\extracolsep{\fill}}l l r r r r r r@{}}\toprule
\multicolumn{2}{@{}l}{Model} & \multirow{2}{*}{$\epsilon_\mu$} & \multirow{2}{*}{$\epsilon_{\rm rms}$} & \multirow{2}{*}{$\epsilon_{uv}$} & \multirow{2}{*}{$\epsilon_{E_x}$} & \multirow{2}{*}{$\epsilon_{E_z}$} & \multirow{2}{*}{$\epsilon_{L^2}$}\\
\cmidrule(r){1-2}
Stage 1 & Stage 2 & & & & & & \\\midrule
\multirow{5}{*}{SC-IFF-Indep} & $-$ & $\boldsymbol{7.49\times10^{-3}}$ & $4.64\times10^{-2}$ & $9.95\times10^{-2}$ & $8.83\times10^{-2}$ & $1.16\times10^{-1}$ & $1.52\times10^{-1}$\\
 & linear-interp & $7.50\times10^{-3}$ & $4.78\times10^{-2}$ & $7.11\times10^{-2}$ & $1.20\times10^{-1}$ & $9.26\times10^{-2}$ & $1.49\times10^{-1}$\\
 & FA-CSF (data only) & $1.04\times10^{-2}$ & $4.52\times10^{-1}$ & $7.12\times10^{-1}$ & $5.55\times10^{-1}$ & $5.60\times10^{-1}$ & $\boldsymbol{1.25\times10^{-1}}$\\
 & MLP (data+stat) & $1.27\times10^{-2}$ & $4.04\times10^{-2}$ & $1.01\times10^{-1}$ & $7.99\times10^{-2}$ & $\boldsymbol{5.46\times10^{-2}}$ & $1.57\times10^{-1}$\\
 & FA-CSF (data+stat) & $8.22\times10^{-3}$ & $\boldsymbol{2.67\times10^{-2}}$ & $\boldsymbol{5.77\times10^{-2}}$ & $\boldsymbol{5.25\times10^{-2}}$ & $9.83\times10^{-2}$ & $1.55\times10^{-1}$\\
\bottomrule\end{tabular*}}
\end{threeparttable}
\end{table}

We replace the PAI module in the spatial branch of FA-CSF with a pointwise MLP whose channel widths are \([48, 64, 128, 256, 256, 128, 64, 48]\). Layer normalization is applied to the input, and GELU activation follows each hidden layer, with no activation after the output layer. The MLP-based correction module has 161,475 parameters, compared with 141,283 for FA-CSF. Both correction modules use the same frozen SC-IFF-Indep predictors and identical training settings. Table~\ref{tab:channel_correction_controls} shows errors of similar magnitude for the two corrections, with lower mean-profile, RMS-profile, shear-stress and streamwise-spectrum errors for FA-CSF, but a lower spanwise-spectrum error for the MLP. These results suggest that FA-CSF learns these flow statistics more effectively at comparable model size in this case.

\FloatBarrier
\label{sec:channel-cost}
Table~\ref{tab:channel_cost_summary} shows that SC-IFF-Indep + FA-CSF generates a complete Channel-DT10 trajectory 2.41 times faster than hierarchical prediction and 3.69 times faster than fixed-stride IFF. In each autoregressive step of $32\Delta T$, it evaluates each of the six stride-specific predictors once and FA-CSF once, compared with 32 predictor evaluations for IFF. It also uses less peak allocated GPU memory than hierarchical prediction, while IFF uses the least.

\begin{table}[!htbp]\centering
\papertablestyle
\begin{threeparttable}
\caption{Time and peak GPU memory for generating a complete Channel-DT10 trajectory.}
\label{tab:channel_cost_summary}
\begin{tabular*}{\linewidth}{@{\extracolsep{\fill}}l r r r@{}}\toprule
 & & \multicolumn{2}{r}{Peak memory (GiB)}\\
\cmidrule(l){3-4}
Model & Time (s) & Allocated & Reserved\\\midrule
IFF & 86.33 & \textbf{0.33} & \textbf{0.51}\\
SC-IFF-Indep & 56.46 & 1.70 & 2.12\\
SC-IFF-Indep + FA-CSF & \textbf{23.42} & 0.66 & 1.00\\
\bottomrule\end{tabular*}
\end{threeparttable}
\end{table}

\endgroup\FloatBarrier

\subsection{Homogeneous isotropic turbulence}\label{sec:hit_results}
We consider forced HIT in a periodic cube of side length $2\pi$, using fDNS velocity fields on a $32^3$ grid obtained from DNS on a $256^3$ grid. With a DNS time step of 0.001, DT50 and DT10 correspond to $\Delta T=0.05$ and 0.01, respectively. Each rollout uses the first snapshot as input and predicts the remaining snapshots of a trajectory containing 1600 (DT50) or 3200 (DT10) snapshots. Time is reported as $t/\tau$, where $\tau$ is the large-eddy turnover time of the reference DNS.

We compare velocity and vorticity RMS histories, spherical-shell energy spectra, and PDFs of longitudinal velocity increments and vorticity magnitude. The velocity RMS at time $t$ is defined as
\begin{equation}\label{eq:hit_velocity_rms}
\bar{u}_{\mathrm{rms}}(t)=\sqrt{\langle|\uvec(t)|^2\rangle_{x,y,z}},
\end{equation}
and the vorticity RMS is defined as
\begin{equation}\label{eq:hit_vorticity_rms}
\bar{\omega}_{\mathrm{rms}}(t)=\sqrt{\langle|\nabla\times\uvec(t)|^2\rangle_{x,y,z}},
\end{equation}
where the brackets denote spatial averaging.

\subsubsection{\textnormal{Homogeneous isotropic turbulence sampled every 50 DNS time steps (HIT-DT50)}}\label{sec:hit-dt50}
On HIT-DT50, SC-IFF-Indep preserves the long-time flow statistics more closely than IFF, and FA-CSF further improves its predictions, although the gains differ among the stride-conditioning strategies. We first compare the relative $L^2$ errors in velocity RMS, vorticity RMS and the time-averaged energy spectrum, then examine the RMS histories and the energy spectra, velocity-increment PDFs and vorticity PDFs at different times measured in large-eddy turnover units.

\begin{samepage}
In Table~\ref{tab:hit_dt50}, SC-IFF-Indep gives the largest reduction in both RMS errors relative to IFF. Stage~2 training reduces all three errors for each SC-IFF variant. The corrected Indep model has the lowest errors in all three metrics and also outperforms DSM.\par
\end{samepage}

\begin{table}[!htbp]\centering
\papertablestyle
\begin{threeparttable}
\caption{Relative $L^2$ errors in velocity and vorticity RMS histories and time-averaged energy spectra for HIT-DT50.}
\label{tab:hit_dt50}
\begin{tabular*}{\linewidth}{@{\extracolsep{\fill}}l r r r@{}}\toprule
Model & $\epsilon_{\bar{u}_{\mathrm{rms}}}$ & $\epsilon_{\bar{\omega}_{\mathrm{rms}}}$ & $\epsilon_E$\\\midrule
DSM & $7.29\times10^{-2}$ & $2.59\times10^{-1}$ & $9.78\times10^{-2}$\\
Fixed-stride IFF & $1.77\times10^{-1}$ & $2.10\times10^{-1}$ & $8.26\times10^{-2}$\\
\midrule
SC-IFF-Concat & $6.23\times10^{-1}$ & $4.20\times10^{-1}$ & $8.27\times10^{-1}$\\
SC-IFF-Indep & $4.28\times10^{-2}$ & $1.15\times10^{-1}$ & $1.29\times10^{-1}$\\
SC-IFF-FiLM & $1.59\times10^{-1}$ & $2.13\times10^{-1}$ & $2.46\times10^{-1}$\\
\midrule
SC-IFF-Concat + FA-CSF & $\underline{4.11\times10^{-1}}$ & $\underline{3.21\times10^{-1}}$ & $\underline{4.90\times10^{-1}}$\\
SC-IFF-Indep + FA-CSF & $\underline{\boldsymbol{3.08\times10^{-2}}}$ & $\underline{\boldsymbol{1.09\times10^{-1}}}$ & $\underline{\boldsymbol{7.24\times10^{-2}}}$\\
SC-IFF-FiLM + FA-CSF & $\underline{8.57\times10^{-2}}$ & $\underline{1.46\times10^{-1}}$ & $\underline{8.75\times10^{-2}}$\\
\bottomrule\end{tabular*}
\end{threeparttable}
\end{table}

\begin{figure}[!htbp]\centering\setlength{\abovecaptionskip}{8pt}
\includegraphics[width=\linewidth]{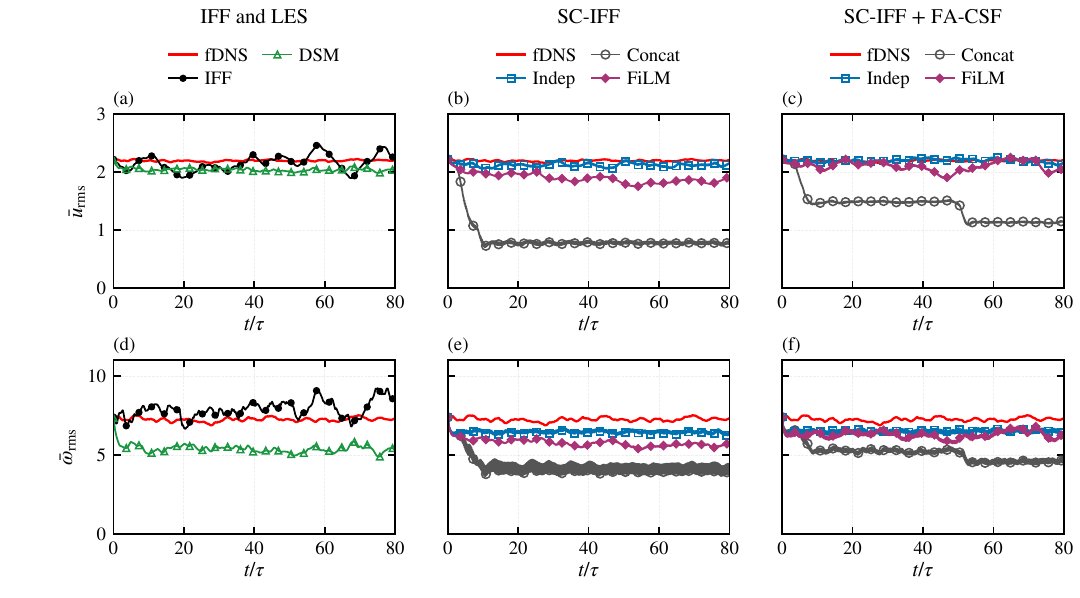}\par
\footnotesize
\caption{Temporal evolutions of velocity and vorticity RMS for HIT-DT50, comparing the fDNS reference, IFF, DSM and the three SC-IFF variants with and without FA-CSF. Panels (a--c) show $\bar{u}_{\mathrm{rms}}$, and (d--f) show $\bar{\omega}_{\mathrm{rms}}$.}
\label{fig:hit_dt50_statistics}\end{figure}

\FloatBarrier
\begin{samepage}
Figure~\ref{fig:hit_dt50_statistics} compares the velocity and vorticity RMS histories, which reflect the kinetic-energy level and rotational intensity. The IFF histories fluctuate more strongly as prediction time increases. Indep stays closest to fDNS. FA-CSF brings FiLM and Concat closer to the reference, although Concat still deviates considerably at late times. DSM and the SC-IFF variants predict lower vorticity RMS than fDNS, with Indep and FiLM closer to the reference than DSM.\par
\end{samepage}

\begin{figure}[!htbp]\centering\setlength{\abovecaptionskip}{8pt}
\includegraphics[width=\linewidth]{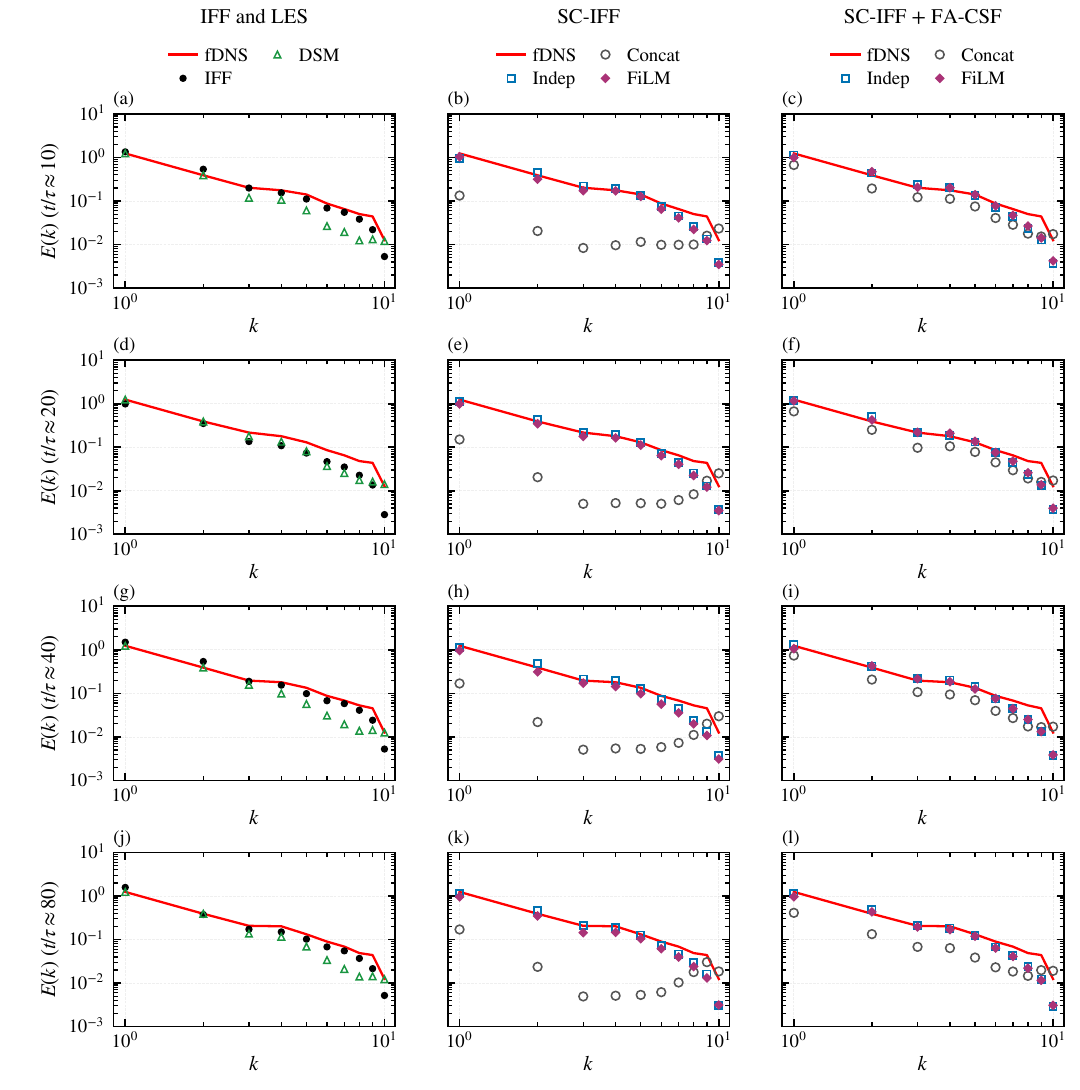}\par
\footnotesize
\caption{Spherical-shell kinetic energy spectra $E(k)$ for HIT-DT50, comparing the fDNS reference, IFF, DSM and the three SC-IFF variants with and without FA-CSF. Panels (a--c), (d--f), (g--i) and (j--l) correspond to $t/\tau\approx10$, 20, 40 and 80, respectively.}
\label{fig:hit_dt50_spectra}\end{figure}

\begin{samepage}
Figure~\ref{fig:hit_dt50_spectra} shows how kinetic energy is distributed across wavenumbers. IFF, Indep and FiLM follow fDNS closely at low and intermediate wavenumbers but underpredict energy at high wavenumbers. For Indep and FiLM, this deficit remains after Stage~2. Concat underpredicts energy more broadly, especially at low and intermediate wavenumbers. FA-CSF partly reduces this bias, but the late-time spectra remain well below fDNS across several wavenumber ranges. This agrees with its low velocity and vorticity RMS in Figure~\ref{fig:hit_dt50_statistics}.\par
\end{samepage}

\FloatBarrier

\begin{figure}[!htbp]\centering\setlength{\abovecaptionskip}{8pt}
\includegraphics[width=\linewidth]{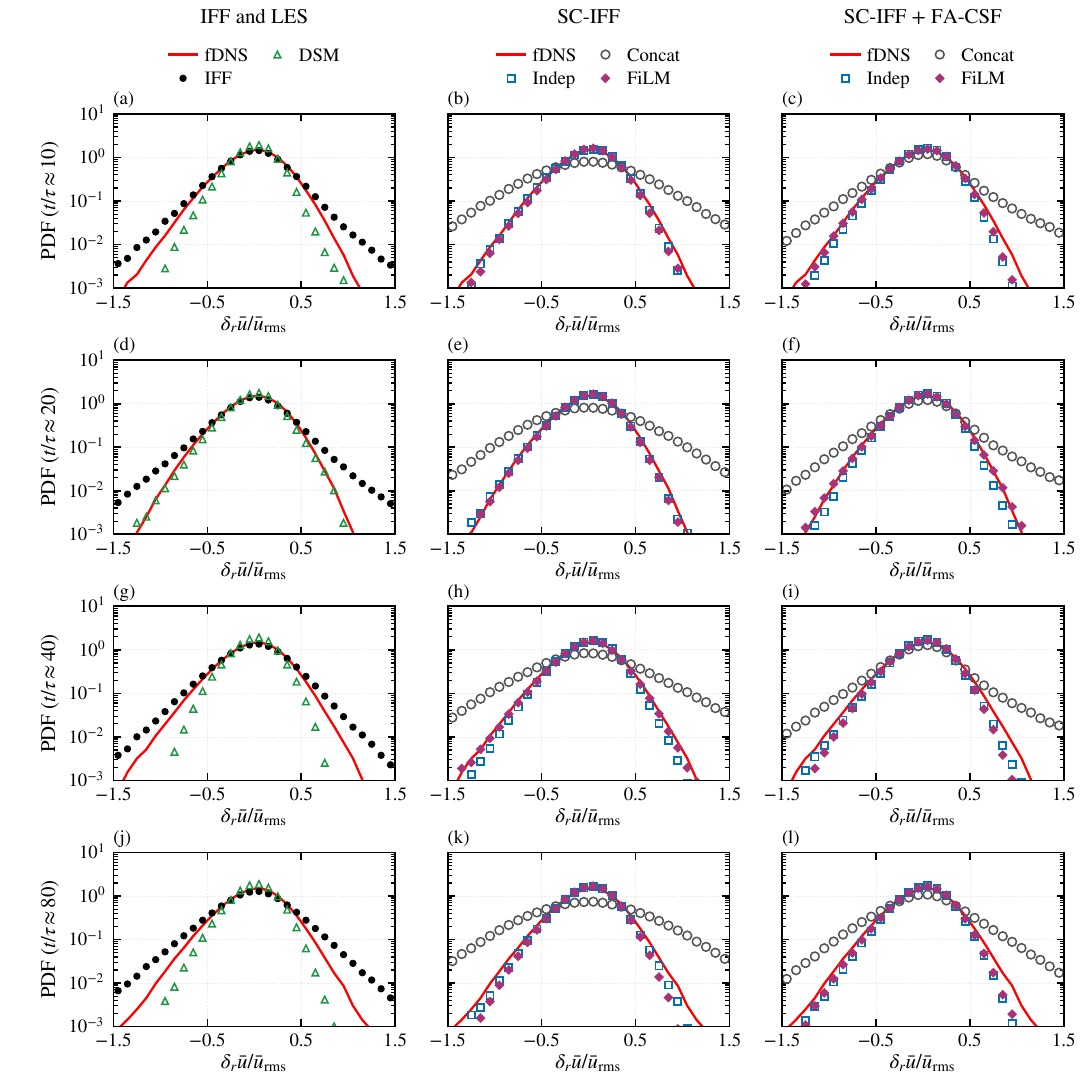}\par
\footnotesize
\caption{PDFs of the normalized velocity increments $\delta_r\bar{u}/\bar{u}_{\mathrm{rms}}$ at a separation of one grid spacing for HIT-DT50, comparing the fDNS reference, IFF, DSM and the three SC-IFF variants with and without FA-CSF. Panels (a--c), (d--f), (g--i) and (j--l) correspond to $t/\tau\approx10$, 20, 40 and 80, respectively.}
\label{fig:hit_dt50_increment_pdfs}\end{figure}

\begin{samepage}
The normalized velocity-increment PDFs in Figure~\ref{fig:hit_dt50_increment_pdfs} describe local velocity differences, with the tails representing large increments. IFF overpredicts the tails, whereas DSM produces a narrower distribution. Indep and FiLM agree more closely with fDNS. FA-CSF narrows Concat's overly broad distribution. At $t/\tau\approx80$, however, uncorrected Indep is closer to fDNS; Stage~2 makes its distribution slightly too narrow. The improvements in Table~\ref{tab:hit_dt50} therefore do not imply a better PDF at every time.\par
\end{samepage}

\FloatBarrier

\begin{figure}[!htbp]\centering\setlength{\abovecaptionskip}{8pt}
\includegraphics[width=\linewidth]{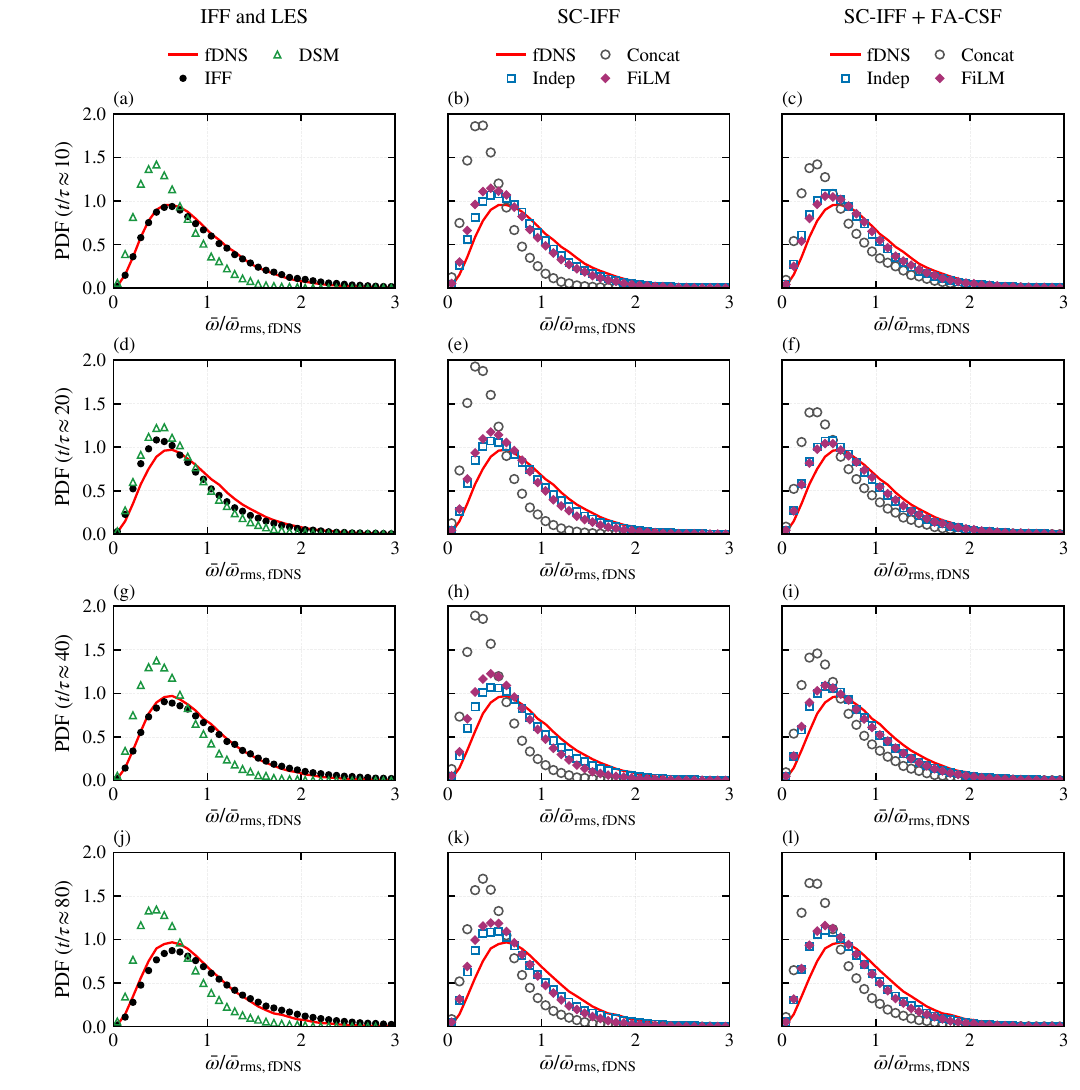}\par
\footnotesize
\caption{PDFs of the normalized vorticity magnitude $\bar{\omega}/\bar{\omega}_{\mathrm{rms,fDNS}}$ for HIT-DT50, comparing the fDNS reference, IFF, DSM and the three SC-IFF variants with and without FA-CSF. Panels (a--c), (d--f), (g--i) and (j--l) correspond to $t/\tau\approx10$, 20, 40 and 80, respectively.}
\label{fig:hit_dt50_vorticity_pdfs}\end{figure}

\begin{samepage}
Figure~\ref{fig:hit_dt50_vorticity_pdfs} shows the distribution of vorticity magnitude and the probability of strong-vorticity events. IFF agrees with the main part of the fDNS distribution but overpredicts the strong-vorticity tail at late times. DSM shifts the distribution towards lower vorticity magnitudes. Indep and FiLM show a similar shift but remain closer to fDNS overall. FA-CSF reduces some of these deviations and also improves Concat, although the gains vary with time. The shift towards lower vorticity is consistent with the slightly low vorticity RMS in Figure~\ref{fig:hit_dt50_statistics}.\par
\end{samepage}

\FloatBarrier

\subsubsection{\textnormal{Homogeneous isotropic turbulence sampled every 10 DNS time steps (HIT-DT10)}}\label{sec:hit-dt10}
On HIT-DT10, all three SC-IFF variants yield lower statistical errors than fixed-stride IFF. Further refinement with FA-CSF does not consistently improve these statistics across variants. We compare statistical errors, RMS histories, energy spectra, velocity-increment PDFs and vorticity-magnitude PDFs.

Table~\ref{tab:hit_dt10_correction} shows that FA-CSF reduces all three errors for Concat. It also reduces the velocity-RMS and mean spectral errors for Indep, but increases its vorticity-RMS error. All three errors increase for the present FiLM checkpoint. Corrected Indep gives the lowest velocity-RMS and mean spectral errors, whereas uncorrected Indep gives the lowest vorticity-RMS error. No single model leads in all three metrics. These results show that stride-conditioned training and prediction substantially improve long-time prediction at fine time steps. Stage~2 provides an additional option for statistical correction, with benefits that depend on the variant and evaluation metric.

\begin{table}[!htbp]\centering
\papertablestyle
\begin{threeparttable}
\caption{Relative $L^2$ errors in velocity and vorticity RMS histories and time-averaged energy spectra for HIT-DT10.}
\label{tab:hit_dt10_correction}
\begin{tabular*}{\linewidth}{@{\extracolsep{\fill}}l r r r@{}}\toprule
Model & $\epsilon_{\bar{u}_{\mathrm{rms}}}$ & $\epsilon_{\bar{\omega}_{\mathrm{rms}}}$ & $\epsilon_E$\\\midrule
DSM & $6.95\times10^{-2}$ & $2.51\times10^{-1}$ & $9.42\times10^{-2}$\\
Fixed-stride IFF & $3.79\times10^{0}$ & $1.23\times10^{1}$ & $1.05\times10^{1}$\\
\midrule
SC-IFF-Concat & $2.82\times10^{-1}$ & $3.84\times10^{-1}$ & $2.44\times10^{-1}$\\
SC-IFF-Indep & $7.43\times10^{-2}$ & $\boldsymbol{2.05\times10^{-1}}$ & $1.45\times10^{-1}$\\
SC-IFF-FiLM & $1.92\times10^{-1}$ & $3.38\times10^{-1}$ & $2.26\times10^{-1}$\\
\midrule
SC-IFF-Concat + FA-CSF & $\underline{1.41\times10^{-1}}$ & $\underline{3.11\times10^{-1}}$ & $\underline{1.38\times10^{-1}}$\\
SC-IFF-Indep + FA-CSF & $\underline{\boldsymbol{6.24\times10^{-2}}}$ & $2.30\times10^{-1}$ & $\underline{\boldsymbol{9.37\times10^{-2}}}$\\
SC-IFF-FiLM + FA-CSF & $2.28\times10^{-1}$ & $3.65\times10^{-1}$ & $2.84\times10^{-1}$\\
\bottomrule\end{tabular*}
\end{threeparttable}
\end{table}
\FloatBarrier

\begin{figure}[!htbp]\centering\setlength{\abovecaptionskip}{8pt}
\includegraphics[width=\linewidth]{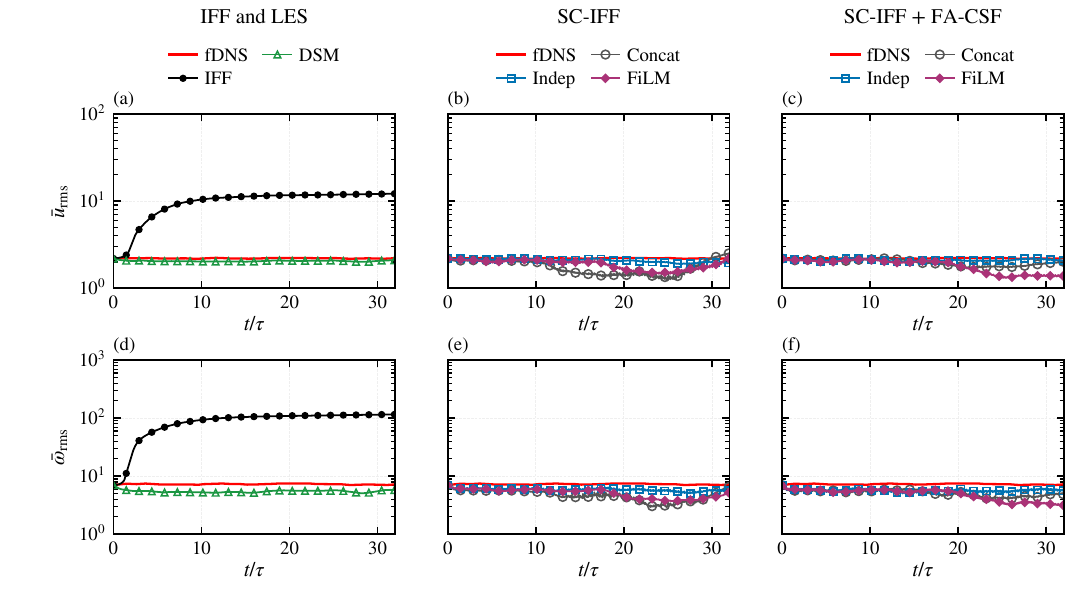}\par
\footnotesize
\caption{Temporal evolutions of velocity and vorticity RMS for HIT-DT10, comparing the fDNS reference, IFF, DSM and the three SC-IFF variants with and without FA-CSF. Panels (a--c) show $\bar{u}_{\mathrm{rms}}$, and (d--f) show $\bar{\omega}_{\mathrm{rms}}$.}
\label{fig:hit_dt10_correction}\end{figure}

The velocity and vorticity RMS histories in Figure~\ref{fig:hit_dt10_correction} reflect the kinetic-energy level and rotational intensity. Both quantities grow substantially in IFF, whereas SC-IFF keeps them closer to the reference. FA-CSF reduces the amplitude deviations of Concat and FiLM during parts of the middle and later stages of the prediction. At longer times, however, both corrected models still underestimate the amplitudes, especially FiLM near the end of the prediction.

\begin{figure}[!htbp]\centering\setlength{\abovecaptionskip}{8pt}
\includegraphics[width=\linewidth]{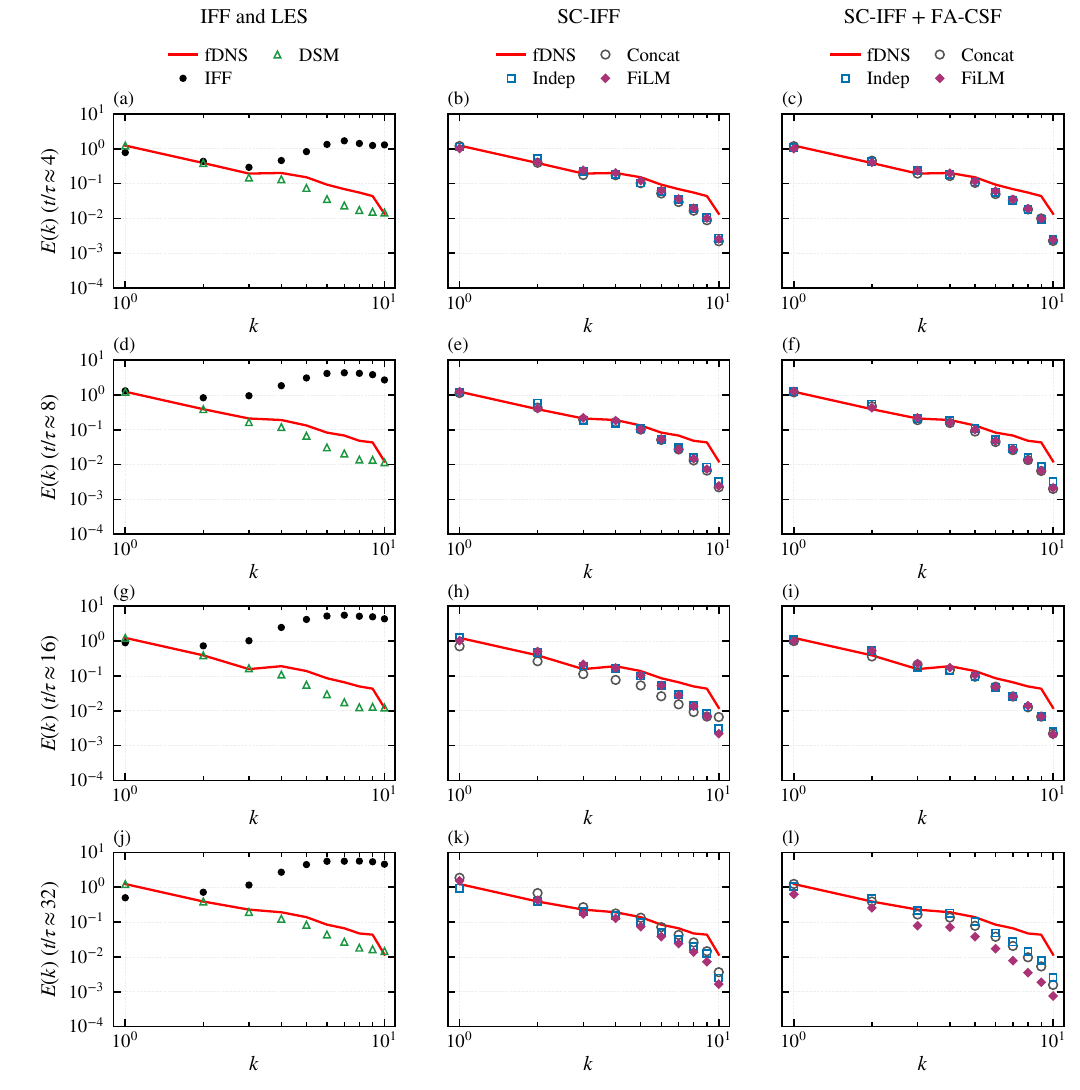}\par
\footnotesize
\caption{Spherical-shell kinetic energy spectra $E(k)$ for HIT-DT10, comparing the fDNS reference, IFF, DSM and the three SC-IFF variants with and without FA-CSF. Panels (a--c), (d--f), (g--i) and (j--l) correspond to $t/\tau\approx4$, 8, 16 and 32, respectively.}
\label{fig:hit_dt10_spectra}\end{figure}

The energy spectra in Figure~\ref{fig:hit_dt10_spectra} show how kinetic energy is distributed across wavenumbers. Repeated autoregressive prediction at this fine sampling interval produces excessive high-wavenumber energy in IFF, while all three SC-IFF variants substantially improve the spectra. At $t/\tau\approx16$, Stage~2 reduces Concat's energy deficit and brings its spectrum closer to the reference. At $t/\tau\approx32$, Concat retains improvements at low wavenumbers but underpredicts energy more strongly at intermediate and high wavenumbers. FiLM shows a broader energy deficit after correction. Stage~2 jointly optimizes the field error along long trajectories and several statistical objectives. A reduction in the combined objective therefore does not guarantee improvement at every time or across every wavenumber range. Compared with the channel-flow results, correction is less consistent in these HIT predictions, and spectral deviations persist at long times.

\begin{figure}[!htbp]\centering\setlength{\abovecaptionskip}{8pt}
\includegraphics[width=\linewidth]{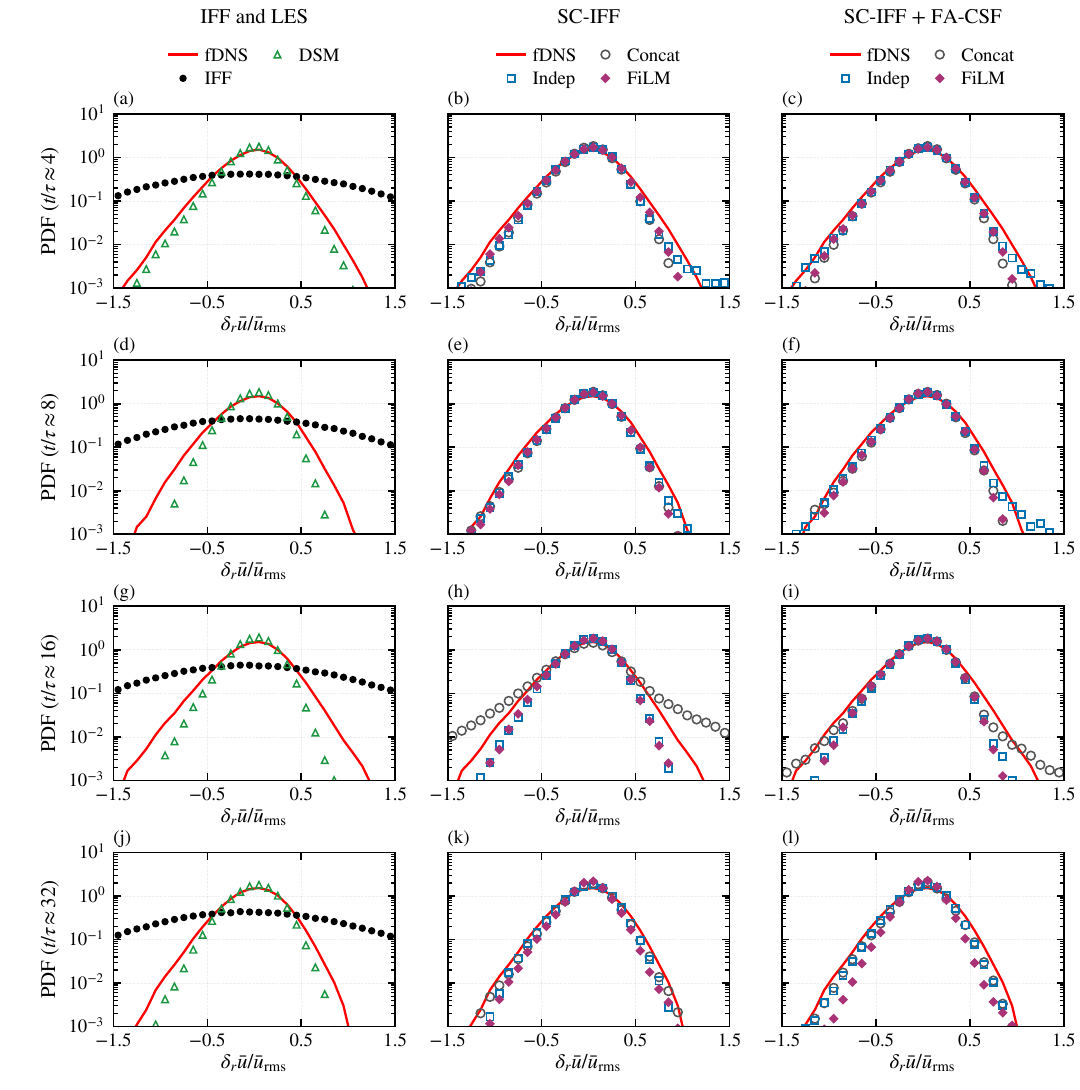}\par
\footnotesize
\caption{PDFs of the normalized velocity increments $\delta_r\bar{u}/\bar{u}_{\mathrm{rms}}$ at a separation of one grid spacing for HIT-DT10, comparing the fDNS reference, IFF, DSM and the three SC-IFF variants with and without FA-CSF. Panels (a--c), (d--f), (g--i) and (j--l) correspond to $t/\tau\approx4$, 8, 16 and 32, respectively.}
\label{fig:hit_dt10_pdfs}\end{figure}

The velocity-increment PDFs in Figure~\ref{fig:hit_dt10_pdfs} describe the distribution of local velocity differences. SC-IFF substantially improves the overly broad IFF distribution. At $t/\tau\approx16$, Stage~2 reduces some of Concat's tail deviations, while its benefit for Indep is limited. FiLM improves slightly at this time but shows larger distribution errors after correction at $t/\tau\approx32$.

\FloatBarrier

\begin{figure}[!htbp]\centering\setlength{\abovecaptionskip}{8pt}
\includegraphics[width=\linewidth]{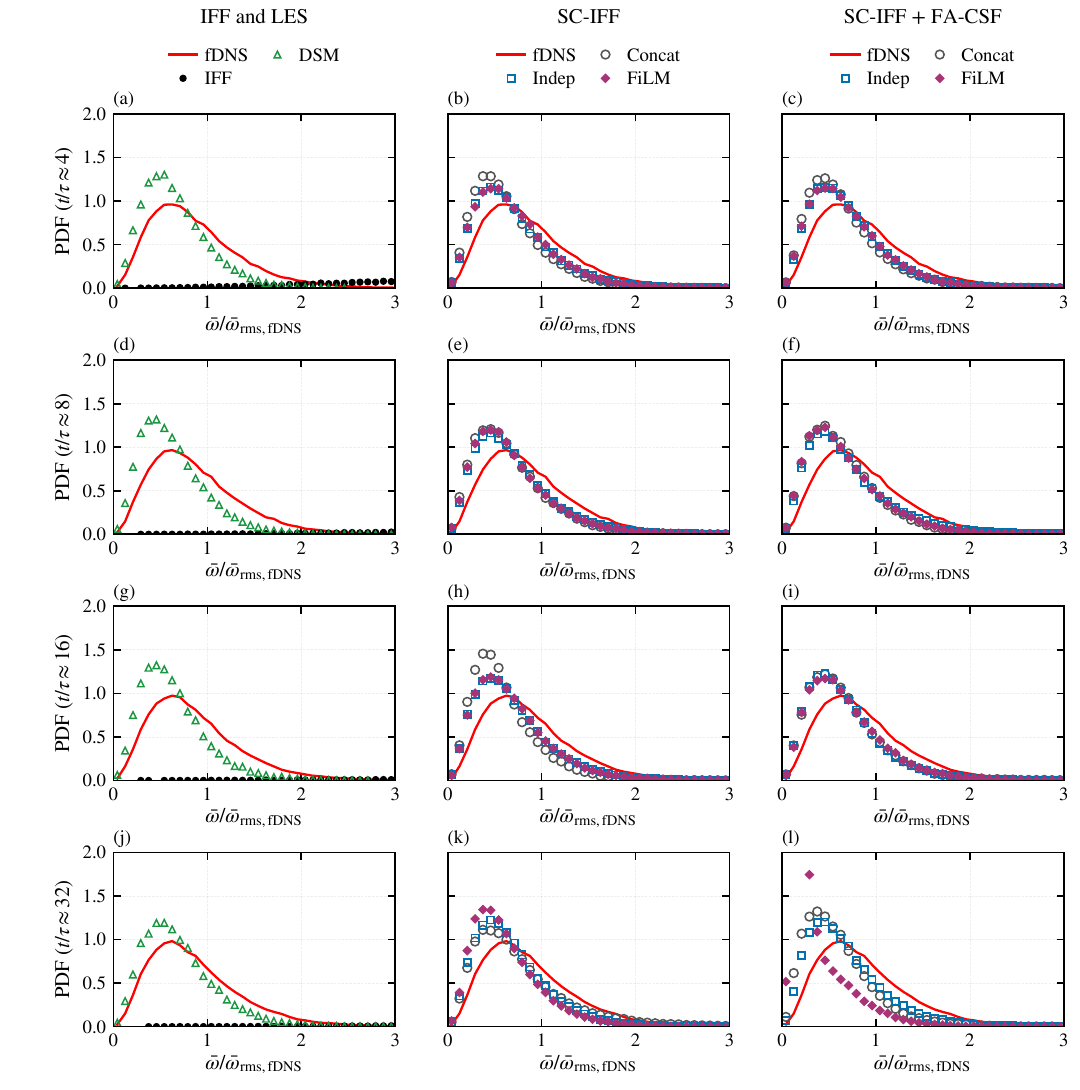}\par
\footnotesize
\caption{PDFs of the normalized vorticity magnitude $\bar{\omega}/\bar{\omega}_{\mathrm{rms,fDNS}}$ for HIT-DT10, comparing the fDNS reference, IFF, DSM and the three SC-IFF variants with and without FA-CSF. Panels (a--c), (d--f), (g--i) and (j--l) correspond to $t/\tau\approx4$, 8, 16 and 32, respectively. About 99\% of IFF samples lie beyond the displayed upper limit at the final time.}
\label{fig:hit_dt10_vorticity_pdfs}\end{figure}

The vorticity-magnitude PDFs in Figure~\ref{fig:hit_dt10_vorticity_pdfs} likewise show that SC-IFF substantially reduces IFF's strong-vorticity bias, with Indep generally closer to the reference. The benefits of Stage~2 remain inconsistent. In particular, corrected FiLM shifts further towards low vorticity at the final time, consistent with its underestimated vorticity RMS.

\FloatBarrier

\Needspace{5\baselineskip}
\section{Conclusions}\label{sec:conclusion}
We have investigated long-time turbulence prediction at fine time steps using SC-IFF predictors trained at power-of-two strides, with FA-CSF fusing and correcting their predictions. FA-CSF reuses the predictions and latent features of the predictors to generate velocity fields at intermediate time steps, and the corrected state at the largest stride initializes the next autoregressive step. Stage~1 pretrains the predictors on short trajectories at multiple strides; Stage~2 freezes them and trains only FA-CSF on longer autoregressive rollouts using field-data and statistical losses. Optimizing this small module reduces the memory required for parameter gradients and optimizer states, thereby supporting gradient propagation along longer trajectories.

In the tested channel-flow and HIT cases, multi-stride training and prediction generally improve long-time flow statistics relative to fixed-stride IFF, with the clearest gains at the finer DT10 sampling interval. SC-IFF-Concat provides a simple baseline that shares one predictor across strides, improving the DT10 statistics in both flows but retaining substantial deviations in some HIT cases. SC-IFF-Indep gives relatively consistent performance across the tested cases. On Channel-DT10, SC-IFF-Indep reduces the error in the RMS velocity fluctuation profiles from 1.42 to 0.0464 and the Reynolds-shear-stress error from 4.18 to 0.0995. On HIT-DT10, it reduces the time-averaged energy-spectrum error from 10.5 to 0.145. Although SC-IFF-Indep has more parameters, sequential transfer learning keeps its training time and peak GPU memory comparable to those of Concat and FiLM, at a small additional storage cost. SC-IFF-FiLM adds only 0.32\% to the IFF parameter count and performs well in channel flow, while its advantage is less consistent in HIT. These results show that multi-stride prediction reduces statistical deviations in long-time prediction at fine time steps, with the choice of conditioning strategy affecting the balance between parameter storage and predictive performance.

FA-CSF further improves several long-time statistics in most tested configurations. For SC-IFF-Indep on Channel-DT10, it reduces the error in the RMS velocity fluctuation profiles from 0.0464 to 0.0267 and the Reynolds-shear-stress error from 0.0995 to 0.0577. On HIT-DT10, it reduces the time-averaged energy-spectrum error from 0.145 to 0.0937. Further controlled comparisons on Channel-DT10 assess the roles of the statistical losses and of the correction architecture. With the predictors frozen, adding statistical losses reduces all five statistical errors relative to training with the field-data loss alone, despite a slightly higher field error. When the factorized-attention module in its spatial branch is replaced by an MLP of comparable size, with the training settings and objective unchanged, FA-CSF yields lower errors in the RMS velocity fluctuation profiles and Reynolds shear stress than the MLP on Channel-DT10. Reusing the predictions at the training strides also accelerates trajectory generation at fine time steps: SC-IFF-Indep with FA-CSF is 3.69 times faster than fixed-stride IFF and 2.41 times faster than hierarchical SC-IFF-Indep prediction on Channel-DT10.

Overall, the proposed framework improves long-time turbulence prediction at fine time steps. In HIT, however, late-time vorticity RMS can be underestimated and small-scale spectral errors can persist or increase after correction. This suggests that Stage~2 refinement in HIT requires careful trade-offs among field accuracy, energy levels and the distribution of energy across scales, together with an appropriate training-window length. The present results are empirical and limited to finite prediction times, with one Reynolds number and one coarse grid per flow. Future work should examine the conditions under which long-time stability can be ensured. Testing generalization across Reynolds numbers, grid resolutions and forcing conditions would support extending the framework to more complex turbulent flows.

\section*{Acknowledgements}

This work was supported by the National Natural Science Foundation of China (NSFC Grant Nos. 12588301 and 12302283), by NSFC Excellence Research Group Program for \lq Multiscale Problems in Nonlinear Mechanics\rq{} (No. 12588201), by the Shenzhen Science and Technology Program (Grant Nos. SYSPG20241211173725008, and KQTD20180411143441009), and by Department of Science and Technology of Guangdong Province (Grant Nos. 2019B21203001, 2020B1212030001, and 2023B1212060001). Additional support was provided by the Innovation Capability Support Program of Shaanxi (Program No. 2023-CXTD-30) and the Center for Computational Science and Engineering of Southern University of Science and Technology.

\end{document}

%% file: figure/method/iff_backbone.tex
\begingroup
\input{figure/method/method_figure_style.tex}
\begin{lrbox}{\MFbox}%
\begin{tikzpicture}[mf]
\node[mfvar] (a-u) at (0.2,0) {\(\uvec^n\)};
\coordinate (a-f0) at (0.56,0);
\node[mffixed,right=3.4mm of a-f0,align=center] (a-norm) {Normalize\\\((\boldsymbol\mu^u,\boldsymbol\sigma^u)\)};
\node[mflearn,right=4.3mm of a-norm] (a-lift) {Lifting};
\node[mfadd,right=5.0mm of a-lift] (a-p1) {\(+\)};
\node[mfenc,align=center] (a-pos) at ($(a-lift)+(0,1.02)$) {Coordinate\\embedding};
\node[mfvar,left=4.5mm of a-pos] (a-xyz) {\((x,y,z)\)};
\coordinate[right=9.9mm of a-p1] (a-f1);
\node[mfattn,right=3.6mm of a-f1] (a-pai) {PAI (b)};
\node[mfadd,right=7.6mm of a-pai] (a-p2) {\(+\)};
\coordinate[right=3.2mm of a-p2] (a-f2);
\node[mffixed,right=10.1mm of a-f2] (a-gn) {GN};
\node[mflearn,right=3.4mm of a-gn,align=center] (a-maps) {Channel\\maps};
\node[mffixed,right=6.0mm of a-maps,align=center] (a-den) {Denormalize\\\((\boldsymbol\mu_s^r,\boldsymbol\sigma_s^r)\)};
\node[mfadd,right=8.2mm of a-den] (a-p3) {\(+\)};
\node[mfvar,right=3.0mm of a-p3] (a-out) {\(\widehat{\uvec}^{\,n+s}\)};
\draw[mfgroup] ($(a-norm.west)+(-.2,-.62)$) rectangle ($(a-p1.east)+(.08,1.98)$);
\draw[mfgroup] ($(a-f1)+(-.28,-.98)$) rectangle ($(a-f2)+(.28,1.98)$);
\draw[mfgroup] ($(a-gn.west)+(-.1,-.62)$) rectangle ($(a-maps.east)+(.08,1.98)$);
\node[mfhead] at ($($(a-norm.west)!.5!(a-p1.east)$)+(0,1.74)$) {Encoding};
\node[mfhead] at ($($(a-f1)!.5!(a-f2)$)+(0,1.74)$) {Implicit iteration};
\node[mfnote] at ($($(a-f1)!.5!(a-f2)$)+(0,1.38)$) {\(L\) iterations, shared parameters};
\node[mfhead] at ($($(a-gn.west)!.5!(a-maps.east)$)+(0,1.74)$) {Projection};
\node[mftitle] at (0,2.8) {(a) IFF prediction over the interval \(s\Delta T\)};
\draw[mfwire] (a-u.east)--(a-f0);
\draw[mfflow] (a-f0)--(a-norm.west);
\draw[mfflow] (a-norm.east)--node[above,mfvar] {\(\widetilde{\uvec}\)} (a-lift.west);
\draw[mfflow] (a-lift.east)--node[below,mfvar,yshift=-1pt] {\(U_{\rm lift}\)} (a-p1.west);
\draw[mfflow] (a-xyz.east)--(a-pos.west);
\draw[mfflow] (a-pos.east)-|node[pos=.3,above,mfvar] {\(U_{\rm pos}\)} (a-p1.north);
\draw[mfwire] (a-p1.east)--node[above,mfvar,pos=.40] {\(U^{(0)}\)} (a-f1);
\draw[mfflow] (a-f1)--(a-pai.west);
\draw[mfflow] (a-pai.east)--node[above,mfvar] {\(\times 1/L\)} (a-p2.west);
\draw[mfflow] (a-f1)--($(a-f1)+(0,.8)$)-|node[pos=.25,above,mfvar] {\(U^{(\ell)}\)} (a-p2.north);
\draw[mfwire] (a-p2.east)--(a-f2);
\draw[mfloop] (a-f2)--($(a-f2)+(0,-.6)$)-|(a-f1);
\node[mfnote,fill=white] at ($($(a-f1)!.5!(a-f2)$)+(0,-.6)$) {\(U^{(\ell+1)}\), \(\ell=0,\ldots,L-1\)};
\MFdot{(a-f1)}\MFdot{(a-f2)}
\draw[mfflow] (a-f2)--node[above,mfvar,pos=.54] {\(U^{(L)}\)} (a-gn.west);
\draw[mfflow] (a-gn.east)--(a-maps.west);
\draw[mfflow] (a-maps.east)--node[above,mfvar] {\(\widetilde{\boldsymbol r}_s\)} (a-den.west);
\draw[mfflow] (a-den.east)--node[above,mfvar,yshift=1pt] {\(\times s\Delta T\)}
  node[below,mfvar,yshift=-1pt] {\(\widehat{\boldsymbol r}_s\)} (a-p3.west);
\draw[mfflow] (a-p3.east)--(a-out.west);
\draw[mfflow] (a-f0)--(0.56,2.32)-|node[pos=.25,above,mfvar] {\(\uvec^n\)} (a-p3.north);
\MFdot{(a-f0)}
\def\Yb{-2.82}
\node[mftitle] at (0,\Yb+1.38) {(b) Parallel axial integration (PAI)};
\node[mfvar] (b-in) at (0.32,\Yb) {\(U^{(\ell)}\)};
\node[mffixed,right=3.6mm of b-in] (b-ln) {LN};
\coordinate[right=4.2mm of b-ln] (b-f);
\node[mflearn,minimum height=.56cm,anchor=west] (b-win) at ($(b-f)+(.36,.4)$) {Linear \(W_{\rm in}\)};
\node[mflearn,minimum height=.56cm,anchor=west] (b-wv) at ($(b-f)+(.36,-.4)$) {Linear \(W_{V}\)};
\draw[mfflow] (b-in.east)--(b-ln.west);
\draw[mfwire] (b-ln.east)--node[above,mfvar,pos=.42] {\(\widetilde U\)} (b-f);
\draw[mfflow] (b-f)|-(b-win.west);
\draw[mfflow] (b-f)|-(b-wv.west);
\MFdot{(b-f)}
\def\Xc{5.1}\def\Wc{2.62}
\foreach \dz in {2,1}{
  \draw[fill=MFAttention,draw=black!45,line width=.5pt,rounded corners=1.5pt]
    ({\Xc+.07*\dz},{\Yb-.66+.12*\dz}) rectangle ({\Xc+\Wc+.07*\dz},{\Yb+.66+.12*\dz});}
\node[mfattn,line width=.75pt,minimum width=\Wc cm,minimum height=1.32cm,anchor=west,align=center] (b-axis) at (\Xc,\Yb)
  {\textbf{Axial branch} \(d\)\\kernel \(\kappa_d\) from \(Z\)\\aggregate \(V\) along \(d\)};
\node[mfnote,anchor=south west] at (\Xc+\Wc+.2,\Yb+.5) {\(d\in\{x,y,z\}\),\\[-1pt]in parallel};
\draw[mfflow] (b-win.east)--node[above,mfvar] {\(Z\)} (\Xc,\Yb+.4);
\draw[mfflow] (b-wv.east)--node[below,mfvar] {\(V\)} (\Xc,\Yb-.4);
\coordinate (b-axo) at (\Xc+\Wc+.14,\Yb);
\node[mffixed,right=7.0mm of b-axo,align=center] (b-cat) {Concatenate\\(axes, heads)};
\node[mffixed,right=3.0mm of b-cat,align=center] (b-in2) {Instance norm\\\((x,y,z)\)};
\node[mflearn,right=3.0mm of b-in2,align=center] (b-ffn) {Pointwise\\feed-forward};
\node[mfvar,right=3.0mm of b-ffn] (b-out) {\(\operatorname{PAI}\bigl(U^{(\ell)}\bigr)\)};
\draw[mfflow] (b-axo)--node[above,mfvar] {\(V^{(d)}\)} (b-cat.west);
\draw[mfflow] (b-cat.east)--(b-in2.west);
\draw[mfflow] (b-in2.east)--(b-ffn.west);
\draw[mfflow] (b-ffn.east)--(b-out.west);
\def\Yc{-5.3}
\draw[mfgroup,fill=black!2] (0,\Yc-1.62) rectangle (16.3,\Yc+1.28);
\draw[black!35,line width=.45pt,dash pattern=on 1.6pt off 1.4pt] (\Xc,\Yb-.66)--(0,\Yc+1.28);
\draw[black!35,line width=.45pt,dash pattern=on 1.6pt off 1.4pt] (\Xc+\Wc,\Yb-.66)--(16.3,\Yc+1.28);
\node[mfhead,anchor=west] at (.12,\Yc+1.02) {\(x\)-axis branch {\normalfont(the \(y\) and \(z\) branches are analogous)}};
\MFcube{(.5,\Yc)}{.17}{MFLatent}
\node[mfvar] at (.58,\Yc+.5) {\(Z\)};
\node[mfshape,anchor=north west] at (.02,\Yc-.38) {\(N_x{\times}N_y{\times}N_z{\times}C\)};
\node[mflearn,align=center,anchor=west] (x-wx) at (1.2,\Yc) {Linear\\\(W_x\)};
\node[mffixed,align=center,right=3.0mm of x-wx] (x-avg) {Average\\over \(y,z\)};
\node[mffixed,right=3.0mm of x-avg] (x-ln) {LN};
\node[mflearn,right=3.0mm of x-ln] (x-mlp) {MLP\(_x\)};
\draw[mfflow] (.86,\Yc)--(x-wx.west);
\draw[mfflow] (x-wx.east)--(x-avg.west);
\draw[mfflow] (x-avg.east)--(x-ln.west);
\draw[mfflow] (x-ln.east)--(x-mlp.west);
\coordinate[right=6.4mm of x-mlp] (x-ux);
\begin{scope}[shift={(x-ux)}]
  \fill[MFLatent] (-.3,-.15) rectangle (.3,.15);
  \draw[black!62,line width=.35pt,step=.1] (-.3,-.15) grid (.3,.15);
  \draw[black!62,line width=.4pt] (-.3,-.15) rectangle (.3,.15);
\end{scope}
\draw[mfflow] (x-mlp.east)--($(x-ux)+(-.33,0)$);
\node[mfvar] at ($(x-ux)+(0,.37)$) {\(U_x\)};
\node[mfshape] at ($(x-ux)+(0,-.38)$) {\(N_x{\times}C\)};
\coordinate (x-f) at ($(x-ux)+(.62,0)$);
\draw[mfwire] ($(x-ux)+(.3,0)$)--(x-f);
\MFdot{(x-f)}
\node[mflearn,align=center,anchor=west] (x-wq) at ($(x-f)+(.3,.54)$) {Linear\\\(W_{Q,x}\)};
\node[mflearn,align=center,anchor=west] (x-wk) at ($(x-f)+(.3,-.54)$) {Linear\\\(W_{K,x}\)};
\node[mfenc,right=3.0mm of x-wq] (x-rq) {RoPE\(_x\)};
\node[mfenc,right=3.0mm of x-wk] (x-rk) {RoPE\(_x\)};
\draw[mfflow] (x-f)|-(x-wq.west);
\draw[mfflow] (x-f)|-(x-wk.west);
\draw[mfflow] (x-wq.east)--(x-rq.west);
\draw[mfflow] (x-wk.east)--(x-rk.west);
\node[mfattn,anchor=west] (x-qk) at ($(x-rq.east)+(.95,-.54)$) {\(Q_xK_x^{T}\)};
\draw[mfflow] (x-rq.east)-|node[pos=.22,above,mfvar] {\(Q_x\)} node[pos=.62,above,mfshape] {\(N_x{\times}C_q\)} (x-qk.north);
\draw[mfflow] (x-rk.east)-|node[pos=.22,below,mfvar] {\(K_x\)} node[pos=.5,below,mfshape] {\(N_x{\times}C_q\)} (x-qk.south);
\coordinate[right=6.8mm of x-qk] (x-kap);
\begin{scope}[shift={(x-kap)}]
  \foreach \i in {0,...,5}{\foreach \j in {0,...,5}{
    \pgfmathtruncatemacro{\MFd}{abs(\i-\j)}
    \pgfmathtruncatemacro{\MFo}{max(8,70-20*\MFd)}
    \fill[MFAttnDark!\MFo!white] ({-.3+.1*\i},{.2-.1*\j}) rectangle ++(.1,.1);}}
  \draw[black!62,line width=.35pt,step=.1] (-.3,-.3) grid (.3,.3);
  \draw[black!62,line width=.4pt] (-.3,-.3) rectangle (.3,.3);
\end{scope}
\draw[mfflow] (x-qk.east)--($(x-kap)+(-.33,0)$);
\node[mfvar] at ($(x-kap)+(0,.52)$) {\(\kappa_x\)};
\node[mfshape] at ($(x-kap)+(0,-.52)$) {\(N_x{\times}N_x\)};
\node[mfattn,anchor=west] (x-sum) at ($(x-kap)+(.68,0)$) {\(\sum_{i'}\kappa_x(i,i')\,V_{i'jk}\)};
\draw[mfflow] ($(x-kap)+(.33,0)$)--(x-sum.west);
\MFcube{($(x-sum.south)+(-.08,-.86)$)}{.15}{MFLatent}
\node[mfvar,anchor=east] at ($(x-sum.south)+(-.3,-.84)$) {\(V\)};
\node[mfshape,anchor=west] at ($(x-sum.south)+(.24,-.86)$) {\(N_x{\times}N_y{\times}N_z{\times}C_v\)};
\draw[mfflow] ($(x-sum.south)+(0,-.62)$)--(x-sum.south);
\coordinate[right=5.4mm of x-sum] (x-vo);
\MFcube{($(x-vo)+(.17,0)$)}{.17}{MFLatent}
\draw[mfflow] (x-sum.east)--(x-vo);
\node[mfvar] at ($(x-vo)+(.25,.5)$) {\(V^{(x)}\)};
\end{tikzpicture}%
\end{lrbox}%
\MFoutput
\endgroup

%% file: figure/method/stride_parameterizations.tex
\begingroup
\input{figure/method/method_figure_style.tex}
\pgfdeclarelayer{MFback}\pgfsetlayers{MFback,main}
\providecommand{\MFskeleton}[4]{%
  \node[mfvar] (#1-in) at (.2,#2) {\(\widetilde{\uvec}\)};
  \ifnum#3=1 \node[mffixed,right=2.2mm of #1-in] (#1-slot) {Concat};
  \else \node[mffixed,draw=none,fill=none,right=2.2mm of #1-in] (#1-slot) {\phantom{Concat}};\fi
  \node[mflearn,align=center,right=3.6mm of #1-slot] (#1-lift) {Lifting\\\(+\,U_{\rm pos}\)};
  \coordinate[right=1.3mm of #1-lift] (#1-fin);
  \coordinate[right=3.0mm of #1-lift] (#1-fork);
  \node[mffixed,right=2.2mm of #1-fork] (#1-ln) {LN};
  \ifnum#4=1 \node[mfenc,right=2.4mm of #1-ln] (#1-filmA) {FiLM};
  \else \node[mfenc,draw=none,fill=none,right=2.4mm of #1-ln] (#1-filmA) {\phantom{FiLM}};\fi
  \node[mfattn,align=center,right=2.4mm of #1-filmA] (#1-rest) {Remaining\\PAI operations};
  \node[mfadd,right=8.0mm of #1-rest] (#1-sum) {\(+\)};
  \coordinate[right=2.6mm of #1-sum] (#1-fout);
  \coordinate (#1-loopout) at ($(#1-fout)+(-.08,0)$);
  \node[mffixed,right=8.0mm of #1-sum] (#1-gn) {GN};
  \ifnum#4=1 \node[mfenc,right=2.4mm of #1-gn] (#1-filmB) {FiLM};
  \else \node[mfenc,draw=none,fill=none,right=2.4mm of #1-gn] (#1-filmB) {\phantom{FiLM}};\fi
  \node[mflearn,align=center,right=2.4mm of #1-filmB] (#1-maps) {Channel\\maps};
  \node[mffixed,align=center,right=7.6mm of #1-maps] (#1-upd) {Denormalize,\\add to \(\uvec^n\)};
  \node[mfvar,right=2.2mm of #1-upd] (#1-out) {\(\widehat{\uvec}^{\,n+s}\)};
  \draw[mfgroup] ($(#1-ln.west)+(-.14,-.45)$) rectangle ($(#1-rest.east)+(.12,.74)$);
  \node[mfnote,anchor=north east,font=\MFsmall\bfseries] at ($(#1-rest.east)+(.1,.72)$) {PAI};
  \draw[mfgroup] ($(#1-gn.west)+(-.14,-.45)$) rectangle ($(#1-maps.east)+(.12,.74)$);
  \node[mfnote,anchor=north east,font=\MFsmall\bfseries] at ($(#1-maps.east)+(.1,.72)$) {Projection};
  \draw[mfwire] (#1-lift.east)--(#1-fork);
  \draw[mfflow] (#1-fork)--(#1-ln.west);
  \ifnum#4=1
    \draw[mfflow] (#1-ln.east)--(#1-filmA.west);
    \draw[mfflow] (#1-filmA.east)--(#1-rest.west);
    \draw[mfflow] (#1-gn.east)--(#1-filmB.west);
    \draw[mfflow] (#1-filmB.east)--(#1-maps.west);
  \else
    \draw[mfflow] (#1-ln.east)--(#1-rest.west);
    \draw[mfflow] (#1-gn.east)--(#1-maps.west);
  \fi
  \draw[mfflow] (#1-rest.east)--node[above,mfvar,pos=.56] {\(\times1/L\)} (#1-sum.west);
  \draw[mfflow] (#1-fork)--($(#1-fork)+(0,-.6)$)-|(#1-sum.south);
  \draw[mfloop] (#1-loopout)--($(#1-loopout)+(0,-.86)$)-|(#1-fin);
  \node[mfnote,fill=white,inner xsep=2pt] at ($($(#1-fin)!.5!(#1-fout)$)+(0,-.86)$) {\(U^{(\ell+1)}\); \(L\) iterations, shared parameters};
  \MFdot{(#1-fork)}\MFdot{(#1-fin)}\MFdot{(#1-loopout)}
  \draw[mfwire] (#1-sum.east)--(#1-fout);
  \draw[mfflow] (#1-fout)--node[above,mfvar,pos=.5,xshift=-1.6mm] {\(U^{(L)}\)} (#1-gn.west);
  \draw[mfflow] (#1-maps.east)--node[above,mfvar,pos=.77] {\(\widetilde{\boldsymbol r}_s\)} (#1-upd.west);
  \draw[mfflow] (#1-upd.east)--(#1-out.west);
}
\providecommand{\MFcard}[3][0]{%
  \begin{pgfonlayer}{MFback}\draw[mfiff] ($(#2-lift.west|-0,#3)+(-.2+.07*#1,-1.1+.12*#1)$) rectangle ($(#2-maps.east|-0,#3)+(.26+.07*#1,.88+.12*#1)$);\end{pgfonlayer}}
\providecommand{\MFcardlabel}[3]{%
  \node[mfnote,anchor=south east,text=MFIFF] at ($(#1-maps.east|-0,#2)+(.18,-1.06)$) {#3};}
\begin{lrbox}{\MFbox}%
\begin{tikzpicture}[mf]
\def\Ya{-1.86}
\node[mftitle] at (0,.19) {(a) SC-IFF-Concat: Append a stride channel to the input};
\MFskeleton{a}{\Ya}{1}{0}
\MFcard{a}{\Ya}
\draw[mfflow] (a-in.east)--(a-slot.west);
\draw[mfflow] (a-slot.east)--(a-lift.west);
\MFcardlabel{a}{\Ya}{one predictor, \(\Theta\) shared by all \(s\in\mathcal S\)}
\node[mfenc,align=center] (a-rho) at ($(a-slot)+(0,1.36)$) {Broadcast\\\(\rho(s)=s/s_{\max}\)};
\draw[mfflow] (a-rho.south)--(a-slot.north);
\def\Yb{-5.12}
\node[mftitle] at (0,\Yb+1.74) {(b) SC-IFF-Indep: Select a separate predictor for each native stride};
\MFskeleton{b}{\Yb}{0}{0}
\MFcard[2]{b}{\Yb}\MFcard[1]{b}{\Yb}\MFcard{b}{\Yb}
\draw[mfflow] (b-in.east)--(b-lift.west);
\MFcardlabel{b}{\Yb}{separate predictors \(\mathcal G_s(\cdot\,;\theta_s)\), \(s\in\mathcal S\)}
\node[mfenc,align=center] (b-sel) at ($(b-slot)+(0,1.12)$) {Select \(\theta_s\)};
\draw[mfflow] (b-sel.south)|-node[pos=.75,above,mfvar] {\(\theta_s\)} ($(b-lift.west|-0,\Yb)+(-.2,.2)$);
\node[mfnote,anchor=north west] at ($(b-lift.west|-0,\Yb)+(-.2,-1.18)$)
  {Initialization: \(\theta_{s_{\max}}\to\theta_{s_{\max}/2}\to\cdots\to\theta_1\); each predictor is then trained on its own stride.};
\def\Yc{-9.3}
\node[mftitle] at (0,\Yc+2.15) {(c) SC-IFF-FiLM: Modulate features with the stride};
\MFskeleton{c}{\Yc}{0}{1}
\MFcard{c}{\Yc}
\draw[mfflow] (c-in.east)--(c-lift.west);
\MFcardlabel{c}{\Yc}{one predictor, \(\Theta\) shared by all \(s\in\mathcal S\)}
\node[mfenc] (c-code) at ($(c-slot)+(0,1.46)$) {\(\boldsymbol c(s)\)};
\node[mfenc,align=center] (c-cond) at ($(c-lift)+(0,1.46)$) {Cond\\(MLP)};
\draw[mfflow] (c-code.east)--(c-cond.west);
\draw[mfflow] (c-cond.east|-0,\Yc+1.25)-|(c-filmA.north);
\draw[mfflow] (c-cond.east|-0,\Yc+1.68)-|(c-filmB.north);
\node[mfvar,anchor=west] at ($(c-filmA.north|-0,\Yc+1.25)+(.1,0)$) {\((\boldsymbol\gamma_s^{\rm FA},\boldsymbol\beta_s^{\rm FA})\), reused in all \(L\) iterations};
\node[mfvar,anchor=west] at ($(c-filmB.north|-0,\Yc+1.68)+(.1,0)$) {\((\boldsymbol\gamma_s^{\rm out},\boldsymbol\beta_s^{\rm out})\)};
\end{tikzpicture}%
\end{lrbox}%
\MFoutput
\endgroup

%% file: figure/method/stride_phase_sequences.tex
\begingroup
\input{figure/method/method_figure_style.tex}
\def\MFxo{2.25}\def\MFdx{1.05}
\def\MFnmax{12}
\providecommand{\MFlane}[3]{%
  \draw[mfgroup,fill=black!2] (1.86,#1-.35) rectangle (16.0,#1+.37);
  \node[mfnote,anchor=east] at (1.76,#1-.04) {\(\tau_0=#3\)};
  \foreach \MFn in {#3,...,\MFnmax}{%
    \pgfmathtruncatemacro{\MFr}{mod(\MFn-#3,#2)}%
    \ifnum\MFr=0
      \pgfmathsetmacro{\MFx}{\MFxo+\MFn*\MFdx}%
      \MFcube{(\MFx,#1-.12)}{.13}{MFField}
      \node[font=\MFsmall,inner sep=0pt] at (\MFx+.03,#1+.2) {\(\uvec^{\MFn}\)};
      \pgfmathtruncatemacro{\MFm}{\MFn+#2}%
      \ifnum\MFm>\MFnmax
        \draw[mfflow,black!70] (\MFx+.28,#1-.12)--(15.3,#1-.12);
        \node at (15.62,#1-.12) {\(\cdots\)};
      \else
        \draw[mfflow,black!80] (\MFx+.27,#1-.12)--(\MFx+#2*\MFdx-.19,#1-.12);
      \fi
    \fi}}
\begin{lrbox}{\MFbox}%
\begin{tikzpicture}[mf]
\node[mfhead,anchor=west] at (0,6.06) {\(s=1\)};
\MFlane{6.1}{1}{0}
\node[mfhead,anchor=west] at (0,4.81) {\(s=2\)};
\MFlane{4.85}{2}{0}
\MFlane{4.07}{2}{1}
\node[mfhead,anchor=west] at (0,2.78) {\(s=4\)};
\MFlane{2.82}{4}{0}
\MFlane{2.04}{4}{1}
\MFlane{1.26}{4}{2}
\MFlane{.48}{4}{3}
\node[font=\MFsmall,text=black!75,inner sep=1pt] at (\MFxo+.5*\MFdx+.03,6.1+.08) {\(\Delta T\)};
\node[font=\MFsmall,text=black!75,inner sep=1pt] at (\MFxo+1.0*\MFdx+.03,4.85+.08) {\(2\Delta T\)};
\node[font=\MFsmall,text=black!75,inner sep=1pt] at (\MFxo+2.0*\MFdx+.03,2.82+.08) {\(4\Delta T\)};
\draw[->,line width=.5pt,black!60] (1.86,-.18)--(16.0,-.18);
\foreach \MFn in {0,...,\MFnmax}{
  \draw[black!60,line width=.5pt] (\MFxo+\MFn*\MFdx,-.18)--++(0,.08);
  \node[font=\MFsmall,text=black!70,anchor=north,inner sep=1.5pt] at (\MFxo+\MFn*\MFdx,-.2) {\MFn};}
\node[mfnote,anchor=east] at (1.76,-.33) {index \(n\)};
\end{tikzpicture}%
\end{lrbox}%
\MFoutput
\endgroup

%% file: figure/method/cross_stride_fusion.tex
\begingroup
\input{figure/method/method_figure_style.tex}
\begin{lrbox}{\MFbox}%
\begin{tikzpicture}[mf]
\def\MFpx{2.3}\def\MFpd{1.42}
\def\MFpy{4.12}
\node[mftitle] at (0,\MFpy+2.62) {(a) Base field and learned correction for one query stride (example: \(s_{\max}=8\), \(s=6\))};
\draw[->,line width=.5pt,black!60] (\MFpx-.25,\MFpy)--(\MFpx+8*\MFpd+.55,\MFpy);
\node[mfnote,anchor=west] at (\MFpx+8*\MFpd+.6,\MFpy) {stride \(s\)};
\foreach \s in {0,...,8}{
  \draw[black!60,line width=.5pt] (\MFpx+\s*\MFpd,\MFpy)--++(0,.08);
  \node[font=\MFsmall,text=black!70,anchor=north,inner sep=1.5pt] at (\MFpx+\s*\MFpd,\MFpy-.02) {\s};}
\draw[black!40,line width=.6pt] (\MFpx+2*\MFpd+.04,\MFpy-.44)--++(0,-.1)-|(\MFpx+4*\MFpd-.04,\MFpy-.44);
\node[mfnote,anchor=north] at (\MFpx+3*\MFpd,\MFpy-.56) {\((s_-,s_+)=(2,4)\) for \(s=3\)};
\draw[MFEncode!55!black,line width=.8pt] (\MFpx+4*\MFpd+.04,\MFpy-.44)--++(0,-.1)-|(\MFpx+8*\MFpd-.04,\MFpy-.44);
\node[mfnote,anchor=north,text=MFEncode!40!black] at (\MFpx+6*\MFpd,\MFpy-.56) {\((s_-,s_+)=(4,8)\) for \(s=5,6,7\)};
\coordinate (p0) at (\MFpx,\MFpy+.95);
\coordinate (p1) at (\MFpx+1*\MFpd,\MFpy+1.3);
\coordinate (p2) at (\MFpx+2*\MFpd,\MFpy+.98);
\coordinate (p4) at (\MFpx+4*\MFpd,\MFpy+1.78);
\coordinate (p8) at (\MFpx+8*\MFpd,\MFpy+.72);
\coordinate (pl) at ($(p4)!.5!(p8)$);
\coordinate (pc) at ($(pl)+(0,.8)$);
\draw[black!55,line width=.7pt,dash pattern=on 3pt off 2pt] (p1)--(p2)--(p4)--(p8);
\draw[densely dotted,black!35,line width=.4pt] (pl)--(\MFpx+6*\MFpd,\MFpy);
\node[font=\MFsmall,text=black!65,inner sep=1pt,align=center] at (\MFpx+3.2*\MFpd,\MFpy+.66) {piecewise-linear\\base field};
\fill (p0) circle (.075);
\node[mfvar,anchor=east] at ($(p0)+(-.1,0)$) {\(\uvec^n\)};
\node[font=\MFsmall,anchor=west,inner sep=1pt,text=black!70] at ($(p0)+(.12,-.28)$) {current field};
\foreach \q/\s/\pos in {p1/1/above,p2/2/below,p4/4/above,p8/8/below}{
  \fill[MFIFF] (\q) circle (.075);
  \node[font=\MFsmall,text=MFIFF,\pos=2pt,inner sep=1pt] at (\q) {\(\mathcal G_{\s}(\uvec^n)\)};}
\draw[black!70,fill=white,line width=.6pt] (pl) circle (.075);
\node[font=\MFsmall,anchor=north east,inner sep=1pt] at ($(pl)+(-.08,-.06)$) {\(\widehat{\uvec}^{\,n+s}_{\rm lin}\)};
\draw[->,draw=MFCSF,line width=.8pt] ($(pl)+(0,.1)$)--($(pc)+(0,-.1)$);
\fill[MFCSF] (pc) circle (.08);
\node[font=\MFsmall,anchor=east,inner sep=1pt,text=MFCSF] at ($(pc)+(-.12,.02)$) {\(\widehat{\uvec}^{\,n+s}\)};
\node[font=\MFsmall,anchor=west,inner sep=1pt,text=MFCSF,align=left] at ($(pl)!.5!(pc)+(.14,0)$)
  {learned correction\\\(s\Delta T\,\boldsymbol\sigma_s^r\odot\widetilde{\boldsymbol r}_s^{\mathrm c}\)};
\coordinate (pe) at ($(p8)+(0,.72)$);
\draw[->,draw=MFCSF,line width=.8pt] ($(p8)+(0,.1)$)--($(pe)+(0,-.15)$);
\draw[black!80,line width=.55pt,fill=white] (pe) circle (.135);
\fill[MFCSF] (pe) circle (.08);
\node[font=\MFsmall,anchor=west,inner sep=1pt,text=MFCSF] at ($(pe)+(.2,0)$) {\(\widehat{\uvec}^{\,n+s_{\max}}\), fed back};
\node[mftitle] at (0,2.92) {(b) Reconstruction of \(\widehat{\uvec}^{\,n+s}\) for a query stride \(s\in\mathcal T\)};
\def\MFyB{1.4}
\def\MFyS{.2}
\def\MFyM{-.75}
\def\MFyT{-1.7}
\def\MFyE{-2.5}
\node[mfvar] (b-s) at (.2,2.2) {\(s\)};
\node[mfenc,align=center,anchor=west] (b-router) at (.6,2.2) {Router\\\(s\mapsto(s_-,s_+)\)};
\draw[mfiff] (.62,-1.2) rectangle (2.86,1.5);
\node[align=center,text=MFIFF] at (1.74,.1) {\textbf{IFF predictors}\\\(\mathcal G_{s_-},\ \mathcal G_{s_+}\)\\[2pt]{\MFsmall frozen \(\Theta^*\)}};
\MFsnow{(2.6,1.24)}{.12}
\draw[mfflow] (b-s.east)--(b-router.west);
\draw[mfflow] (b-router.south-|1.74,0)--node[right,mfvar] {\(s_\pm\)} (1.74,1.5);
\node[mfvar] (b-u) at (.2,-.3) {\(\uvec^n\)};
\coordinate (b-uf) at (.42,-.3);
\draw[mfwire] (b-u.east)--(b-uf);
\draw[mfflow] (b-uf)--(.62,-.3);
\MFdot{(b-uf)}
\draw[mfcsf,fill=none] (3.0,-3.1) rectangle (16.3,1.95);
\node[mfhead,anchor=west,text=MFCSF,fill=white,inner xsep=3pt] at (3.2,1.95) {FA-CSF};
\node[font=\MFsmall,anchor=south west,inner sep=1pt] at (3.04,\MFyB+.02) {\(\mathcal G_{s_\pm}(\uvec^n)\)};
\node[font=\MFsmall,anchor=south west,inner sep=1pt] at (3.04,\MFyS+.02) {\(U^{(L)}_{s_\pm}\)};
\node[font=\MFsmall,anchor=south west,inner sep=1pt] at (3.04,\MFyM+.02) {\(\widetilde{\boldsymbol r}_{s_\pm}\)};
\node[mffixed,align=center,anchor=west] (b-int) at (4.16,\MFyB) {Linear\\interpolation};
\draw[mfflow] (2.86,\MFyB)--(b-int.west);
\draw[mfflow] (b-router.east)-|node[pos=.3,above,mfvar] {\(s,\ s_\pm\)} (b-int.north);
\node[mfadd] (b-add) at (15.62,\MFyB) {\(+\)};
\draw[mfflow] (b-int.east)--node[above,mfvar,pos=.5] {\(\widehat{\uvec}^{\,n+s}_{\rm lin}\)} (b-add.west);
\node[mfnote,anchor=north] at ($(b-int.east)!.72!(b-add.west)+(0,-.06)$) {parameter-free base path};
\node[mfvar,anchor=south] (b-out) at (15.62,2.2) {\(\widehat{\uvec}^{\,n+s}\)};
\draw[mfflow] (b-add.north)--(b-out.south);
\node[mffixed,anchor=west] (b-lnz) at (4.16,\MFyS) {LN};
\node[mflearn,right=2.6mm of b-lnz] (b-proj) {Linear};
\node[mffixed,align=center,anchor=west] (b-norm) at (4.16,\MFyT) {Normalize};
\node[mffixed,minimum height=2.4cm,anchor=west] (b-cat) at (6.42,-.75) {Concat};
\draw[mfflow] (2.86,\MFyS)--(b-lnz.west);
\draw[mfflow] (b-lnz.east)--(b-proj.west);
\draw[mfflow] (b-proj.east)--(b-cat.west|-0,\MFyS);
\draw[mfflow] (2.86,\MFyM)--(b-cat.west|-0,\MFyM);
\draw[mfflow] (b-uf)|-(b-norm.west);
\draw[mfflow] (b-norm.east)--node[above,mfvar] {\(\widetilde{\uvec}\)} (b-cat.west|-0,\MFyT);
\coordinate (b-f) at (8.3,\MFyM);
\draw[mfwire] (b-cat.east|-0,\MFyM)--node[above,mfvar,pos=.48] {\(U_{\rm route}\)} (b-f);
\MFdot{(b-f)}
\node[mflearn,align=center,anchor=west] (b-sl) at (8.6,\MFyS) {Linear,\\GELU};
\node[mfattn,right=3.2mm of b-sl] (b-pai) {PAI};
\node[mffixed,anchor=west] (b-tc) at (8.6,\MFyT) {Concat};
\node[mflearn,align=center,anchor=west] (b-tl) at (b-pai.west|-0,\MFyT) {Linear,\\GELU};
\node[mfenc,anchor=west] (b-emb) at (b-tc.west|-0,\MFyE) {MLP\((\boldsymbol c(s))\), broadcast};
\node[mfvar] (b-s2) at (7.55,\MFyE) {\(s\)};
\draw[mfflow] (b-s2.east)--(b-emb.west);
\draw[mfflow] (b-f)|-(b-sl.west);
\draw[mfflow] (b-f)|-(b-tc.west);
\draw[mfflow] (b-sl.east)--(b-pai.west);
\draw[mfflow] (b-tc.east)--(b-tl.west);
\draw[mfflow] (b-emb.north-|b-tc)--(b-tc.south);
\node[mfnote,anchor=south] at ($(b-sl.north)!.5!(b-pai.north)+(0,.04)$) {spatial branch};
\node[mfnote,anchor=west] at ($(b-emb.east)+(.12,.12)$) {temporal branch};
\node[mfadd] (b-sum) at (11.55,\MFyM) {\(+\)};
\draw[mfflow] (b-pai.east)-|node[pos=.3,above,mfvar] {\(U_{\rm sp}\)} (b-sum.north);
\draw[mfflow] (b-tl.east)-|node[pos=.3,below,mfvar] {\(U_{\rm tm}\)} (b-sum.south);
\node[mffixed,right=3.2mm of b-sum] (b-ln2) {LN};
\node[mflearn,align=center,right=2.6mm of b-ln2] (b-lin) {Linear};
\node[mffixed,align=center,anchor=center] (b-scale) at (15.62,\MFyM) {\(\times\,s\Delta T\)\\\(\times\,\boldsymbol\sigma_s^r\)};
\draw[mfflow] (b-sum.east)--(b-ln2.west);
\draw[mfflow] (b-ln2.east)--(b-lin.west);
\draw[mfflow] (b-lin.east)--node[above,mfvar,pos=.64] {\(\widetilde{\boldsymbol r}_s^{\mathrm c}\)} (b-scale.west);
\draw[mfflow] (b-scale.north)--(b-add.south);
\node[mfnote,anchor=north] (b-zi) at (b-lin.south) {zero-initialized};
\draw[mfgroup] (3.98,\MFyE-.45) rectangle ($(b-zi.east|-0,.95)+(.12,0)$);
\node[mfnote,anchor=north east,font=\MFsmall\bfseries] at ($(b-zi.east|-0,.93)+(.08,0)$) {Learned correction (\(\phi\))};
\end{tikzpicture}%
\end{lrbox}%
\MFoutput
\endgroup

%% file: figure/method/autoregressive_trajectory.tex
\begingroup
\input{figure/method/method_figure_style.tex}
\def\MFxo{.62}\def\MFdx{.92}
\colorlet{MFIFFlight}{MFIFF!38}
\providecommand{\MFnode}[4]{%
  \pgfmathsetmacro{\MFx}{\MFxo+(#1)*\MFdx}%
  \ifnum#4=1 \draw[black!80,line width=.55pt,fill=white] (\MFx,#2) circle (.135);\fi
  \fill[#3] (\MFx,#2) circle (.085);}
\providecommand{\MFarc}[5][MFIFF]{%
  \pgfmathsetmacro{\MFxa}{\MFxo+(#2)*\MFdx}\pgfmathsetmacro{\MFxb}{\MFxo+(#3)*\MFdx}%
  \draw[->,draw=#1,line width=.6pt] (\MFxa+.04,#4+.1) .. controls (\MFxa+.12,#4+#5) and (\MFxb-.12,#4+#5) .. (\MFxb-.02,#4+.12);}
\providecommand{\MFaxis}[1]{%
  \draw[line width=.5pt,black!55] (\MFxo-.2,#1)--(\MFxo+16*\MFdx+.3,#1);
  \foreach \MFn in {0,...,4,6,7,...,12,14,15,16}{\draw[black!55,line width=.45pt] (\MFxo+\MFn*\MFdx,#1)--++(0,-.07);}
  \foreach \MFn in {5,13}{\node[fill=white,inner xsep=1.5pt,inner ysep=1pt,text=black!60] at (\MFxo+\MFn*\MFdx,#1) {\(\cdots\)};}
  \node[black!60] at (\MFxo+16*\MFdx+.62,#1) {\(\cdots\)};}
\begin{lrbox}{\MFbox}%
\begin{tikzpicture}[mf]
\def\Ya{-.98}\def\Yb{-3.7}\def\Yc{-7.04}
\foreach \MFn in {8,16}{\foreach \MFya/\MFyb in {\Ya-.2/\Ya+.62,\Yb-.2/\Yb+1.42,\Yc/\Yc+.38}{
  \draw[black!30,line width=.5pt,dash pattern=on 2pt off 2pt] (\MFxo+\MFn*\MFdx,\MFya)--(\MFxo+\MFn*\MFdx,\MFyb);}}
\node[mftitle,anchor=south west,align=left,text width=16.25cm] at (0,-.16) {(a) Fixed-stride IFF: One evaluation per output; every output is fed back (\(s_{\max}\) IFF evaluations per \(s_{\max}\Delta T\))};
\MFaxis{\Ya}
\foreach \MFb in {0,8}{
  \foreach \MFn in {0,1,2,3,6,7}{\MFarc{\MFb+\MFn}{\MFb+\MFn+1}{\Ya}{.42}}
  \MFarc[MFIFFlight]{\MFb+4}{\MFb+4.72}{\Ya}{.36}
  \MFarc[MFIFFlight]{\MFb+5.28}{\MFb+6}{\Ya}{.36}}
\node[font=\MFsmall,text=MFIFF,anchor=south,inner sep=1pt] at (\MFxo+.5*\MFdx,\Ya+.36) {\(\mathcal G_1\)};
\MFnode{0}{\Ya}{black}{0}
\foreach \MFn in {1,2,3,4,6,7,...,12,14,15,16}{\MFnode{\MFn}{\Ya}{MFIFF}{1}}
\node[mftitle,anchor=south west,align=left,text width=16.25cm] at (0,\Yb+1.59) {(b) Hierarchical SC-IFF: Finer strides start from predicted fields; feedback every \(s_{\max}\Delta T\)\\(\(s_{\max}\) IFF evaluations per \(s_{\max}\Delta T\))};
\MFaxis{\Yb}
\foreach \MFb in {0,8}{
  \MFarc{\MFb}{\MFb+8}{\Yb}{1.42}
  \MFarc{\MFb}{\MFb+4}{\Yb}{.98}
  \MFarc{\MFb}{\MFb+2}{\Yb}{.66}
  \MFarc[MFIFFlight]{\MFb+5.12}{\MFb+6}{\Yb}{.56}
  \foreach \MFk in {0,2}{\MFarc{\MFb+\MFk}{\MFb+\MFk+1}{\Yb}{.4}}
  \MFarc[MFIFFlight]{\MFb+4}{\MFb+4.72}{\Yb}{.36}
  \MFarc{\MFb+6}{\MFb+7}{\Yb}{.4}}
\node[font=\MFsmall,text=MFIFF,anchor=south,inner sep=1pt] at (\MFxo+4*\MFdx,\Yb+1.16) {\(\mathcal G_{s_{\max}}\)};
\node[font=\MFsmall,text=MFIFF,anchor=south,inner sep=1pt] at (\MFxo+3.1*\MFdx,\Yb+.72) {\(\mathcal G_4\)};
\node[font=\MFsmall,text=MFIFF,anchor=south,inner sep=1pt] at (\MFxo+5.56*\MFdx,\Yb+.46) {\(\mathcal G_2\)};
\node[font=\MFsmall,text=MFIFF,anchor=south,inner sep=1pt] at (\MFxo+6.5*\MFdx,\Yb+.34) {\(\mathcal G_1\)};
\MFnode{0}{\Yb}{black}{0}
\foreach \MFn in {1,2,3,4,6,7,9,10,11,12,14,15}{\MFnode{\MFn}{\Yb}{MFIFF}{0}}
\foreach \MFn in {8,16}{\MFnode{\MFn}{\Yb}{MFIFF}{1}}
\node[mftitle,anchor=south west,align=left,text width=16.25cm] at (0,\Yc+2.34) {(c) SC-IFF + FA-CSF: Each native predictor once from the current field; feedback every \(s_{\max}\Delta T\)\\(\(|\mathcal S|\) IFF evaluations per \(s_{\max}\Delta T\))};
\MFaxis{\Yc}
\foreach \MFb in {0,8}{
  \pgfmathsetmacro{\MFxs}{\MFxo+\MFb*\MFdx}
  \draw[mfcsf,fill=MFCSFFill,line width=.6pt,rounded corners=1.5pt]
    (\MFxs+.84*\MFdx,\Yc+.42) rectangle (\MFxs+8.1*\MFdx,\Yc+.82);
  \node[font=\MFsmall,text=MFCSF,inner sep=0pt] at (\MFxs+4.47*\MFdx,\Yc+.62) {FA-CSF: interpolate and correct for every \(s\in\mathcal T\)};
  \draw[MFCSF,line width=.5pt] (\MFxs+.07,\Yc+.07)--(\MFxs+.84*\MFdx,\Yc+.62);
  \foreach \MFk in {1,2,3,4,6,7,8}{
    \draw[->,draw=MFCSF,line width=.5pt] (\MFxs+\MFk*\MFdx,\Yc+.42)--(\MFxs+\MFk*\MFdx,\Yc+.12);}
  \foreach \MFs/\MFh in {1/1.62,2/1.78,4/1.98,8/2.16}{
    \pgfmathsetmacro{\MFxe}{\MFxs+\MFs*\MFdx}
    \draw[->,draw=MFIFF,line width=.6pt] (\MFxs+.1,\Yc+.1) .. controls (\MFxs+.12+.1*\MFs*\MFdx,\Yc+\MFh) and (\MFxe,\Yc+\MFh) .. (\MFxe,\Yc+1.24);
    \draw[MFIFF,line width=.45pt] (\MFxe,\Yc+1.08)--(\MFxe,\Yc+.82);
    \draw[MFIFF,line width=.6pt,fill=white] (\MFxe,\Yc+1.16) circle (.075);}
  \node[text=MFIFF,inner sep=0pt] at (\MFxs+5*\MFdx,\Yc+1.16) {\(\cdots\)};}
\foreach \MFs in {1,2,4}{
  \node[font=\MFsmall,text=MFIFF,anchor=west,inner sep=1pt] at (\MFxo+\MFs*\MFdx+.08,\Yc+1.12) {\(\mathcal G_{\MFs}\)};}
\node[font=\MFsmall,text=MFIFF,anchor=east,inner sep=1pt] at (\MFxo+8*\MFdx-.1,\Yc+1.12) {\(\mathcal G_{s_{\max}}\)};
\MFnode{0}{\Yc}{black}{0}
\foreach \MFn in {1,2,3,4,6,7,9,10,11,12,14,15}{\MFnode{\MFn}{\Yc}{MFCSF}{0}}
\foreach \MFn in {8,16}{\MFnode{\MFn}{\Yc}{MFCSF}{1}}
\node[font=\MFsmall,anchor=north,text=black!70,inner sep=1pt] at (\MFxo,\Yc-.1) {\(t_n\)};
\node[font=\MFsmall,anchor=north,text=black!70,inner sep=1pt] at (\MFxo+8*\MFdx,\Yc-.1) {\(t_n+s_{\max}\Delta T\)};
\node[font=\MFsmall,anchor=north,text=black!70,inner sep=1pt] at (\MFxo+16*\MFdx,\Yc-.1) {\(t_n+2s_{\max}\Delta T\)};
\draw[{Bar[width=1.6mm]Latex[length=1.1mm,width=.9mm]}-{Latex[length=1.1mm,width=.9mm]Bar[width=1.6mm]},black!60,line width=.45pt] (\MFxo,\Yc-.52)--(\MFxo+\MFdx,\Yc-.52);
\node[font=\MFsmall,anchor=west,text=black!70,inner sep=1pt] at (\MFxo+\MFdx+.04,\Yc-.52) {\(\Delta T\)};
\def\Yl{\Yc-1.0}
\fill[black] (.18,\Yl) circle (.085);
\node[mfnote,anchor=west] at (.32,\Yl) {current field at \(t_n\)};
\fill[MFIFF] (2.62,\Yl) circle (.085);
\node[mfnote,anchor=west] at (2.76,\Yl) {output of an IFF predictor};
\fill[MFCSF] (6.06,\Yl) circle (.085);
\node[mfnote,anchor=west] at (6.2,\Yl) {output of FA-CSF};
\draw[MFIFF,line width=.6pt,fill=white] (8.5,\Yl) circle (.075);
\node[mfnote,anchor=west] at (8.64,\Yl) {native prediction used by FA-CSF};
\draw[black!80,line width=.55pt] (13.12,\Yl) circle (.135);\fill[black!45] (13.12,\Yl) circle (.085);
\node[mfnote,anchor=west] at (13.32,\Yl) {fed back as next input};
\end{tikzpicture}%
\end{lrbox}%
\MFoutput
\endgroup

%% file: figure/method/training_prediction_workflow.tex
\begingroup
\input{figure/method/method_figure_style.tex}
\tikzset{mfstage/.style={draw=black!30,line width=.55pt,rounded corners=3pt},
  mfstagetitle/.style={anchor=west,font=\MFtitlefont,fill=white,inner xsep=3pt,inner ysep=1pt}}
\providecommand{\MFblock}[4]{%
  \draw[mfgroup,fill=white] (#2-1.18,#3-.82) rectangle (#2+1.18,#3+.86);
  \node[mfiff,minimum width=2.06cm,minimum height=.62cm,inner sep=1pt,text=MFIFF] (#1-g) at (#2,#3+.4) {\hspace*{3.2mm}\(\mathcal G(\cdot,s;\Theta^*)\)};
  \MFsnow{($(#1-g.west)+(.27,0)$)}{.11}
  \node[mfcsf,minimum width=2.06cm,minimum height=.62cm,inner sep=1pt,text=MFCSF] (#1-c) at (#2,#3-.42) {FA-CSF \((\phi)\)};
  \draw[mfflow] (#1-g.south)--(#1-c.north);
  \node[mfnote,anchor=south west] at (#2-1.18,#3+.88) {#4};}
\begin{lrbox}{\MFbox}%
\begin{tikzpicture}[mf]
\draw[mfstage] (0,4.8) rectangle (16.3,8.0);
\node[mfstagetitle] at (.2,8.0) {Stage 1: SC-IFF predictors on \(M\)-step input--output pairs};
\node[mfnote,anchor=north east] at (16.15,7.82) {update \(\Theta\); select \(\Theta^*\) on validation rollouts by \(\mathcal E_{\rm spec}\) and \(\mathcal E_{\rm ac}\)};
\def\Yo{6.32}
\node[mfdata,align=center,minimum height=1.0cm] (s1-data) at (1.45,\Yo) {Training trajectories\\sampled every \(\Delta T\)};
\node[mffixed,align=center,minimum height=1.0cm] (s1-re) at (4.42,\Yo) {Reindex by stride\\Construct datasets};
\node[mfiff,align=center,minimum height=1.0cm,inner sep=4pt,text=MFIFF] (s1-g) at (7.95,\Yo) {\textbf{SC-IFF rollout}\\\(\mathcal G(\cdot,s;\Theta)\), \(M\) steps};
\node[mfdata,align=center,minimum height=1.0cm] (s1-loss) at (11.55,\Yo) {Relative field loss\\\(\mathcal L_{\rm data}\) or \(\mathcal L_{{\rm data},s}\)};
\draw[mfflow] (s1-data.east)--(s1-re.west);
\draw[mfflow] (s1-re.east)--node[above,mfvar] {\((\uvec^n,s)\)} node[below,font=\MFsmall,inner sep=1pt] {\(s\in\mathcal S\)} (s1-g.west);
\draw[mfflow] (s1-g.east)--node[above,mfvar] {\(\widehat{\uvec}^{\,n+ms}\)} (s1-loss.west);
\draw[mfflow] (s1-re.north)--++(0,.42)-|node[pos=.2,above,font=\MFsmall,inner sep=1pt] {reference outputs \(\uvec^{n+ms}\), \(m=1,\ldots,M\)} (s1-loss.north);
\draw[mfgrad] (s1-loss.south)--++(0,-.5)-|node[pos=.25,below,font=\MFsmall,text=MFGrad,inner sep=1pt] {\(\nabla_\Theta\)} ($(s1-g.south)+(.6,0)$);
\node[mfnote,anchor=west,align=left] at (13.0,\Yo-.05)
  {Concat, FiLM: one \(\Theta\),\\mixed-stride minibatches\\[2pt]Indep: \(\theta_{s_{\max}}\to\cdots\to\theta_1\),\\one stride at a time};
\draw[mfstage] (0,-.42) rectangle (16.3,4.52);
\node[mfstagetitle] at (.2,4.52) {Stage 2: FA-CSF on autoregressive rollouts};
\node[mfnote,anchor=north east] at (16.15,4.34) {update only \(\phi\); select \(\phi^*\) on validation rollouts by \(\mathcal E_{\rm spec}\) and \(\mathcal E_{\rm ac}\)};
\def\Yt{2.42}
\node[mfdata,align=center,minimum height=1.0cm] (s2-data) at (1.45,\Yt) {Training window\\\((\uvec^0,\uvec^1,\ldots,\uvec^{N_{\rm win}})\)};
\MFblock{b1}{4.45}{\Yt}{step 1}
\MFblock{b2}{7.9}{\Yt}{step 2}
\MFblock{bk}{11.9}{\Yt}{last step}
\draw[mfflow] (s2-data.east)--node[above,mfvar] {\(\uvec^0\)} (3.27,\Yt);
\draw[mfflow] (5.63,\Yt)--node[above,mfvar] {\(\widehat{\uvec}^{\,s_{\max}}\)} (6.72,\Yt);
\draw[mfflow] (9.08,\Yt)--(9.68,\Yt);
\node at (9.9,\Yt) {\(\cdots\)};
\draw[mfflow] (10.12,\Yt)--(10.72,\Yt);
\draw[mfflow,draw=MFIFF] ($(s1-g.south)+(-.6,0)$)--(7.35,3.72);
\draw[MFIFF,line width=.6pt] (4.7,3.72)--(12.15,3.72);
\foreach \MFx in {4.7,8.15,12.15}{\draw[mfflow,draw=MFIFF] (\MFx,3.72)--(\MFx,3.12);}
\MFdot{(7.35,3.72)}
\node[font=\MFsmall,anchor=east,text=MFIFF,inner sep=1pt] at (7.27,4.06) {validation-selected \(\Theta^*\), frozen};
\node[mfdata,minimum width=9.86cm,minimum height=.5cm] (s2-traj) at (7.45,.62)
  {predicted trajectory \(\widehat{\uvec}^{\,1},\ldots,\widehat{\uvec}^{\,N_{\rm win}}\), one field every \(\Delta T\)};
\foreach \MFx in {4.45,7.9,11.9}{\draw[mfflow,mfbridge] (\MFx-.45,\Yt-.82)--(\MFx-.45,.87);}
\node[mfdata,align=center,minimum height=.62cm] (s2-loss) at (14.35,.62)
  {\(\lambda_{\rm traj}\mathcal L_{\rm traj}+\mathcal L_{\rm stat}\)};
\draw[mfflow] (s2-traj.east)--(s2-loss.west|-s2-traj.east);
\draw[mfflow] (s2-data.south)|-(14.35,.08)--(s2-loss.south);
\node[font=\MFsmall,anchor=north,inner sep=1pt] at (7.45,.06) {reference \(\uvec^1,\ldots,\uvec^{N_{\rm win}}\)};
\draw[mfgrad,-,preaction={draw=white,solid,line width=2.6pt}] (s2-loss.north)--(14.35,1.3)--(4.95,1.3);
\foreach \MFx in {4.45,7.9,11.9}{\draw[mfgrad] (\MFx+.5,1.3)--(\MFx+.5,\Yt-.73);}
\node[font=\MFsmall,anchor=north,text=MFGrad,inner sep=1pt,fill=white] at (9.0,1.25) {\(\nabla_\phi\) through all steps};
\end{tikzpicture}%
\end{lrbox}%
\MFoutput
\endgroup

%% file: Revised_Manuscript__Clean_Version_.bbl
\begin{thebibliography}{99}

\bibitem{Pope2000TurbulentFlows}
  S.B. Pope,
  Turbulent Flows,
  Cambridge University Press, Cambridge, 2000.

\bibitem{BruntonNoackKoumoutsakos2020MLFluid}
  S.L. Brunton, B.R. Noack, P. Koumoutsakos,
  Machine learning for fluid mechanics,
  Annu. Rev. Fluid Mech. 52 (1) (2020) 477--508.

\bibitem{DuraisamyIaccarinoXiao2019AgeData}
  K. Duraisamy, G. Iaccarino, H. Xiao,
  Turbulence modeling in the age of data,
  Annu. Rev. Fluid Mech. 51 (1) (2019) 357--377.

\bibitem{BeckKurz2021Perspective}
  A. Beck, M. Kurz,
  A perspective on machine learning methods in turbulence modeling,
  GAMM-Mitt. 44 (1) (2021) e202100002.

\bibitem{YeungPope1988ParticleTracking}
  P.K. Yeung, S.B. Pope,
  An algorithm for tracking fluid particles in numerical simulations of homogeneous turbulence,
  J. Comput. Phys. 79 (2) (1988) 373--416.

\bibitem{FossellaBiferale2025MSDA}
  F. Fossella, L. Biferale, A. Carrassi, M. Cencini, V. Gupta,
  Multiscale data assimilation in turbulent models,
  Phys. Rev. E 113 (2) (2026) 024208.

\bibitem{Bieker2020DeepMPC}
  K. Bieker, S. Peitz, S.L. Brunton, J.N. Kutz, M. Dellnitz,
  Deep model predictive flow control with limited sensor data and online learning,
  Theor. Comput. Fluid Dyn. 34 (2020) 577--591.

\bibitem{KovachkiLiLiuAzizzadenesheliBhattacharyaStuartAnandkumar2023}
  N. Kovachki, Z. Li, B. Liu, K. Azizzadenesheli, K. Bhattacharya, A.M. Stuart, A. Anandkumar,
  Neural operator: learning maps between function spaces with applications to PDEs,
  J. Mach. Learn. Res. 24 (89) (2023) 1--97.

\bibitem{LuJinPangZhangKarniadakis2021}
  L. Lu, P. Jin, G. Pang, Z. Zhang, G.E. Karniadakis,
  Learning nonlinear operators via DeepONet based on the universal approximation theorem of operators,
  Nat. Mach. Intell. 3 (3) (2021) 218--229.

\bibitem{LiKovachkiAzizzadenesheliLiuBhattacharyaStuartAnandkumar2021}
  Z. Li, N. Kovachki, K. Azizzadenesheli, B. Liu, K. Bhattacharya, A.M. Stuart, A. Anandkumar,
  Fourier neural operator for parametric partial differential equations,
  in: International Conference on Learning Representations (ICLR), 2021.

\bibitem{Li2022Fourier}
  Z. Li, W. Peng, Z. Yuan, J. Wang,
  Fourier neural operator approach to large eddy simulation of three-dimensional turbulence,
  Theor. Appl. Mech. Lett. 12 (6) (2022) 100389.

\bibitem{Luo2024CompressibleRT}
  T. Luo, Z. Li, Z. Yuan, W. Peng, T. Liu, L. Wang, J. Wang,
  Fourier neural operator for large eddy simulation of compressible Rayleigh--Taylor turbulence,
  Phys. Fluids 36 (7) (2024) 075165.

\bibitem{ParkChoi2021Toward}
  J. Park, H. Choi,
  Toward neural-network-based large eddy simulation: application to turbulent channel flow,
  J. Fluid Mech. 914 (2021) A16.

\bibitem{GuanChattopadhyaySubelHassanzadeh2022Stable}
  Y. Guan, A. Chattopadhyay, A. Subel, P. Hassanzadeh,
  Stable a posteriori LES of 2D turbulence using convolutional neural networks: backscattering analysis and generalization to higher Re via transfer learning,
  J. Comput. Phys. 458 (2022) 111090.

\bibitem{Zhao2025LESnets}
  S. Zhao, Z. Li, B. Fan, Y. Wang, H. Yang, J. Wang,
  LESnets (large-eddy simulation nets): physics-informed neural operator for large-eddy simulation of turbulence,
  J. Comput. Phys. 537 (2025) 114125.

\bibitem{Peng2022AttentionEnhanced}
  W. Peng, Z. Yuan, J. Wang,
  Attention-enhanced neural network models for turbulence simulation,
  Phys. Fluids 34 (2) (2022) 025111.

\bibitem{Peng2023LinearAttentionFNO}
  W. Peng, Z. Yuan, Z. Li, J. Wang,
  Linear attention coupled Fourier neural operator for simulation of three-dimensional turbulence,
  Phys. Fluids 35 (1) (2023) 015106.

\bibitem{HaoWangSuYingDongLiuChengSongZhu2023GNOT}
  Z. Hao, Z. Wang, H. Su, C. Ying, Y. Dong, S. Liu, Z. Cheng, J. Song, J. Zhu,
  GNOT: a general neural operator transformer for operator learning,
  in: Proceedings of the 40th International Conference on Machine Learning, PMLR, vol. 202, 2023, pp. 12556--12569.

\bibitem{Li2023Scalable}
  Z. Li, D. Shu, A. Barati Farimani,
  Scalable transformer for PDE surrogate modeling,
  in: Advances in Neural Information Processing Systems, vol. 36, 2023, pp. 28010--28039.

\bibitem{WuLuoWangWangLong2024Transolver}
  H. Wu, H. Luo, H. Wang, J. Wang, M. Long,
  Transolver: a fast transformer solver for PDEs on general geometries,
  in: Proceedings of the 41st International Conference on Machine Learning, PMLR, vol. 235, 2024, pp. 53681--53705.

\bibitem{Li2024TransformerLES}
  Z. Li, T. Liu, W. Peng, Z. Yuan, J. Wang,
  A transformer-based neural operator for large-eddy simulation of turbulence,
  Phys. Fluids 36 (6) (2024) 065167.

\bibitem{DuKrishnapriyan2025EddyFormer}
  Y. Du, A.S. Krishnapriyan,
  EddyFormer: accelerated neural simulations of three-dimensional turbulence at scale,
  in: Advances in Neural Information Processing Systems, vol. 38, 2025, pp. 114474--114505.

\bibitem{LaiChenYangWangWangXu2026DynFormer}
  P. Lai, Y. Chen, D. Yang, R. Wang, F. Wang, H. Xu,
  From complex dynamics to DynFormer: rethinking transformers for PDEs,
  arXiv preprint arXiv:2603.03112 (2026).

\bibitem{Yang2024An}
  H. Yang, Z. Li, X. Wang, J. Wang,
  An implicit factorized transformer with applications to fast prediction of three-dimensional turbulence,
  Theor. Appl. Mech. Lett. 14 (6) (2024) 100527.

\bibitem{Yang2026Implicit}
  H. Yang, Y. Wang, J. Wang,
  Implicit factorized transformer approach to fast prediction of turbulent channel flows,
  Sci. China Phys. Mech. Astron. 69 (1) (2026) 214606.

\bibitem{Zou2025Uncertainty}
  X. Zou, Z. Li, Y. Wang, H. Yang, J. Wang,
  Uncertainty quantification and stability of neural operators for prediction of three-dimensional turbulence,
  J. Comput. Phys. 549 (2026) 114640.

\bibitem{Wang2024Prediction}
  Y. Wang, Z. Li, Z. Yuan, W. Peng, T. Liu, J. Wang,
  Prediction of turbulent channel flow using Fourier neural operator-based machine-learning strategy,
  Phys. Rev. Fluids 9 (8) (2024) 084604.

\bibitem{McCabe2023Towards}
  M. McCabe, P. Harrington, S. Subramanian, J. Brown,
  Towards stability of autoregressive neural operators,
  Trans. Mach. Learn. Res. (2023).

\bibitem{GonzalezDemoulinBernard2023Towards}
  F. Gonzalez, F.-X. Demoulin, S. Bernard,
  Towards long-term predictions of turbulence using neural operators,
  arXiv preprint arXiv:2307.13517 (2023).

\bibitem{Li2023LongTermIUFNO}
  Z. Li, W. Peng, Z. Yuan, J. Wang,
  Long-term predictions of turbulence by implicit U-Net enhanced Fourier neural operator,
  Phys. Fluids 35 (7) (2023) 075145.

\bibitem{WuZhangHe2025NOForcing}
  C. Wu, X.-L. Zhang, G. He,
  Neural operator-based stochastic forcing for resolvent prediction of space-time turbulence statistics in channel flows,
  J. Fluid Mech. 1024 (2025) A1.

\bibitem{Bengio2015ScheduledSampling}
  S. Bengio, O. Vinyals, N. Jaitly, N. Shazeer,
  Scheduled sampling for sequence prediction with recurrent neural networks,
  in: Advances in Neural Information Processing Systems, vol. 28, 2015, pp. 1171--1179.

\bibitem{Lamb2016ProfessorForcing}
  A.M. Lamb, A. Goyal, Y. Zhang, S. Zhang, A. Courville, Y. Bengio,
  Professor forcing: a new algorithm for training recurrent networks,
  in: Advances in Neural Information Processing Systems, vol. 29, 2016, pp. 4601--4609.

\bibitem{Chen2025Neural}
  C. Chen, J.-L. Wu,
  Neural dynamical operator: continuous spatial-temporal model with gradient-based and derivative-free optimization methods,
  J. Comput. Phys. 520 (2025) 113480.

\bibitem{Liu2022Hierarchical}
  Y. Liu, J.N. Kutz, S.L. Brunton,
  Hierarchical deep learning of multiscale differential equation time-steppers,
  Philos. Trans. R. Soc. A 380 (2229) (2022) 20210200.

\bibitem{Linot2023Stabilized}
  A.J. Linot, J.W. Burby, Q. Tang, P. Balaprakash, M.D. Graham, R. Maulik,
  Stabilized neural ordinary differential equations for long-time forecasting of dynamical systems,
  J. Comput. Phys. 474 (2023) 111838.

\bibitem{Abueidda2026Time}
  D.W. Abueidda, M. Nonna, P. Pantidis, M.E. Mobasher,
  Time resolution independent operator learning,
  Comput. Methods Appl. Mech. Eng. 450 (2026) 118586.

\bibitem{Nguyen2024Stormer}
  T. Nguyen, R. Shah, H. Bansal, T. Arcomano, R. Maulik, V. Kotamarthi, I. Foster, S. Madireddy, A. Grover,
  Scaling transformer neural networks for skillful and reliable medium-range weather forecasting,
  in: Advances in Neural Information Processing Systems, vol. 37, 2024, pp. 68740--68771.

\bibitem{Herde2024Poseidon}
  M. Herde, B. Raoni\'{c}, T. Rohner, R. K\"{a}ppeli, R. Molinaro, E. de B\'{e}zenac, S. Mishra,
  Poseidon: efficient foundation models for PDEs,
  in: Advances in Neural Information Processing Systems, vol. 37, 2024, pp. 72525--72624.

\bibitem{Leinonen2024ModAFNO}
  J. Leinonen, B. Bonev, T. Kurth, Y. Cohen,
  Modulated adaptive Fourier neural operators for temporal interpolation of weather forecasts,
  arXiv preprint arXiv:2410.18904 (2024).

\bibitem{LippeEtAl2023}
  P. Lippe, B.S. Veeling, P. Perdikaris, R.E. Turner, J. Brandstetter,
  PDE-Refiner: achieving accurate long rollouts with neural PDE solvers,
  in: Advances in Neural Information Processing Systems, vol. 36, 2023, pp. 67398--67433.

\bibitem{HuangPerdikaris2025PhysicsCorrect}
  X. Huang, P. Perdikaris,
  PhysicsCorrect: a training-free approach for stable neural PDE simulations,
  in: Proceedings of the AAAI Conference on Artificial Intelligence, vol. 40 (26), 2026, pp. 22057--22065.

\bibitem{Cao2025SpectralRefiner}
  S. Cao, F. Brarda, R.P. Li, Y. Xi,
  Spectral-Refiner: accurate fine-tuning of spatiotemporal Fourier neural operator for turbulent flows,
  in: International Conference on Learning Representations (ICLR), 2025.

\bibitem{MoinSquiresCabotLee1991Dynamic}
  P. Moin, K. Squires, W. Cabot, S. Lee,
  A dynamic subgrid-scale model for compressible turbulence and scalar transport,
  Phys. Fluids A 3 (11) (1991) 2746--2757.

\bibitem{NicoudDucros1999WALE}
  F. Nicoud, F. Ducros,
  Subgrid-scale stress modelling based on the square of the velocity gradient tensor,
  Flow Turbul. Combust. 62 (3) (1999) 183--200.

\bibitem{Smagorinsky1963}
  J. Smagorinsky,
  General circulation experiments with the primitive equations. I. The basic experiment,
  Mon. Weather Rev. 91 (3) (1963) 99--164.

\bibitem{Tancik2020FourierFeatures}
  M. Tancik, P.P. Srinivasan, B. Mildenhall, S. Fridovich-Keil, N. Raghavan, U. Singhal, R. Ramamoorthi, J.T. Barron, R. Ng,
  Fourier features let networks learn high frequency functions in low dimensional domains,
  in: Advances in Neural Information Processing Systems, vol. 33, 2020, pp. 7537--7547.

\bibitem{Su2024RoFormer}
  J. Su, M. Ahmed, Y. Lu, S. Pan, W. Bo, Y. Liu,
  RoFormer: enhanced transformer with rotary position embedding,
  Neurocomputing 568 (2024) 127063.

\bibitem{Perez2018FiLM}
  E. Perez, F. Strub, H. de Vries, V. Dumoulin, A. Courville,
  FiLM: visual reasoning with a general conditioning layer,
  in: Proceedings of the AAAI Conference on Artificial Intelligence, vol. 32 (1), 2018, pp. 3942--3951.

\bibitem{Jiang2023Invariant}
  R. Jiang, P.Y. Lu, E. Orlova, R. Willett,
  Training neural operators to preserve invariant measures of chaotic attractors,
  in: Advances in Neural Information Processing Systems, vol. 36, 2023, pp. 27645--27669.

\bibitem{Jacobs1991Adaptive}
  R.A. Jacobs, M.I. Jordan, S.J. Nowlan, G.E. Hinton,
  Adaptive mixtures of local experts,
  Neural Comput. 3 (1) (1991) 79--87.

\bibitem{Shazeer2017Outrageously}
  N. Shazeer, A. Mirhoseini, K. Maziarz, A. Davis, Q.V. Le, G.E. Hinton, J. Dean,
  Outrageously large neural networks: the sparsely-gated mixture-of-experts layer,
  in: International Conference on Learning Representations (ICLR), 2017.

\bibitem{Wang2025MixtureOfExperts}
  H. Wang, H. Xin, J. Wang, X. Yang, F. Zha, H. Dong, Y. Jiang,
  Mixture-of-experts operator transformer for large-scale PDE pre-training,
  in: Advances in Neural Information Processing Systems, vol. 38, 2025, pp. 35566--35595.

\bibitem{Novati2021Automating}
  G. Novati, H. Lascombes de Laroussilhe, P. Koumoutsakos,
  Automating turbulence modelling by multi-agent reinforcement learning,
  Nat. Mach. Intell. 3 (1) (2021) 87--96.

\end{thebibliography}
